\documentclass[11pt,twoside,a4paper,pdf]{article}
\pdfoutput=1
\usepackage[pdftex]{graphicx,color}
\usepackage{times}
\usepackage{xspace}
\usepackage{booktabs}
\usepackage[english]{babel}
\usepackage{subcaption}
\usepackage{amssymb}
\usepackage{cite}
\usepackage{xspace}
\usepackage{a4wide}
\usepackage{feynmp}
\usepackage{amsmath,multicol}
\usepackage{multirow}
\usepackage{jheppub}
\usepackage{float}
\usepackage{afterpage}

\newcommand{\mrm}[1]{\ensuremath{\mathrm{#1}}\xspace}
\newcommand{\mbf}[1]{\ensuremath{\mathbf{#1}}\xspace}
\newcommand{\ttt}[1]{\texttt{#1}}

\newcommand{\code}[1]{{\small\ttt{#1}}\xspace}

\newcommand{\eqRef}[1]{eq.~(\ref{#1})\xspace}

\newcommand{\eqsRef}[1]{eq.~(\ref{#1})\xspace}

\newcommand{\secRef}[1]{section~\ref{#1}\xspace}

\newcommand{\tabRef}[1]{tab.~\ref{#1}\xspace}

\newcommand{\figRef}[1]{fig.~\ref{#1}\xspace}

\newcommand{\inst}[1]{$^{#1}$}
\renewcommand{\and}{, }

\begin{document}
\unitlength = 1mm

\vspace*{-1.8cm}\begin{minipage}{\textwidth}
\flushright\footnotesize
COEPP-MN-15-1\\
LU-TP-15-16\\
MCNET-15-09\\
\end{minipage}
\vskip1.25cm
{\Large\bf
\begin{center}
String Formation Beyond Leading Colour
\end{center}}
\vskip5mm
{\begin{center}
{\large 
Jesper~R.~Christiansen\inst{1,2}\and 
Peter~Z.~Skands\inst{2,3}
}\end{center}\vskip3mm
\begin{center}
\parbox{0.9\textwidth}{
\inst{1}: Department of Astronomy and Theoretical Physics, Lund University, S\"olvegatan 14, Lund, Sweden\\
\inst{2}: Theoretical Physics, CERN, CH-1211,
Geneva 23, Switzerland\\
\inst{3}: School of Physics and Astronomy, Monash University, VIC-3800, Australia
}
\end{center}
\vskip5mm
\begin{center}
\parbox{0.85\textwidth}{
\begin{center}
\textbf{Abstract}
\end{center}\small
We present a new model for the hadronisation of multi-parton 
systems, in which colour correlations beyond leading $N_C$ are allowed
 to influence the formation of confining potentials (strings). The
 multiplet structure of $SU(3)$ is combined with a minimisation of the string
 potential energy, to decide between which partons strings should form,
 allowing also for  ``baryonic''  
 configurations (e.g., two colours can combine coherently to form an
 anticolour).    
In $e^+e^-$collisions, modifications to the leading-colour picture are
small, suppressed by both colour and kinematics factors. But in $pp$
collisions, multi-parton interactions increase the number of
 possible subleading connections, counteracting their naive $1/N_C^2$
 suppression. Moreover, those that reduce the overall string
lengths are kinematically favoured. The model, which we have
implemented in the PYTHIA 8 generator, is capable of
 reaching agreement not only with the important 
 $\left<p_\perp\right>(n_\mathrm{charged})$ distribution but also  
with measured rates (and
ratios) of kaons and hyperons, in both $ee$ and $pp$ 
collisions. Nonetheless, the shape of their $p_\perp$ spectra remains
challenging to explain.
} 
\end{center}
\vspace*{1cm}

\section{Introduction}

The description of hadronic final states at high-energy colliders 
involves a complicated cocktail of physics effects, dominated by
QCD~\cite{Dissertori:2003pj,Skands:2012ts,Buckley:2011ms}. For the
calculation of \emph{inclusive}  
hard-scattering cross sections, factorisation  allows 
most of the 
complicated long-distance physics to be represented in the
form of universal parton distribution
functions (PDFs)~\cite{Ball:2012wy}, while the short-distance parts can be
calculated perturbatively. Perturbative aspects, such as hard-process matrix elements, 
parton showers, and decay (chains) of short-lived resonances, are
generally coming under
increasingly good control, due to a combination of advances:
better amplitude calculations (including better automation and
better
interfaces~\cite{Boos:2001cv,Alwall:2006yp,vanHameren:2009dr,Cascioli:2011va,Alwall:2011uj,Alioli:2013nda,Cullen:2014yla,Alwall:2014hca}),   
better parton-shower 
algorithms (e.g.\ ones based on 
QCD
dipoles~\cite{Gustafson:1987rq,Sjostrand:2004ef,Nagy:2005aa,Giele:2007di,Dinsdale:2007mf,Schumann:2007mg,Platzer:2009jq}),
and better techniques for how to combine them 
(matching and merging,
see~\cite{Buckley:2011ms,Hamilton:2012np,Lonnblad:2012ix,Hartgring:2013jma,Alwall:2014hca} and
references therein).  
These successes build on an extensive
prior experience with perturbative approximations to QCD at both fixed and
infinite order, and the tractable nature of the perturbative expansions
themselves.

To describe the full (\emph{exclusive}) event structure, however, several
additional soft-physics effects must be accounted 
for, such as hadronisation, 
multiple parton interactions (MPI), Bose-Einstein correlation effects,
and beam remnants. These are connected with the rich structure of QCD beyond
perturbation theory and are vital, each in their own way, to 
the understanding of issues such as underlying-event/pileup effects on
isolation and accurate jet calibrations, and the interpretation of
identified-particle rates and spectra.  
 
For these aspects,
explicit calculations can only be performed in the context of
simplified phenomenological models, constructed so as to capture the
essential features of full (nonperturbative) QCD.   
An example relevant to this paper is the Lund string model of
hadronisation~\cite{Andersson:1998tv,Andersson:1983ia}, whose cornerstone is the observation 
that the static QCD potential between a quark 
and an antiquark in an overall colour-singlet state grows linearly with the
distance between them, for distances larger than about 0.5
fm~\cite{Bali:1992ab}. This is interpreted as a consequence of the
gluon field between the charges forming a high-tension ``string''
(with tension $\kappa\sim 1\,\mrm{GeV}/\mrm{fm}$), which subsequently
fragments into hadrons.  

While the details of the string-breaking process may be complicated
(the Lund model invokes quantum
tunnelling to describe this aspect~\cite{Andersson:1998tv}), the first
question that any hadronisation model needs to address is 
therefore simply: \emph{between which partons do confining potentials
  arise?} In string-based models, this is equivalent to answering the
question between which partons string pieces should be
formed. Traditionally, Monte Carlo event generators make use of the
leading-colour (LC) approximation to trace the colour flow on an
event-by-event basis (see~\cite{Buckley:2011ms,pdg2012}), leading to
partonic final states in which each quark is colour-connected to a
single (unique) other parton in the event (equivalent to a leading-colour QCD
dipole~\cite{Gustafson:1986db}). Gluons are represented as carrying both a
colour and an anticolour charge, and are hence each connected to two other
partons. At the level of strings, this is interpreted as 
gluons forming
transverse ``kinks'' on strings whose endpoints are quarks and
antiquarks~\cite{Andersson:1998tv}. Studies at 
$ee$ colliders show this to be a quite reasonable approximation in
that environment, and the traditional Lund string model, implemented in
PYTHIA~\cite{Sjostrand:2006za,Sjostrand:2007gs,Sjostrand:2014zea}, is capable of delivering a good
description of the vast majority of $ee$ collider data (for recent
studies, see,
e.g.,~\cite{Buckley:2009bj,Firdous:2013noa,Fischer:2014bja,Skands:2014pea}). 

The question of \emph{colour reconnections} (CR) --- broadly, whether other
string topologies than the LC one could lead to non-negligible
corrections with respect to the LC picture --- 
was studied at
LEP~\cite{Abbiendi:2003ri,Achard:2003pe,Achard:2003ik,Siebel:2005uw,Abbiendi:2005es,Schael:2006ns,Abdallah:2006uq,Schael:2006mz},
chiefly in the context of CR uncertainties on $W$ mass determinations in $ee\to 
WW$~\cite{Sjostrand:1993hi}, with conclusion that excluded the very aggressive
models and disfavoured the no CR scenario at 2.8 standard
deviation~\cite{Schael:2013ita}. The uncertainty on the $W$ mass from this 
source ended 
up at $\Delta m_W\sim 35\,\mrm{MeV}$, corresponding to 
about 0.05\%. 

There are strong physical reasons to think that
CR effects \emph{should} be highly suppressed at LEP, however. Firstly, there
is a ``trivial'' parametric suppression of  
beyond-LC effects of order $1/N_C^2\sim 10\%$. Secondly, 
the two $W$ decay systems are separate
colour-singlet systems, with a space-time separation of order of the inverse $W$
width, $\Gamma_W\sim 2\,\mrm{GeV}$. This separation implies that 
interference effects between the two systems should be highly
suppressed for wavelengths shorter than $1/\Gamma_W$, i.e., there can
be essentially no perturbative cross-talk between them. 
This line of argument motivated the phrasing of CR models that operate
only at the non-perturbative level  
as the most physically reasonable~\cite{Sjostrand:1993hi}, an
observation that we shall also adhere to in the present work. 
Thirdly, the QCD coherence of
perturbative parton cascades implies that, inside each $W$ (or $Z$)
decay system, angles of successive QCD emissions tend to be
ordered from large to small~\cite{Marchesini:1983bm}, so that there is
very little 
space-time overlap between the QCD dipoles inside each system. 
This means that,
even if one were to allow to set up confining potentials between
non-LC-connected partons, these would tend to correspond to
\emph{larger} opening angles and therefore they would have a higher
total potential energy (longer strings) than the equivalent LC
ones. The LC topology should therefore also be \emph{dynamically} favoured
over any possible non-LC ones. 
All these factors contribute to an
expectation of quite small effects, at least in the context of $e^+e^-$
collisions. 

Moving to $pp$ collisions (and using $pp$ as a shorthand to
  for any generic hadron-hadron collision, including in particular
  also $p\bar{p}$ ones), the situation changes
dramatically. Trivially, one must now include coloured initial-state
partons, with associated coloured beam remnants. But more importantly, 
the modern understanding of the underlying event (UE) and of
 soft-inclusive (minimum-bias/pileup) physics in general, especially at high
 particle multiplicities, is that
they are dominated by contributions from multiple parton interactions
(MPI)~\cite{Sjostrand:1987su}. In a $pp$ event that contains several MPI 
systems, there is a non-negligible possibility of phase-space overlaps between
final states from different MPI systems. Moreover, since 
the MPI scattering centres must all reside within the proton radius, which is
of the same order as the transverse size of QCD strings, the 
initial-state (beam) jets will all ``sit'' right on top of each other, a
situation which should affect the fragmentation especially at high
rapidities. Finally, unlike the case for
angular-ordered partons inside a jet, there is no perturbative principle that 
predisposes colour-connected partons from different MPI or
beam-remnant systems to have small opening angles; indeed a recent
study~\cite{Gieseke:2012ft}  
found that such ``inter-MPI/remnant'' invariant masses (denoted
$i$-type and $n$-type in~\cite{Gieseke:2012ft}) tend
to be among the largest in the events, corresponding to a high 
potential energy in a string context, and hence with the most to gain
from potential reconnections. For these reasons, we expect qualitatively larger
effects in $pp$ collisions. 

There are also tantalising hints from hadron-collider data that
nontrivial physics effects are present at the hadronisation stage in
$pp$ collisions. The most important such clue is furnished by 
the dependence of the average (charged) particle $p_\perp$ on the
particle multiplicity,
$\left<p_\perp\right>(n_\mrm{ch})$. Measurements of this quantity in 
minimum-bias events, first made at the ISR~\cite{Breakstone:1983up}
and since by UA1~\cite{Albajar:1989an},
CDF~\cite{Acosta:2001rm,Aaltonen:2009ne} and the LHC  
experiments~\cite{Khachatryan:2010nk,Aad:2010ac,Abelev:2013bla}, reveal that
$\left<p_\perp\right>$ grows with 
$n_\mrm{ch}$, as can be seen in the plots 
in \figRef{fig:avgptofnch} (from
\href{http://mcplots.cern.ch}{mcplots.cern.ch}~\cite{Karneyeu:2013aha}). 
\begin{figure}[t]
  \captionsetup[subfigure]{skip=3pt}
\centering
\subcaptionbox{\label{fig:avgtofncha}}{\!\includegraphics*[scale=0.26]{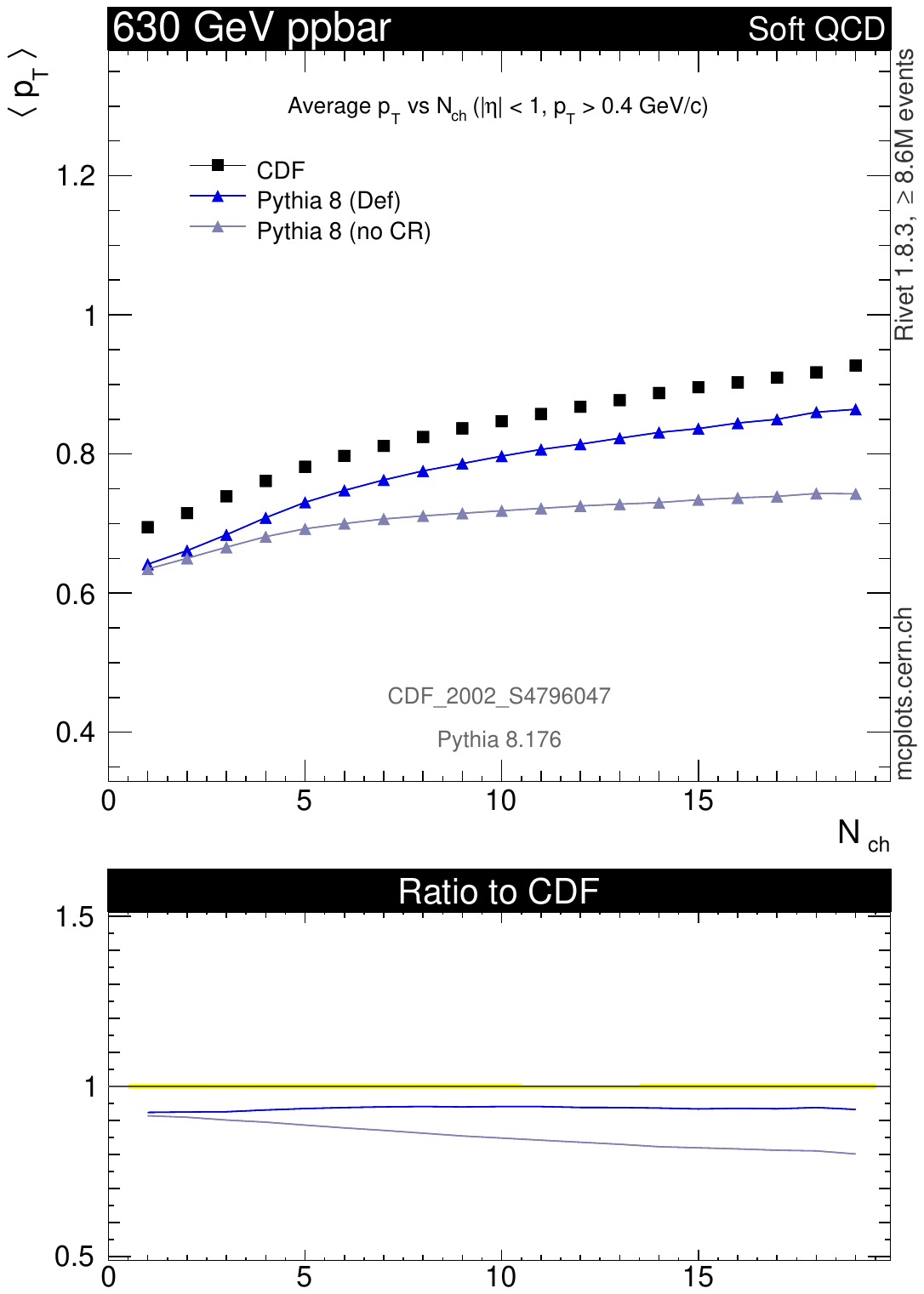}\!}
\subcaptionbox{\label{fig:avgtofnchb}}{\!\includegraphics*[scale=0.26]{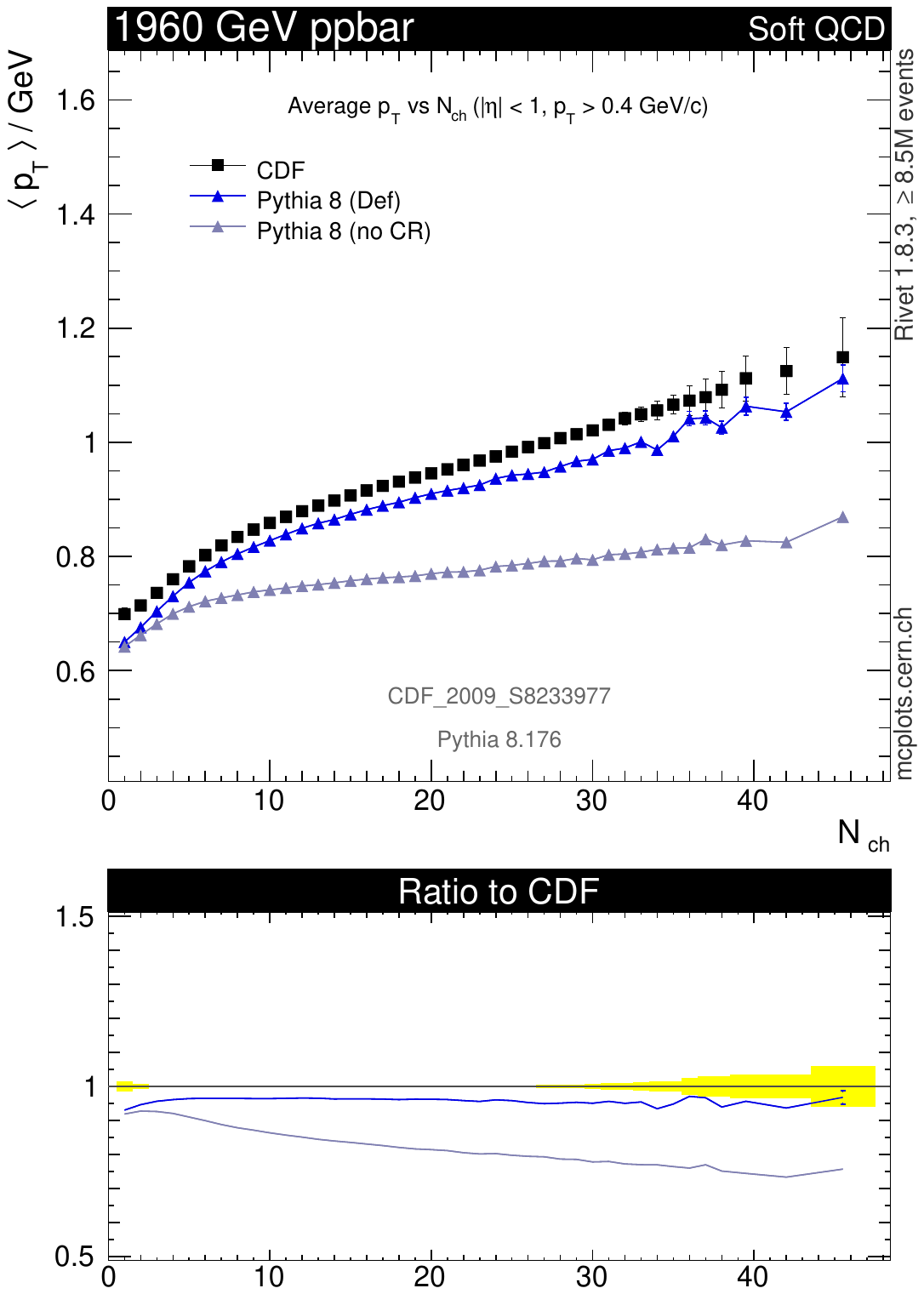}\!}
\subcaptionbox{\label{fig:avgtofnchc}}{\!\includegraphics*[scale=0.26]{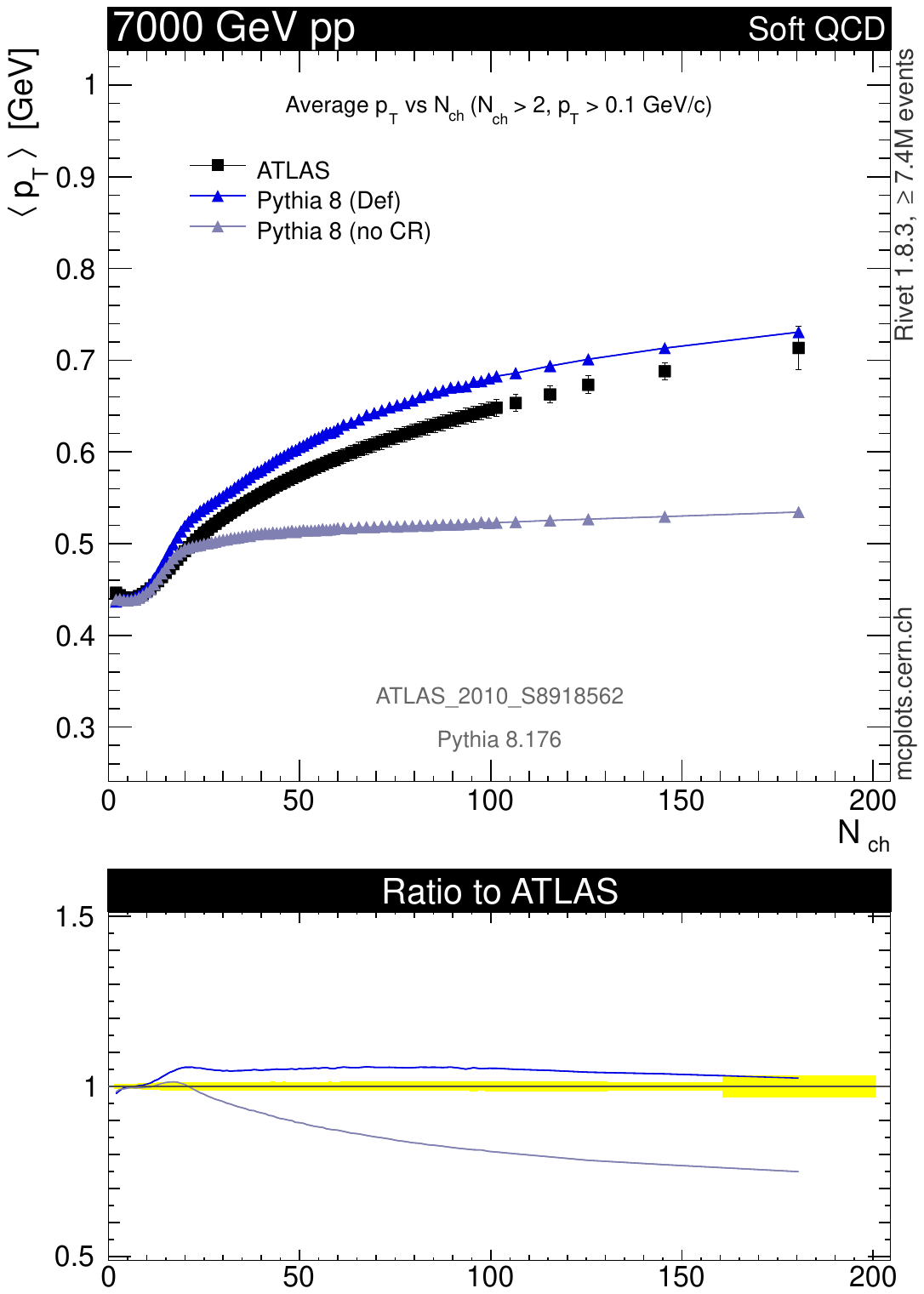}\!}
\caption{Measurements of $\left<p_\perp\right>(n_{Ch})$ in
  minimum-bias events at 630 GeV~\cite{Acosta:2001rm} (\subref{fig:avgtofncha}), 1960
  GeV~\cite{Aaltonen:2009ne} (\subref{fig:avgtofnchb}), and 7000
  GeV~\cite{Aad:2010ac} (\subref{fig:avgtofnchc}),
  compared to PYTHIA 8.175~\cite{Sjostrand:2007gs} (tune
  4C~\cite{Corke:2010yf}), with and without colour reconnections switched
  on. (Plots from \href{http://mcplots.cern.ch}{mcplots.cern.ch}~\cite{Karneyeu:2013aha}.)
\label{fig:avgptofnch}}
\end{figure} 
This cannot be accounted for by independently hadronising MPI systems,
for which the expectation would be that
$\left<p_\perp\right>(n_\mrm{ch})$ should be almost flat, as is also
illustrated by the ``no CR'' curves in \figRef{fig:avgptofnch}. (If
each MPI hadronises 
independently, then per-particle quantities such as 
$\left<p_\perp\right>$ should be independent of the number of MPI,
which is correlated with $n_\mrm{Ch}$~\cite{Sjostrand:1987su}.) 
The observation that $\left<p_\perp\right>$ increases with
$n_\mrm{Ch}$ therefore strongly suggests that some form of collective
hadronisation phenomenon is at play, correlating partons from
different MPI systems. 

Given these arguments, and the
realisation~\cite{Skands:2007zg} that precision 
kinematic extractions of the top quark mass at hadron colliders (see
e.g.,~\cite{Aaltonen:2012va,Chatrchyan:2012cz,Aaltonen:2013wca,Chatrchyan:2013xza,ATLAS:2014wva,Abazov:2014dpa,Aad:2015nba}
for experimental methods and \cite{Juste:2013dsa} for a recent
phenomenology review) can be significantly
affected by colour reconnections\footnote{For completeness we note
  that, similarly to above, much
  smaller effects are expected in $e^+e^-$
  environments~\cite{Khoze:1994fu,Khoze:1999up}.}, several toy models have
appeared~\cite{Rathsman:1998tp,Sandhoff:2005jh,Buttar:2006zd,Skands:2007zg,Gieseke:2012ft,Argyropoulos:2014zoa},
relying mainly   
on potential-energy minimisation arguments to reconfigure the
partonic colour connections for hadronisation. Although these models
have had some success in describing the
$\left<p_\perp\right>(n_\mrm{Ch})$ distribution
(as e.g., in~\figRef{fig:avgptofnch}), the lack of
rigorous underpinnings have implied
that large uncertainties remain, which still contribute about a 500
MeV uncertainty on the hadronic top mass
extraction~\cite{Juste:2013dsa,ATLAS:2014wva,Argyropoulos:2014zoa}. In
this paper, we take a first step 
towards creating a more realistic model, combining the earlier
string-length minimisation arguments with selection rules based on 
the colour algebra of $SU(3)$. Our treatment amounts to taking the LC
connections produced by the shower as a starting point, complemented
by an $SU(3)$-weighted randomization over the set of possible
subleading topologies that would have been present in a 
full-colour treatment. 
The missing colour information should thereby be restored,
at least in a statistical sense. 

An alternative line of argument, pursued in particular in the EPOS
model~\cite{Pierog:2013ria}, 
invokes the notion of hydrodynamic collective
flow to explain the 
$\left<p_\perp\right>(n_\mrm{Ch})$ distribution (as well as the
so-called CMS ``ridge
effect''~\cite{Khachatryan:2010gv,Werner:2010ss} and a host of other
$pp$ observables~\cite{Pierog:2013ria}). 
Certainly, the presence of hydro effects in $pp$ is a hypothesis that,
if confirmed, would have far-reaching consequences, and it will be an important
task for future 
experimental and phenomenological studies to find ways of
disentangling CR effects from hydro ones. In this context, our paper
should therefore also be viewed as an attempt to see how far one can
get \emph{without} postulating genuine 
(pressure-driven) collective-flow effects in $pp$. Within this
context, it is important to note that CR can mimic flow effects to
some extent, via the creation of boosted
strings~\cite{Ortiz:2013yxa}. Alternatively, it is possible that the
effective string tension could be rising, as in the idea of colour
ropes~\cite{Andersson:1991er}, with recent work along these lines
reported on by the Lund group~\cite{Bierlich:2014xba}. 
Finally, we note that non-hydro
rescattering has also been proposed~\cite{Corke:2009tk} 
as a potential mechanism
contributing to the rise of $\left<p_\perp\right>(n_\mrm{Ch})$, though
the explicit model of parton-parton rescattering effects presented
in~\cite{Corke:2009tk} found only very small effects. The possibility
of Boltzmann-like elastic (or even inelastic) final-state hadron-hadron
rescattering is still open. As usual, nature's solution is likely to 
involve an interplay of effects at different levels. Nevertheless,
before exploring further effects at the hadron level, we believe it
makes good sense to first examine the hadronisation process itself,
which is the topic of this work. 

Finally, we note that colour flows beyond LC
have also been invoked in the context of $J/\psi$
formation~\cite{Fritzsch:1977ay,Ali:1978kn,Fritzsch:1979zt,Eriksson:2008tm},  
and as a potential mechanism to generate diffractive topologies in $ep$ and $pp$
collisions~\cite{Buchmuller:1995qa,Edin:1995gi}.

In \secRef{sec:model}, we briefly recapitulate the treatment
of colour space for the existing MPI models in PYTHIA, and present the
new model that we have developed, combining the minimisation of the string potential with
the multiplet structure of QCD.
 In \secRef{sec:tuning}, we constrain 
the resulting free model parameters on a selection of both $ee$ and $pp$
data, discussing the physics consequences of the new colour-space
treatment as we go along. In \secRef{sec:top}, we consider
implications for precision extractions of the top quark mass at hadron
colliders. Finally, in \secRef{sec:summary}, we summarise and give an
outlook. 

\section{The Model \label{sec:model}}

In this section, we present the colour-space model that we have
developed, which allows strings to form not only between
LC-connected partons, but also between specific non-LC-connected ones,
following combination rules that approximate the multiplet structure
of full-colour QCD. 
We begin with a brief summary of the current modelling, in
\secRef{sec:existingModels}. We then turn to a general discussion of
coherence effects beyond leading $N_C$ in \secRef{sec:beyondLC}. 
Finally, in \secRef{sec:newModel}, we present the detailed
implementation of the new model. 

We emphasise that there is a conceptual difference between 
colour-space \emph{ambiguities}, such as those explored in this work,
and physical colour \emph{reconnections}. The subleading-colour effects
we discuss here arise naturally in ``full-colour'' $SU(3)$ and 
do not involve any physical exchange of colour or
momentum (although explicit algorithms may of course still employ an
iterative-reconnection scheme to find the potential-energy minimum).
Strictly speaking, the term colour \emph{reconnections} should be
reserved to describe effects related to
\emph{dynamical} reconfigurations of the colour/string space that involve
explicit exchange of colours and momentum, 
via  perturbative gluon 
exchanges or non-perturbative string interactions. Effects of this type
are not explored directly in this work, instead we refer the
interested reader to the SK string-interaction models presented in
\cite{Sjostrand:1993hi,Khoze:1994fu,Khoze:1999up}. Somewhat sloppily, we
follow the entrenched 
convention in the field and use the acronym ``CR'' for effects of either
kind here.

\subsection{Existing MPI Models and Colour Space 
\label{sec:existingModels}}

In a naive LC picture, each MPI scattering system is 
viewed as separate and distinct from all other systems in colour
space. The very simplest colour-space options in the old PYTHIA 6 MPI 
model~\cite{Sjostrand:1987su} and the first HERWIG (and HERWIG++) MPI
models~\cite{Butterworth:1996zw,Gieseke:2010zz} go a step further,
representing each MPI final state  
as two quarks (or gluons), colour-connected directly to each other,
i.e., treating each MPI system as a separate hadronising colour-singlet
system. However, this ignores that the incoming partons are coloured,
and hence that the total colour charge of each MPI scattering system
is in general non-zero. These particular models therefore violate colour
conservation and are unphysical. 

To be LC-correct one must take into account that 
each MPI-initiator parton should cause one or two strings to be
stretched to its remnant (one for quarks, two for gluons). This
conserves colour, but still has the implication that 
no strings would be stretched \emph{between} different MPI systems. 
This situation 
is illustrated in \figRef{fig:remnantsa}. 
\DeclareGraphicsRule{*}{mps}{*}{}
\begin{figure}[t]
\centering
\captionsetup[subfigure]{skip=20pt}
\subcaptionbox{\label{fig:remnantsa}}{
\begin{fmffile}{remnantcolora}
\begin{fmfgraph*}(52,22)
\fmfstraight
\fmftop{mpi1l,mpi1c,mpi1r,mpi2l,mpi2c,mpi2r,mpi3l,mpi3c,mpi3r}
\fmfbottom{br1l,br1c,br1r,br2l,br2c,br2r,br3l,br3c,br3r}
\fmfv{d.shape=circle,d.fill=empty,label.dist=0,d.siz=32,label=MPI 1}{mpi1c}
\fmfv{d.shape=circle,d.fill=empty,label.dist=0,d.siz=32,label=MPI 2}{mpi2c}
\fmfv{d.shape=circle,d.fill=empty,label.dist=0,d.siz=32,label=MPI 3}{mpi3c}
\fmf{dbl_plain,label=Hadron Remnant}{br1l,br3r}
\fmf{gluon}{br1c,mpi1c}
\fmf{gluon}{br2c,mpi2c}
\fmf{gluon}{br3c,mpi3c}
\fmffreeze
\fmfi{plain,fore=red}{vpath (__br1c,__mpi1c) shifted (thick*(-3,1))}
\fmfi{plain,fore=red}{vpath (__br1c,__mpi1c) shifted (thick*(5,1))}
\fmfi{plain,fore=red}{vpath (__br2c,__mpi2c) shifted (thick*(-3,1))}
\fmfi{plain,fore=red}{vpath (__br2c,__mpi2c) shifted (thick*(5,1))}
\fmfi{plain,fore=red}{vpath (__br3c,__mpi3c) shifted (thick*(-3,1))}
\fmfi{plain,fore=red}{vpath (__br3c,__mpi3c) shifted (thick*(5,1))}
\end{fmfgraph*}
\end{fmffile}
}
\hspace*{1.5cm}
\subcaptionbox{\label{fig:remnantsb}}{
\begin{fmffile}{remnantcolorb}
\begin{fmfgraph*}(52,22)
\fmfstraight
\fmftop{mpi1l,mpi1c,mpi1r,mpi2l,mpi2c,mpi2r,mpi3l,mpi3c,mpi3r}
\fmfbottom{br1l,br1c,br1r,br2l,br2c,br2r,br3l,br3c,br3r}
\fmfv{d.shape=circle,d.fill=empty,label.dist=0,d.siz=32,label=MPI 1}{mpi1c}
\fmfv{d.shape=circle,d.fill=empty,label.dist=0,d.siz=32,label=MPI 2}{mpi2c}
\fmfv{d.shape=circle,d.fill=empty,label.dist=0,d.siz=32,label=MPI 3}{mpi3c}
\fmf{dbl_plain,label=Hadron Remnant}{br1l,br3r}
\fmf{gluon}{br1c,mpi1c}
\fmf{gluon}{br2c,mpi2c}
\fmf{gluon}{br3c,mpi3c}
\fmffreeze
\fmf{phantom}{br1r,br2c}
\fmf{phantom}{br2r,br3c}
\fmfi{plain,fore=red}{vpath (__br1c,__mpi1c) shifted (thick*(-3,1))}
\fmfi{plain,fore=red}{vpath (__br1c,__mpi1c) shifted (thick*(5,3))}
\fmfi{plain,fore=red}{vpath (__br2c,__mpi2c) shifted (thick*(-3,3))}
\fmfi{plain,fore=red}{vpath (__br2c,__mpi2c) shifted (thick*(5,3))}
\fmfi{plain,fore=red}{vpath (__br3c,__mpi3c) shifted (thick*(-3,3))}
\fmfi{plain,fore=red}{vpath (__br3c,__mpi3c) shifted (thick*(5,1))}
\fmfi{plain,fore=red}{vpath (__br1r,__br2c) shifted (thick*(-3.5,3))}
\fmfi{plain,fore=red}{vpath (__br2r,__br3c) shifted (thick*(-3.5,3))}
\fmffreeze
\end{fmfgraph*}
\end{fmffile}
}

\caption{(\subref{fig:remnantsa}): in a strict LC picture, each MPI initiator gluon
 increases the ``colour charge'' of the beam remnant by two
 units. (\subref{fig:remnantsb}): allowing different MPI initiators to be connected
 in a colour chain reduces the total colour charge of the remnants. Here
 for example, no strings will be stretched directly between the shown
 remnant and the final states of MPI 2.
(Note that the colour assignments shown are for illustration only, and
  would be represented by Les Houches Colour Tags~\cite{Boos:2001cv,Alwall:2006yp} in a real event generator.) \label{fig:remnants}} 
\end{figure}
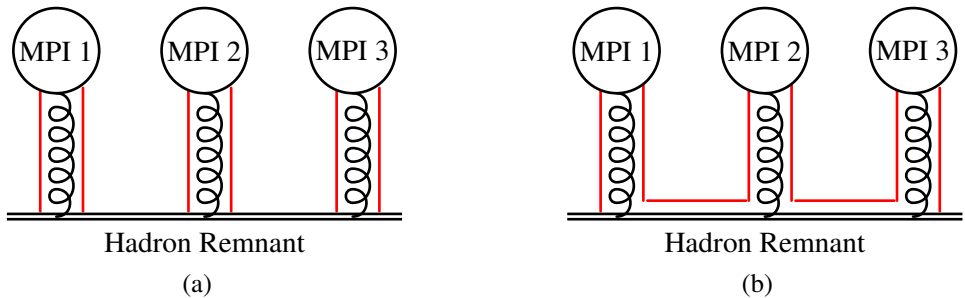
Physically, this can lead to  arbitrarily many
strings being stretched across the central rapidity region, one or two
for each MPI (corresponding to adding their total colour chargers
together as scalar quantities, rather than as $SU(3)$ vectors). 

However, already in the context of earlier
works~\cite{Sjostrand:1987su,Sjostrand:2004pf}, it was 
noted that even this picture cannot be quite physically correct. 
Since all the MPI initiators on each side are extracted from one and 
the same (colour-singlet) beam particle, 
and since they are extracted at a rather low scale of order the perturbative
evolution cutoff $p_{\perp 0} \sim \mbox{one to a few GeV}$, 
there is presumably some overlap and accompanying saturation
effects, implying that they are not completely independent. 
Not knowing
the exact form of the correlations, a pragmatic solution is to minimise
the total colour charge of the 
remnant (and hence the number of strings stretched to it), by allowing
the different MPI systems to be colour-connected to each other along a
``chain'' in colour space, as illustrated in \figRef{fig:remnantsb}. 
Variations of this are used in the current forms
of both the 
PYTHIA~\cite{Sjostrand:1987su,Sjostrand:2004pf} and 
HERWIG++~\cite{Gieseke:2012ft} MPI models, reducing the number of
strings/clusters especially in the remnant-fragmentation region at high
rapidities. It is, however, 
still fundamentally ambiguous 
exactly which systems to connect and how. In the example of
\figRef{fig:remnants}, it is arbitrary that it happens to be  
the colour of MPI 1 and the anticolour of MPI 3 which end up connected
to the remnant. For a more detailed discussion of this aspect, see
e.g.~\cite{Sjostrand:2004pf}. An interesting physics point is that, in
this picture, the particle production at very forward
rapidities is controlled essentially by how large one allows
the colour charge of the remnant to become, which in turn depends on the
number of MPI and their mutual colour correlations. This could
presumably be revealed by studies correlating the particle production 
in the central region (sensitive to the number of MPI) 
with that in the forward region (sensitive to the total charge of the
remnant).  

In the absence of any further CR effects, the relationship between the
number of MPI and the average particle multiplicity at
central rapidities is still approximately linear. Consequently, the
per-particle spectra in high-multiplicity events 
(with many MPI) are similar to those in (non-diffractive) low-multiplicity
events (with few MPI)\footnote{For very low multiplicities, well-understood 
 bias effects cause the average particle $p_\perp$ to increase (if the
 event is required to contain only one particle, then that particle
 must be carrying all the scattered energy), while for high
  multiplicities,  the contribution
  from hard-jet fragmentation also generates slightly harder
  spectra.}. This is what leads to the simple expectation of the flat
$\left<p_\perp\right>(n_\mrm{Ch})$ spectrum exhibited by the ``no CR''
curves that were shown in \figRef{fig:avgptofnch} in the previous
section. However, as was also remarked on there, the experimental data
convincingly rule out such a constant behaviour. 
This observation is the main reason additional non-trivial final-state
CR effects have been included in both HERWIG++ and PYTHIA. 

In the original (non-interleaved) MPI model in PYTHIA
6~\cite{Sjostrand:1987su,Sjostrand:2006za}, the  
parameters \ttt{PARP(85)} and \ttt{PARP(86)} allowed to force a fraction
of the MPI final states to be two gluons colour-connected to their
nearest neighbours in momentum space.
The physical picture was that the hardest
interaction built up a ``skeleton'' of string pieces, onto which a fraction
of the gluons from MPI were grafted  (by brute
force) in the places where they caused
the least amount of change of string length. This effectively minimised the increase in
string length from those gluons. An important factor contributing to
the revival of the question of CR in
hadron collisions was the tuning studies of this model, carried out by
Rick Field on underlying-event and minimum-bias data from the
CDF experiment at the
Tevatron~\cite{Abe:1988yu,Acosta:2001rm,Affolder:2001xt}. His
resulting ``Tune A'' and 
related tunes~\cite{Field:2002vt,Field:2005sa} were the first to give good fits to the
available data at the time, but the surprising conclusion was that
in order to do so this ``colour-space grafting'' had to be done nearly all
the time.  

An alternative set of CR models, which relied on physical analogies
with overlapping strings in superconductors, were developed only in 
the context of $e^+e^-$
collisions~\cite{Sjostrand:1993hi,Khoze:1994fu,Khoze:1999up}, 
chiefly with the aim of studying potential CR uncertainties on the $W$
mass, see  \cite{Sjostrand:2013cya} and references therein. As far as
we are aware, this class of models has not yet been applied in the
context of the more complex environment of hadron-hadron collisions. 

In the new (interleaved) MPI model in PYTHIA
6~\cite{Sjostrand:2004ef,Sjostrand:2006za}, showers and MPI were
carried out in parallel, with physical colour flows. This was too
complicated to handle with the old CR model. A new 
``colour annealing'' CR scenario was
developed~\cite{Sandhoff:2005jh,Skands:2007zg,Wicke:2008iz} which,
after the shower evolution had finished, allowed for a fraction of
partons to ``forget'' their LC colour 
connections,
with new ones determined based on the string area law
(shorter strings are preferred), following a simplified annealing-like
algorithm, in a similar spirit to an earlier model by 
Rathsman, called the ``Generalized Area-Law'' (GAL)
model~\cite{Rathsman:1998tp}. The fraction of partons that forgot
their LC colour connections was assumed to grow with the
number of MPI, with a per-MPI probability given by the parameter
\texttt{PARP(78)}. A further parameter, \texttt{PARP(77)}, allowed to
suppress the reconnection probability for fast-moving partons. 
Although still intended as a toy model, the new colour-annealing models
 obtained good agreement with the Tevatron
minimum-bias and underlying-event data, e.g.\ in the form of the Perugia
family of tunes~\cite{Skands:2009zm,Skands:2010ak}. The most
recent incarnations, the Perugia 2011 and 2012 tunes, also included LHC
data and were among the main reference tunes used during Run 1 of the
LHC~\cite{Skands:2010ak}. However, a study comparing independent
MPI+CR tunings at different collider energies revealed different
preferred CR parameter values at different CM
energies~\cite{Schulz:2011qy}, implying that the modelling of this aspect, or at least  
its energy dependence, was still inadequate.

In PYTHIA 8, the default MPI colour-space treatment is similar to that
of the original PYTHIA 6 model, although starting out from a more detailed
modelling of the colour flow in each MPI. With a certain probability, controlled by the
parameter \texttt{ColourReconnection:range}, all the gluons of each lower-$p_\perp$ 
interaction can be inserted onto the colour-flow dipoles of a
higher-$p_\perp$ one, in such a way as to minimise the total string
length~\cite{Argyropoulos:2014zoa}. The effects of this model was already illustrated in
\figRef{fig:avgptofnch}. A set of alternative CR scenarios was also
presented in~\cite{Argyropoulos:2014zoa}, but were still mostly
intended as toy models in the context of estimating uncertainties on
the top-quark mass. 

Finally, in the most recent developments of the HERWIG++ MPI model, an
explicit scenario for colour reconnections has likewise been
introduced~\cite{Gieseke:2012ft}, based on a simulated-annealing
algorithm that minimises (sums of) cluster masses. 
In the context of the cluster hadronisation
model~\cite{Webber:1983if}, the minimisation of cluster masses fulfils a
similar function as the minimisation of string lengths above. The two
minimisations differ in that the string length measure is closely related
to the product of the invariant masses rather than the sum used in the cluster
model. 
The main model parameter is 
the probability to accept a favourable reconnection,
$p_{\mathrm{reco}}$. The study in~\cite{Gieseke:2012ft} emphasised in
particular that the largest pre-reconnection cluster masses are
spanned between hard partons and the remnants (denoted $n$-type
clusters), with inter-MPI ones (spanned directly between partons from different
MPI systems and denoted $i$-type) having the second-largest masses. The former
again indicates that there is a non-trivial interplay with the
non-perturbative hadronisation of the beam remnant, while the latter
reflects the lack of a priori knowledge about the colour
correlations \emph{between} different MPI systems. Similarly to the
qualitative conclusions made with the PYTHIA CR models, the HERWIG++
study found that quite large values of $p_\mathrm{reco}\sim 0.5$ were
required to describe hadron-collider data.

\subsection{Beyond Leading Colour \label{sec:beyondLC}}

To illustrate the colour-space ambiguity between different MPI
systems, and between them and the beam remnant, let us take the simple case of 
double-parton scattering (DPS), with all the initiator partons being
gluons. What happens in colour space when we extract \emph{two} gluons from a
proton? Even if we imagine that the two gluons are completely
uncorrelated, QCD gives several possibilities for their superpositions:
\begin{equation}
\mbf{8} \otimes \mbf{8} = \mbf{27} \oplus \mbf{10} \oplus
\overline{\mbf{10}} \oplus \mbf{8} \oplus \mbf{8} \oplus \mbf{1}~.
\label{eq:88}
\end{equation}
The highest-charge multiplet, here the \mbf{27} (a
``viginti-septet''), effectively represents the LC term:  
incoherent addition of the two gluons, each carrying two units of (LC)
colour charge (one colour and one anticolour), for a total of 4 string
pieces required to be attached to the remnant\footnote{Assuming each string can only carry
  one unit of flux, or equivalently a 4-unit ``colour rope'',
  see~\cite{Bierlich:2014xba}.}. 
However, note that the
probability for this to occur is 
\begin{equation}
P_\mathrm{LC} = \frac{27}{64} < 50\%~,
\end{equation}
hence the naive expectation that subleading topologies
should be suppressed by $1/N_C^2$ is badly broken already in this very
simple case\footnote{In general, the highest-charge multiplet in the
  combination of $k$ gluons represents a fraction $(k+1)^3/8^k$ of the
  possibilities.}. 
The decuplets (octets) correspond to
coherent-superposition topologies with a lower total colour charge and 
consequently only three (two) string pieces attached to the remnant. 
The singlet represents the special case in
which the two MPI-initiator gluons have identical and opposite colours, with 
total colour charge 0 (generating a diffractive-looking topology from
the point of view of the remnant). In
QCD, for two random (uncorrelated) gluons, there is a 1/64 probability
for this to happen purely by chance. 

The other possible two-parton combinations are:
\begin{eqnarray}
\mbf{3} \otimes \mbf{8}  & = & \mbf{15} \oplus \overline{\mbf{6}} \oplus
\mbf{3}~,\label{eq:38}\\
\mbf{3} \otimes \overline{\mbf{3}}  & = & \mbf{8} \oplus \mbf{1}~,\label{eq:33bar}\\
\mbf{3} \otimes \mbf{3}  & = & \mbf{6} \oplus
\overline{\mbf{3}}~,\label{eq:33}
\end{eqnarray}
where strict LC would correspond to populating only the \mbf{15} (quindecuplet), \mbf{8} (octet), and \mbf{6}
(sextet), respectively. 
\begin{figure}[t]
\centering
\includegraphics[scale=0.535]{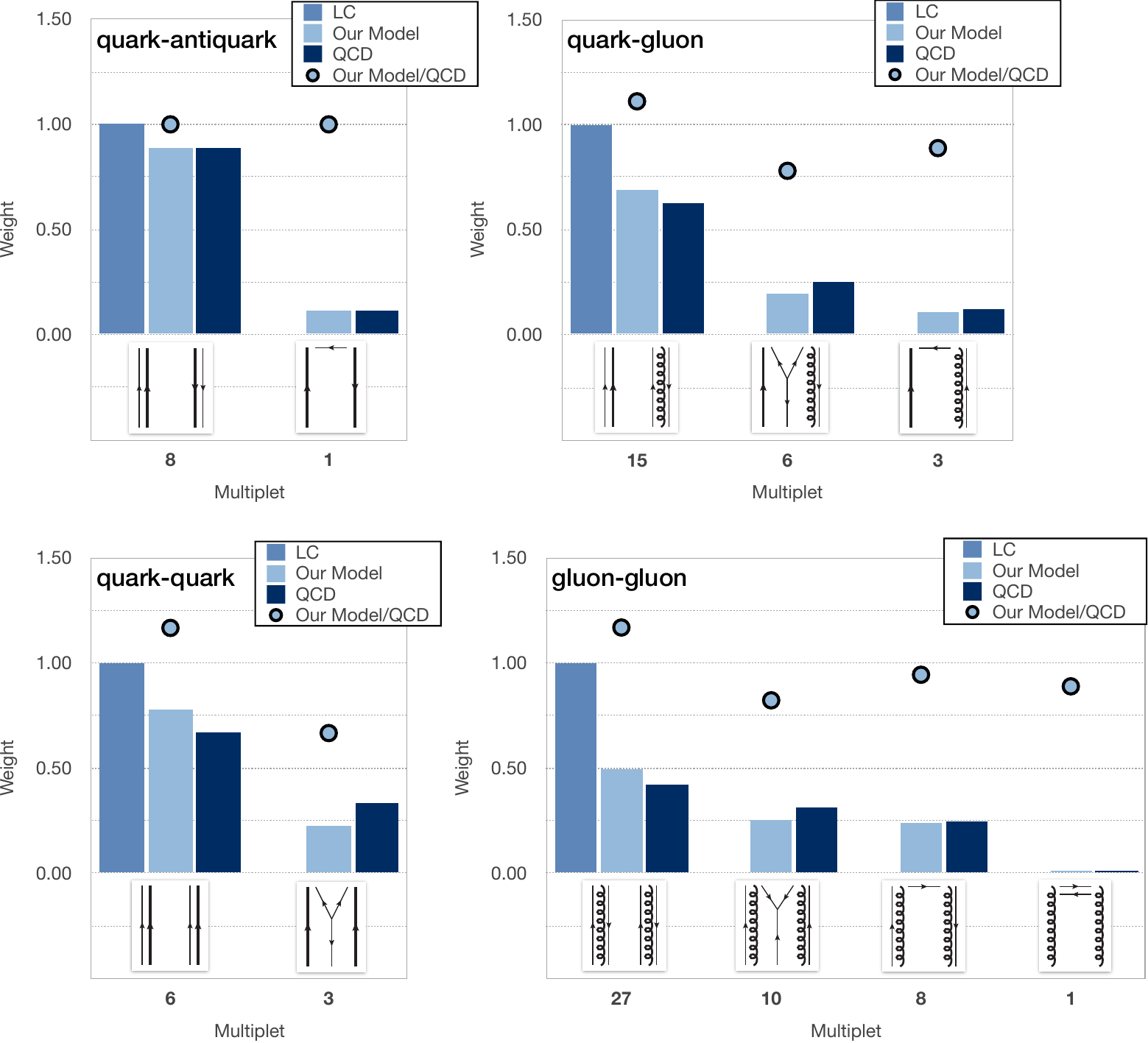}
\caption{Illustration of the possible colour states of two
  random (uncorrelated) partons. In strict LC, only the completely incoherent
  superposition is populated. Our model (described below)
  gives a systematically better approximation. Filled circles show the
  ratio between our model and full $SU(3)$. 
  The diagrams below the
  histograms attempt to illustrate corresponding colour-flow
  configurations, with thick and thin lines denoting partons and colour-flow
  lines, respectively. 
\label{fig:QCDmultiplet}}
\end{figure}
The relative weights (probabilities) for each multiplet in each of
these combinations are illustrated in \figRef{fig:QCDmultiplet},
along with diagrams exemplifying corresponding colour flows (with thick
lines indicating partons, thin ones colour-flow lines).
For each multiplet, three vertical bars indicate the probability
associated with that multiplet in strict Leading Colour (LC), in our model (defined below),
and in $SU(3)$ (QCD), respectively.  The filled circles 
represent the \emph{ratio} between our model and QCD, so for those
\emph{unity} indicates perfect agreement. Note that, since the
subleading multiplets are absent in LC, only two non-zero bars appear
for them. Below, we shall also consider the probability for three
uncorrelated triplets to form an overall singlet, which  is $1/27$ in QCD:
\begin{eqnarray}
\mbf{3} \otimes \mbf{3} \otimes \mbf{3}  & = & \mbf{10} \oplus \mbf{8}
\oplus  \mbf{8} \oplus \mbf{1}~\label{eq:333}~,
\end{eqnarray}
while in our simplified model it will come out to be $2/81 = (1 - \frac13)/27$. 

We emphasise that we only use these composition rules for
colour-unconnected partons, 
which in the context of our model we approximate as
being totally uncorrelated. LC-connected partons are
always in a singlet with respect to each other, and colour neighbours
(e.g., the two colour lines of a gluon or those of a $q\bar{q}$ pair produced
by a $g\to q\bar{q}$ splitting) are never in a singlet with respect to
each other.

The approximation of colour-unconnected partons being totally
uncorrelated, combined with a set of specific colour-space
parton-parton composition rules, such as those of $SU(3)$ or the
simplified ones defined below, allow us to build up an
approximate picture of the possible colour-space correlations that a
complicated parton system can have, including randomised coherence
effects beyond Leading Colour. Due to the subleading correlations,
there are many possible string topologies that could represent such a 
parton system, including but not limited to the LC one. 
The selection principle that determines how the system collapses 
into a specific string configuration will be furnished by the minimisation of
the string potential, as we shall return to below.

Our model thus consists of two stages. First, we generate 
an approximate
picture of the possible colour states of a parton system. Then, we
select a specific realisation of that state in terms of explicit
string connections. This is done at the time when the system is
prepared for hadronisation, i.e., after parton showering but before
string fragmentation.

By maintaining the structure of the (LC) showers unchanged, we neglect
any possibility of reconnections occurring already at the perturbative
level. Though perturbative gluon exchanges and/or full-colour shower
effects might mediate such effects in nature, we expect their 
consequences to be suppressed relative to the non-perturbative
ones considered here. This is partly due to the coherence and
collinear-enhancement properties already acting to
minimise the mass of LC dipoles inside each perturbative cascade, and partly due to the
space-time separation between different systems (be they different MPI
systems, which are typically separated by transverse distances of order
$1/\Lambda_\mrm{QCD}$ inside the proton, or different resonance-decay
systems separated by $1/\Gamma_\mrm{res}$). Thus, at high $Q\gg
\Lambda_\mrm{QCD}$ or $Q\gg \Gamma_\mrm{res}$ we don't
expect any cross-talk between different MPI or different resonance
systems, respectively. The case can be made that perturbative
reconnection effects could still be active at longer 
wavelengths, but we expect that such semi-soft effects can presumably
be absorbed in the non-perturbative modelling without huge mistakes. 

\subsection{The New Model \label{sec:newModel}}

Our simplified colour-space model is defined as follows. Rather than
attempting to capture the full correlations (which we
have emphasised are not a priori known anyway and would require a
cumbersome matrix-based formalism), we note that the 
main subleading parton-parton combination possibilities
of real QCD can be encoded in a single ``colour index'', running from 1 to
9 (with corresponding indices for anti-colours).

Quarks are assigned a single such colour index, antiquarks a single
anticolour index, and gluons 
have one of each, with the restriction that their colour and anticolour
indices cannot be the same. Thus, formally our model has 9 different
quark colour states and 72 kinds of gluon states.
 We emphasise that these 
indices should not be confused with the ordinary 3-dimensional $SU(3)$
quark colour indices (red, green, and blue); rather, our index labels
the possible colour states of two-parton (and in some cases
three-parton) \emph{combinations}. Thus, for example, a quark and an
antiquark are in an overall colour-singlet state if the colour index of
the former equals the anticolour index of the latter, otherwise they
are in an octet state, cf.~\eqRef{eq:33bar}.  We note that 
a similar index was used already in the models of ``dipole
swing'' presented in
~\cite{Avsar:2006jy,Avsar:2007xh}, though here we generalise to parton
combinations involving colour-epsilon structures as 
well, cf.~\figRef{fig:QCDmultiplet}.

Confining potentials will be allowed to form 
between any two partons that have matching colour and anticolour
indices. Since LC-connected partons are forced to have
matching colour and anticolour indices, the ``original'' (LC) string
topology always remains 
possible, but now further possibilities also exist involving partons
that \emph{accidentally} have matching indices, illustrated in
\figRef{fig:crex1}.

\begin{figure}[t]
\centering\vskip7mm
\begin{fmffile}{crex}
\begin{fmfgraph*}(40,20)
\fmfforce{-0.15w,1.15h}{v1}
\fmfv{label.dist=0,label=\textbf{\large (A)}}{v1}
\fmftop{t1,tc,t2}
\fmfleft{l1,l2}
\fmfright{r1,rc,r2}
\fmfbottom{b1,g12,bd,b2}
\fmfv{d.shape=circle,d.fill=empty,label.dist=0,d.siz=20,label=$q_1$}{b1}
\fmfv{d.shape=circle,d.fill=empty,label.dist=0,d.siz=20,label=$g_{12}$}{g12}
\fmfv{d.shape=circle,d.fill=empty,label.dist=0,d.siz=20,label=$\bar{q}_2$}{b2}
\fmfv{d.shape=circle,d.fill=empty,label.dist=0,d.siz=20,label=$q'_2$}{r2}
\fmfv{d.shape=circle,d.fill=empty,label.dist=0,d.siz=20,label=$\bar{q}'_2$}{l2}
\fmf{plain,fore=red}{b1,g12}
\fmf{plain,fore=red}{g12,b2}
\fmf{plain,fore=red}{l2,r2}
\end{fmfgraph*}\hskip1.5cm\raisebox{1cm}{\textbf{vs.}}\hskip1.5cm
\begin{fmfgraph*}(40,20)
\fmfforce{-0.15w,1.15h}{v1}
\fmfv{label.dist=0,label=\textbf{\large (B)}}{v1}
\fmftop{t1,tc,t2}
\fmfleft{l1,l2}
\fmfright{r1,rc,r2}
\fmfbottom{b1,g12,bd,b2}
\fmfv{d.shape=circle,d.fill=empty,label.dist=0,d.siz=20,label=$q_1$}{b1}
\fmfv{d.shape=circle,d.fill=empty,label.dist=0,d.siz=20,label=$g_{12}$}{g12}
\fmfv{d.shape=circle,d.fill=empty,label.dist=0,d.siz=20,label=$\bar{q}_2$}{b2}
\fmfv{d.shape=circle,d.fill=empty,label.dist=0,d.siz=20,label=$q'_2$}{r2}
\fmfv{d.shape=circle,d.fill=empty,label.dist=0,d.siz=20,label=$\bar{q}'_2$}{l2}
\fmf{plain,fore=red}{b1,g12}
\fmf{plain,fore=red}{g12,l2}
\fmf{plain,fore=red}{b2,r2}
\end{fmfgraph*}
\end{fmffile}
\vskip5mm
\caption{Illustration of a multi-parton state with a rather simple colour-space
  ambiguity. Subscripts indicate colour-space indices. {\sl
    (A):} the ``original'' (LC) string topology. {\sl(B):} an
  alternative string topology, allowed by the accidentally matching
  ``2'' indices. \label{fig:crex1}}
\end{figure}
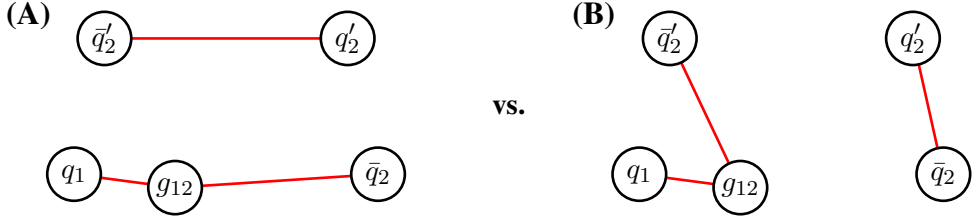

Furthermore, two
\emph{colour} indices are allowed to sum coherently to a single
\emph{anticolour} index within three separate closed index groups: 
[1,4,7], [2,5,8], and [3,6,9]. E.g., two quarks carrying indices 2 and
5 respectively, are allowed to appear to the rest of the event as
carrying a
single combined anti-8 index.  
These index combinations represent the
antisymmetric $\varepsilon_{ijk}$ colour combinations that were
pictorially represented as Y-shaped ``colour junctions'' in
\figRef{fig:QCDmultiplet}. A junction can therefore be interpreted as the
string extension of a baryon with the baryon number ($\varepsilon_{ijk}$)
located in the centre of the Y-shape~\cite{Sjostrand:2002ip}. 
An explicit example of a parton system
whose colour state includes such a possibility is shown in
\figRef{fig:crex2}.  
\begin{figure}[t]
\centering\vskip7mm
\begin{fmffile}{crex2}
\begin{fmfgraph*}(40,20)
\fmfforce{-0.15w,1.15h}{v1}
\fmfv{label.dist=0,label=\textbf{\large (C)}}{v1}
\fmftop{t1,tc,t2}
\fmfleft{l1,l2}
\fmfright{r1,rc,r2}
\fmfbottom{b1,g12,bd,b2}
\fmfv{d.shape=circle,d.fill=empty,label.dist=0,d.siz=20,label=$q_1$}{b1}
\fmfv{d.shape=circle,d.fill=empty,label.dist=0,d.siz=20,label=$g_{12}$}{g12}
\fmfv{d.shape=circle,d.fill=empty,label.dist=0,d.siz=20,label=$\bar{q}_2$}{b2}
\fmfv{d.shape=circle,d.fill=empty,label.dist=0,d.siz=20,label=$\bar{q}'_5$}{r2}
\fmfv{d.shape=circle,d.fill=empty,label.dist=0,d.siz=20,label=$q'_5$}{l2}
\fmf{plain,fore=red}{b1,g12}
\fmf{plain,fore=red}{g12,b2}
\fmf{plain,fore=red}{l2,r2}
\end{fmfgraph*}\hskip1.5cm\raisebox{1cm}{\textbf{vs.}}\hskip1.5cm
\begin{fmfgraph*}(40,20)
\fmfforce{-0.15w,1.15h}{v1}
\fmfv{label.dist=0,label=\textbf{\large (D)}}{v1}
\fmftop{t1,tc,t2}
\fmfleft{l1,l2}
\fmfright{r1,rc,r2}
\fmfbottom{b1,g12,bd,b2}
\fmfv{d.shape=circle,d.fill=empty,label.dist=0,d.siz=20,label=$q_1$}{b1}
\fmfv{d.shape=circle,d.fill=empty,label.dist=0,d.siz=20,label=$g_{12}$}{g12}
\fmfv{d.shape=circle,d.fill=empty,label.dist=0,d.siz=20,label=$\bar{q}_2$}{b2}
\fmfv{d.shape=circle,d.fill=empty,label.dist=0,d.siz=20,label=$\bar{q}'_5$}{r2}
\fmfv{d.shape=circle,d.fill=empty,label.dist=0,d.siz=20,label=$q'_5$}{l2}
\fmf{plain,fore=red}{b1,g12}
\fmf{plain,fore=red}{g12,j1,l2}
\fmf{plain,fore=red}{b2,j2,r2}
\fmf{plain,fore=red}{j1,j2}
\end{fmfgraph*}
\end{fmffile}
\vskip5mm
\caption{Illustration of a similar multi-parton state as in
  \figRef{fig:crex1}, but now with index assignments resulting 
  in a junction-type colour-space ambiguity. The orientation of the top
  $q'_{5}-\bar{q}'_5$ dipole has also been reversed relative to
  \figRef{fig:crex1}. 
  {\sl (C):} the original (LC) string topology. {\sl (D):} an
  alternative string topology with a junction and an antijunction,
  allowed by the cyclically matching ``2'' and ``5'' indices.\label{fig:crex2}}
\end{figure}
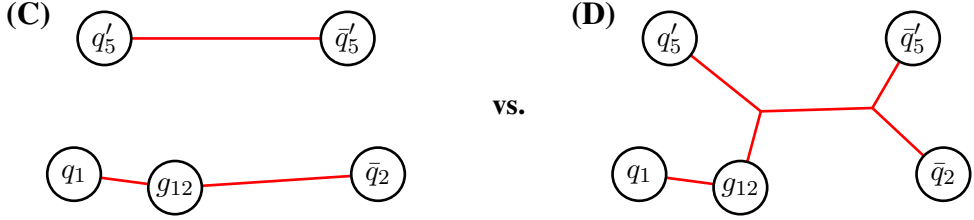
A model for string hadronisation of such
 topologies was developed in~\cite{Sjostrand:2002ip} and has
subsequently also been applied to the modelling of baryon beam
remnants~\cite{Sjostrand:2004pf}. We reuse it here for 
hadronisation of junction-type colour-index combinations.

The new model can be divided into two main parts: a new treatment of
the colour flow in  
the beam remnant and a new CR scheme. The two models are independent and can
therefore in principle be combined both with each other as well as with other
models. (Note, however, that the old PYTHIA 8 CR scheme is inextricably
linked with the colour treatment of the beam remnant and therefore
only works together with the old 
beam-remnant 
model.) Both of the models occasionally result in complicated multi-junction configurations that the
existing PYTHIA hadronisation cannot handle. Rather than attempting
to address these somewhat pathological topologies in detail, this
problem is circumvented by a 
clean-up method that simplifies the structure of the resulting systems
to a level that PYTHIA can handle.

The next two sections describe respectively the details of the new
beam-remnant model and the new CR scheme, including technical aspects
and the algorithmic implementation. Afterwards the junction clean-up
method is described.  

\subsubsection{Colour Flow in the Beam Remnant \label{sec:remnant}}
It being an inherently non-perturbative object, we
do not expect to be able to use perturbative QCD to understand the
structure of the beam remnant. Instead, we rely on conservation laws; 
the partons making up the beam remnant 
must, together with those that have been kicked out by MPI, sum up to
the total energy and momentum of the beam particle, be in an overall
colour-singlet state, with unit baryon number (for a proton beam),
carrying the appropriate total valence content for each quark flavour,
with equal numbers of sea quarks and antiquarks. 
The machinery used to conserve all these quantities should be consistent
with whatever knowledge of QCD we possess, such as the standard
single-parton-inclusive PDFs to which our framework reduces in the case
of single-parton scattering.  

In this work the focus is on the formation of colour-singlet states,
including the use of $SU(3)$ epsilon tensors. This naturally leads to a  
modification of the treatment of baryon number conservation, due to the
close link between baryons and the epsilon tensors in $SU(3)$. The modelling of
energy/momentum and flavour conservation is not touched relative to
the existing modelling of those aspects, and thus only a small review
is presented here (for more details see \cite{Sjostrand:2004pf}). 

The overall algorithm can be structured as follows:
\begin{enumerate}
  \item Determine the colour structure of the already scattered partons.
  \item Add the minimum amount of partons needed for flavour conservation.
  \item Add the minimum amount of gluons required to obtain a colour-singlet state.
  \item Connect all colours.
  \item With all the partons determined find their energy fractions.
\end{enumerate}
The conservation of baryon number is not listed as a separate point, but 
naturally follows from the formation of junctions. Let us now consider each 
of these points individually starting from the top.

To calculate the colour structure of the beam remnant, let us return to 
the DPS example of earlier. With a probability of $27/64$, the two
gluons form a completely incoherent  
state, leaving four colour charges to be 
compensated for in the beam remnant (two colours and two
anticolours). However the three valence quarks alone are insufficient
to build up a \mbf{27} (\eqRef{eq:333}), and therefore a minimum of one additional
gluon is needed. Then two of the quarks can combine to form a
$\overline{\mbf{3}}$, which can form an \mbf{8} with the remaining
valence quark, which then can enter in a \mbf{27} with the added
gluon. Conversely, if 
the two gluons had been in an octet state instead, the additional
gluon would not have been needed. Thus, in order to determine
the minimal number of gluons needed in the beam remnant, we need to know the
overall colour representation of all the MPI initiators combined.

\begin{figure}[t]
\centering\subcaptionbox{$Q^2\gg \Lambda_\mrm{QCD}^2$\label{fig:resHi}}{
\includegraphics[scale=0.4]{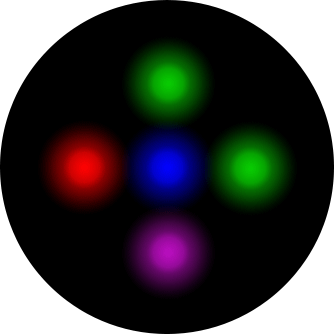}} \hspace*{1cm}
\subcaptionbox{$Q^2\sim \Lambda_\mrm{QCD}^2$\label{fig:resLo}}{\includegraphics[scale=0.4]{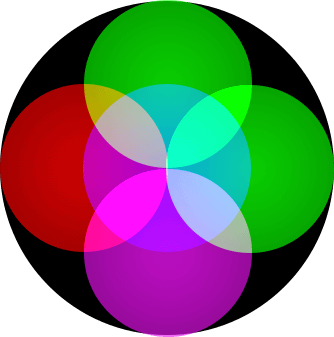}}
\caption{A proton (black) with five distinct colour sources (e.g.,
  four MPI-initiator partons plus one representing the 
  beam remnant), chosen so that they add to a singlet (with magenta =
  antigreen). Shown are two different 
  resolution scales representative of (\subref{fig:resHi}) the
  perturbative stage, during which the MPI systems are considered as
  being uncorrelated in colour space, and (\subref{fig:resLo}) the  
  nonperturbative stage, at which the beam remnant is
  considered and we assign higher weights to states with lower total
  QCD charge in order to mimic saturation effects. \label{fig:remnantColour}} 
\end{figure}
While it could be possible to choose this representation purely
statistically, based on the (simplified or full) $SU(3)$ weights, we
note two reasons that a lower total
beam-remnant charge is likely to be preferred in nature. Firstly, to
determine the 
most preferred 
configuration the string length needs to  be considered. Since the beam
remnants reside in the very forward regime, strings spanned between
the remnant and the scattered gluons tend to be  
long, and as such a good approximation is to minimise the number of strings 
spanned to the beam. This corresponds to preferring a low-charge
colour-multiplet state for the remnant, and as a consequence also
minimises the number of additional gluons  
required. Secondly, a purely stochastic selection
corresponds to the assumption that the scattered partons are 
uncorrelated in colour space. For hard MPIs (at $Q\gg
\Lambda_\mrm{QCD}$), this is presumably a good 
approximation, since the typical space-time separation of the 
collisions are such that two independent interactions do not have time to 
communicate. This is illustrated by \figRef{fig:resHi}. 
However, after the initial-state radiation is added, the lower 
evolution scale implies larger spatial wavefunctions, allowing for
interference between different interactions, illustrated by
\figRef{fig:resLo}. An additional argument is that at a low evolution scale the
number of partons is low, thus to combine to an overall singlet the
correlation between the few individual 
partons needs to be large. To provide a complete
description of this cross-talk,  
multi-parton densities for arbitrarily many partons would be needed,
ideally including colour correlations and saturation effects. Although
correlations in double-parton densities has been the topic of several
recent
developments~\cite{Gaunt:2009re,Flensburg:2011kj,Blok:2011bu,Diehl:2011yj,Manohar:2012jr,Manohar:2012pe,Chang:2012nw,Blok:2013bpa,Snigirev:2014eua}, 
the field is still not at a stage at which it 
would be straightforward to combine explicit double-parton distributions
with the standard single-parton-inclusive ones (which a code like PYTHIA
must be compatible with), nor to generalise them to
arbitrarily many partons. The only formalism
we are aware of that addresses all of these issues (in particular
flavour and momentum correlations for arbitrarily many partons) while reducing
to the single-parton ones for the hardest interaction remains the one
developed in the context of the current PYTHIA
beam-remnant model~\cite{Sjostrand:2004pf}. In this study, we supplement the 
momentum- and flavour-correlation model of \cite{Sjostrand:2004pf} with
a simple model of colour-space saturation effects appropriate to the
$SU(3)$-multiplet language used in this work. Noting that
saturation should lead to a suppression of higher-multiplet states, 
we use a simple ansatz of exponential suppression with multiplet size,
$M$:
\begin{equation}
  p(M) = \exp \left ( - M / k_\text{saturation} \right )
\end{equation}
where $p$ is the probability to accept a multiplet of size $M$
and $k_\text{saturation}$ is a free parameter that controls the amount of suppression. 

Everything stated above for the two-gluon case can be generalised to
include quarks and an arbitrary number of MPI. The calculation just 
extends to slightly more complicated expressions:
\begin{equation}
  \mbf{8} \otimes \mbf{3} \otimes \mbf{\bar{3}} \otimes \mbf{8} \otimes \ldots ~=~ \ldots
\end{equation}
where quarks enter as triplets, antiquarks as antitriplets and gluons as octets.
The statistical probability to choose any specific multiplet can be calculated in a 
similar fashion, either in full or simplified QCD. There is however
still an ambiguity in how the colours are connected. 
For instance consider two quarks and one antiquark forming an overall
triplet state. 
The colour calculation will not tell us which of the quarks are in a
singlet with the antiquark. In the case of
ambiguities, the implementation is to choose randomly. The CR
algorithm applied later may anyhow change the initial colour topology, 
 lessening the effect of the above choice.

At the  level of the technical implementation, the choice of colour state for
 the scattered partons is transferred to the final 
 state particles of the event using the LC structure of the MPI and
 PS. For example,  if the two gluons are in the octet state, one of the
 LHE colour tags (see~\cite{Boos:2001cv,Alwall:2006yp}) is changed 
accordingly, and is propagated through to the colour of
 the final state particles.

As a special case the overall colour structure of the beam remnant is not
allowed to be a  
singlet. This is to avoid double-counting between diffractive and
non-diffractive events. In the DPS example, if the gluons form a colour
singlet, they essentially make up (part of) a pomeron, and thus should
fall under the single-diffractive description. It would be interesting
to look into the interplay between MPI and diffraction in more detail
using the colour-multiplet language developed here,
but this would require its own dedicated study, beyond the scope of
this work.  

The conservation of flavour is relative straightforward and
follows~\cite{Sjostrand:2004pf}. The principle used 
is to add the minimum needed flavour. For example if only an $\text{s}$
quark is scattered from a proton, the remnant will consist of 
an $\bar{\text{s}}$ plus the three valence quarks.

With the  flavour structure and colour multiplet of the beam remnant
known, it is  
now possible to calculate explicitly how many gluons need to be added to
obtain a colour-singlet state. Again the idea is to add the 
minimum number of gluons to the beam remnant. Colour-junction
structures, which we have argued can arise naturally in the colour structure
of the scattered partons, complicate this calculation slightly. 
To achieve an overall colour-singlet state the number of 
junctions minus the number of antijunctions has to match that of the
beam particle. 
Taking this into account the minimal number of gluons is given by
\begin{equation}
  N_{\mrm{gluons}} = \text{max}\left (0,\frac{(N_{\mrm{colour}} - N_{\mrm{quarks}} +
 \| N_{\mrm{junctions}} - N_{\mrm{antijunctions}} - b\|)}{2} \right );
\end{equation}
where $b$ is the beam baryon number (1 if the beam is a baryon, 0 if the beam 
is a meson, and -1 if the is an antibaryon). The division by two is due to the 
gluons carrying two colour lines. Without junctions, the number of 
gluons is simply the number of colour lines to the remnant minus
the number of available quarks to connect those colour lines to. It is
easiest to understand how  
junctions change this, by noting that the creation of a junction
basically takes two colour charges and turn into one anticolour. Thus
the number of required connections goes down by one for each additional
junction needed. 

After the gluons are added, all the colour connections and junction structures
are assigned randomly between the remaining colours, with one exception: 
if the beam particle is a baryon and a junction needs to be constructed
(similarly for an antibaryon and an antijunction), two of the valence 
quarks will be used to form the junction structure (possibly embedded in
a diquark), if they have not already been scattered in the MPIs. 

With finally the full parton structure known,
including both flavours and explicit colours, the last step of the
construction of the beam remnant is the assignment of energy fractions
($x$ values) to each remnant parton, according to modified PDFs. 
To obtain overall energy-momentum conservation, the individual partons
are scaled by an overall factor. The scaling  
becomes slightly more complicated by the introduction of primordial $k_\perp$. 
Details on the modified PDF versions and the scaling can be found 
in~\cite{Sjostrand:2004pf}.
\subsubsection{Colour Flow in the Whole Event} \label{sec:CR}
As discussed above, the CR model is applied after the parton-shower
evolution has finished (and after inclusion of the beam-remnant
partons as described in the preceding subsection), just before the
hadronisation. The model builds on two main principles: 
a simplified $SU(3)$ structure of QCD, based on indices from 1 to 9, 
to tell which configurations are possible; and the potential energy of
the resulting string systems, as measured by the so-called $\lambda$
measure~\cite{Andersson:1998tv}, to choose between the 
allowed configurations.

\begin{figure}[tp]\centering%
\subcaptionbox{Type I: ordinary dipole-style reconnection\label{fig:posReconI}}{
    \includegraphics[width=0.7\textwidth]{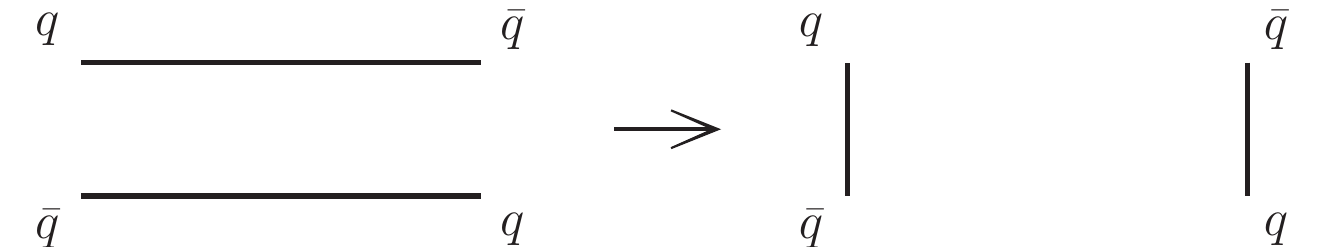} }\\[1cm]
\subcaptionbox{Type II: junction-style reconnection\label{fig:posReconII}}{
    \includegraphics[width=0.7\textwidth]{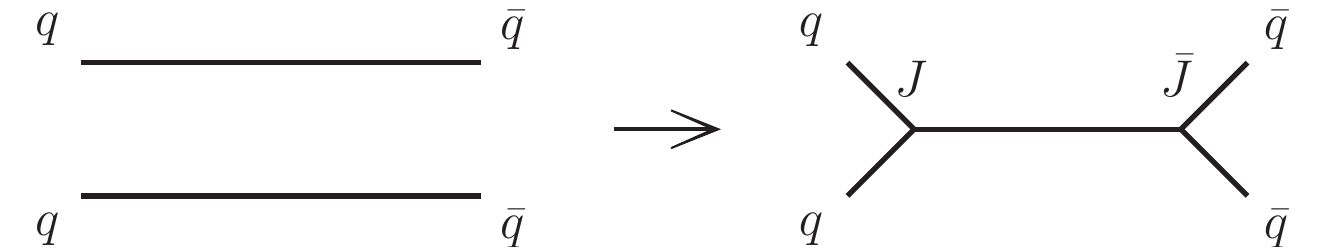} }\\[1cm]
\subcaptionbox{Type III: baryon-style junction reconnection\label{fig:posReconIII}}{
    \includegraphics[width=0.7\textwidth]{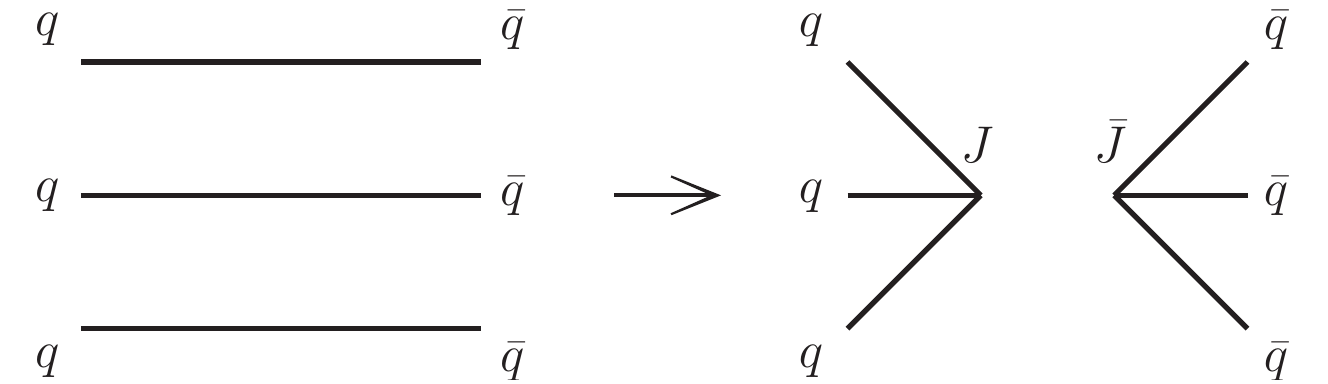} } \\[1cm]
\subcaptionbox{Type IV: zipper-style junction reconnection\label{fig:posReconIV}}{
    \includegraphics[width=0.7\textwidth]{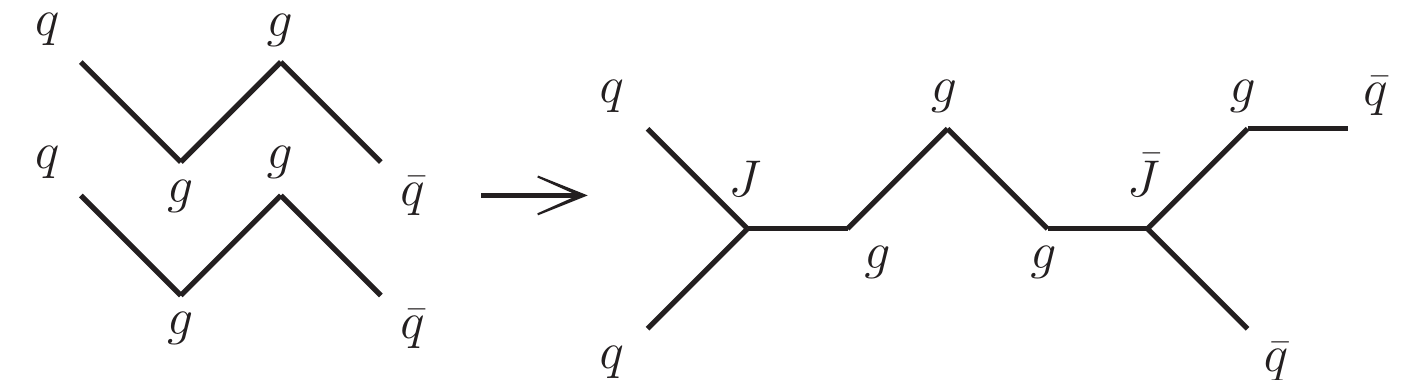}}
  \caption{\label{fig:posRecon} The four different allowed reconnection types. 
    Type I (\subref{fig:posReconI}) is the ordinary string reconnection. Type II (\subref{fig:posReconII}) is the formation of
    a connected junction antijunction pair. Type III (\subref{fig:posReconIII}) is the formation of 
    junction and antijunction, which are not directly connected. Type IV (\subref{fig:posReconIV}) is 
    similar to type II except that it allows for gluons to be added between the two junctions.}
\end{figure}

The starting point for the model is the LC configuration emerging from
the showers + beam remnants.
Thus between each LC-connected pair of partons a tentative dipole is 
constructed. This configuration is then changed by allowing two
(or three) dipoles to reconnect, and this procedure is iterated until no more
reconnections occur. In each step of the algorithm, four different
types of reconnections can occur, illustrated in fig.
\ref{fig:posRecon}:
\begin{enumerate}
\item simple dipole-type reconnections involving two dipoles that
  exchange endpoints (\figRef{fig:posReconI});
\item two dipoles can form a junction-antijunction 
structure (\figRef{fig:posReconII});
\item three dipoles can form 
a junction-antijunction structure (\figRef{fig:posReconIII});
\item two multi-parton string systems can form a junction and an
  antijunction at different points along the string and connect them
  via their gluons (\figRef{fig:posReconIV}). 
\end{enumerate}
Note that, although mainly dipoles between 
quarks are shown in the illustrations, all dipoles
($q$-$\bar{q}$, $q$-$g$,  
$g$-$\bar{q}$ and $g$-$g$) are treated in the same manner in the
implementation. Within an LC dipole, 
the quark and antiquark are assumed to be completely colour coherent,
so that the probabilities for two dipoles to be in a colour-coherent
state can be found by the standard $SU(3)$ products. In full QCD, the
probabilities for type I (dipole) and II (junction) reconnections for $q$-$\bar{q}$
dipoles are given by \eqsRef{eq:33bar} 
and \eqref{eq:33} as $P^{q\bar{q}}_\mathrm{I}=1/9$ and
$P^{qq}_\mathrm{II}=1/3$, respectively. For $gg$ dipoles, the
calculation is complicated slightly by the fact that \eqRef{eq:88}
takes into account both the colour and anticolour charges of both of the
gluons. With a probability of $P^{gg}_\mathrm{I}=8/64=1/8$ each, either ``side''
(colour or anticolour) of the gluons are allowed to reconnect (for a
$1/64$ probability that CR is allowed on both sides). And with a total
probability of $P^{gg}_\mathrm{II}=20 / 64=5/16$ either one \emph{or} the
other side is allowed a junction-type reconnection (both sides would
be equivalent to a dipole-style reconnection already counted
above). For simplicity, the index rules described in the beginning of
this section have been defined to treat $q\bar{q}$, $qg$, and $gg$ dipoles all
on an equal footing. The result is a compromise of
$P_\mathrm{I}=1/9$ for all dipole-type reconnections and $P_\mathrm{II}=2/9$
for all junction-type reconnections. Differences between $q\bar{q}$
and $gg$ combinations still arise due to gluons being prevented from
having the same colour and anticolour indices, and since the
combination of two type-II reconnections is equivalent to a type-I
reconnection. A comparison between the weights resulting from our
simplified treatment and the multiplet weights in full QCD for the
simple case of two-parton combinations was  illustrated in
\figRef{fig:QCDmultiplet}. Note also that the probability for a
type-III reconnection among three uncorrelated $q\bar{q}$ dipoles 
(essentially creating a baryon from three uncorrelated quarks) is
$P^{qqq}_\mathrm{III}=1/27$ in full QCD  (among the 27 different ways to combine 3
quarks, only one is a singlet) while it is only 2/3 as large in our 
model; $P_\mathrm{III}=2/9\times 1/9=2/81$. Although our model should
be a significant step in the right direction, we therefore still
expect a tendency to underestimate baryon production. When tuning the
model below, we shall see that we are able to compensate for this by letting
junction-type reconnections appear somewhat more energetically
favourable compared with dipole-type reconnections. 

To recapitulate the model implementation, each dipole is assigned a
random index value between 1 and 9. Two dipoles are allowed to do a type-I  
reconnection if the two numbers are equal, providing the $1/9$ 
probability. Type-II reconnections are allowed if the two 
numbers modulo three are equal and the indices are different (e.g. 1-4 and 1-7), 
thereby providing the $2/9$ probability. Three dipoles are allowed to do the 
type-III reconnection if they all have the same index modulo three and are 
all different (1-4-7). The type-IV reconnection follows the same principles: 
a dipole needs to have the same index, and a junction needs to have different-but-equal-under-modulo-three index. The exact probability for type-IV 
reconnections depend on the number of gluons in the string. 

The number of allowed colour indices can in principle be changed (the
9 above), e.g.\ to vary the strength of CR. 
However, the type-II, -III, and -IV reconnections rely on the use of modulo, thus 
care should be taken if junction formation is allowed. A different method to 
control the strength of the CR will be discussed below.

The above colour considerations only tell which new colour configurations are
\emph{allowed} and not whether they are \emph{preferable}. 
To determine this, we invoke a minimisation of the $\lambda$ string-length measure. The
$\lambda$ measure can be interpreted as the potential energy of a string, more detailed it
is the area spanned by the string prior to hadronisation. It is closely connected
to the total rapidity span of the string, and thereby also its total particle
production. The minimisation is carried out by only allowing reconnections
that lower the $\lambda$ measure, which ensures that a local minimum is reached.

A further complication is that, while the $\lambda$-measure for 
a quark-antiquark system with any number of gluon kinks in between 
is neatly defined by an iterative procedure~\cite{Andersson:1998tv},
the measure defined there did not include junction structures. The first 
extension to handle these were achieved by starting from the simple measure 
between a quark antiquark dipole \cite{Sjostrand:2002ip}:
\begin{equation}
  \lambda^{q\bar{q}} ~= ~\ln \left (1+\frac{s_{q\bar{q}}}{2m_0^2} \right) 
\label{eq:lambda1}
\end{equation}
where $s_{q\bar{q}}$ is the dipole mass squared and $m_0$ is a constant with dimensions
of energy, of order $\Lambda_\mrm{QCD}$. For high dipole masses, the
``$1$'' in \eqRef{eq:lambda1} 
can be neglected, splitting the $\lambda$-measure neatly into two parts: 
one from the quark and one from the antiquark end 
(in the dipole rest frame):
\begin{equation}
\lambda^{q\bar{q}} ~\stackrel{s\gg
    m_0^2}{\to}~ \ln\left(\frac{s}{2m_0^2}\right) ~= ~\ln
  \frac{\sqrt{2}E_q}{m_0} + \ln\frac{\sqrt{2}E_{\bar{q}}}{m_0}~.
\end{equation}
 The extension to handle a junction system
used the same method, going to the junction rest-frame and adding up the
``$\lambda$-measures'' from all the three (anti-)quark ends. The end result became
\begin{equation}
  \lambda^{q_1q_2q_3} = \ln \frac{\sqrt{2}E_1}{m_0}+\ln \frac{\sqrt{2}E_2}{m_0}
  + \ln \frac{\sqrt{2}E_3}{m_0}
\end{equation}
where the energies are calculated in the junction rest
frame\footnote{Note: we use a slightly  
different definition of $m_0$ here compared to the original paper 
\cite{Sjostrand:2002ip}}. This procedure
worked well in the scenarios considered in that study, since all the dipoles 
had a relative large mass. However, in the context of our CR model, we
will often be considering dipoles that have quite small masses. In
that case, continuing to ignore the ``$1$'' in \eqRef{eq:lambda1} 
can lead to arbitrarily large  negative $\lambda$ measures. Among
other things, such a behaviour could allow soft  
particles with vanishing string lengths to have 
a disproportionately large impact on dipoles with a
large invariant mass. Generalising this behaviour to soft junction
structures results in similar effects, namely that soft  
particles can have a disproportionately large effect.

An alternative measure is here proposed to remove the problem with
negative string lengths,
\begin{equation}
\lambda' = \ln \left (1 + \frac{\sqrt{2}E_1}{m_0} \right ) + 
          \ln \left (1 + \frac{\sqrt{2}E_2}{m_0} \right )
\end{equation}
where the energies are calculated in the rest-frame of the
dipoles. This measure is always positive definite. 
In the case of massless particles the $\lambda'$-measure can be rewritten to
\begin{equation}
  \lambda' = \ln \left( 1+ \frac{s}{2m_0^2}+\frac{\sqrt{2s}}{m_0} \right )
\end{equation}
where again $s$ is the invariant mass squared of the dipole and $m_0$ is
a constant. The  
two measures agree in the limit of large invariant masses ($s \gg m_0$). 
The implementation includes a few alternative 
measures as options, but the above is
chosen as the default measure and therefore also the one that the parameters 
are tuned for.

A final complication regarding the $\lambda$ measure is that the form
above cannot be used to describe the distance between  
two directly connected junctions. Instead the same measure as described in 
\cite{Sjostrand:2002ip} 
is also used in this study ($\lambda = \beta_1 \beta_2 + 
\sqrt{(\beta_1\beta_2)^2-1}$, where $\beta_1$ and $\beta_2$ are the 4-velocities
of the two junction systems).

Since the $\lambda$-measure for junctions introduces additional approximations, 
a tuneable parameter is added to control the junction production. 
Several options for this parameter are possible and we settled on a
$m_{0\mrm{j}}\neq m_0$ in the $\lambda$-measure for junctions. A higher $m_{0\mrm{j}}$ means a lower 
$\lambda$ measure, resulting in an enhancement of the junction production.
We cast the free parameter as the ratio, 
\begin{equation}
\mbox{\ttt{junctionCorrection}~~~:~~~}C_j = m_{0\mrm{j}}/m_{0}~,~~~
\end{equation}
thus a value $C_j$ above unity indicates an  enhancement in 
junction production, and vice versa. 
The possibility of a junction enhancement can be seen as providing a 
crude mechanism to compensate for the intrinsic suppression of junction 
topologies in the colour-space model. Indeed, in the section on tuning 
below we find that values above unity are preferred in order to fit the 
observed amounts of baryon production.

\begin{figure}[tp]
  \centering
  \subcaptionbox{Ordinary pseudoparticle\label{fig:pseudoa}}{\includegraphics[width=0.4\textwidth]{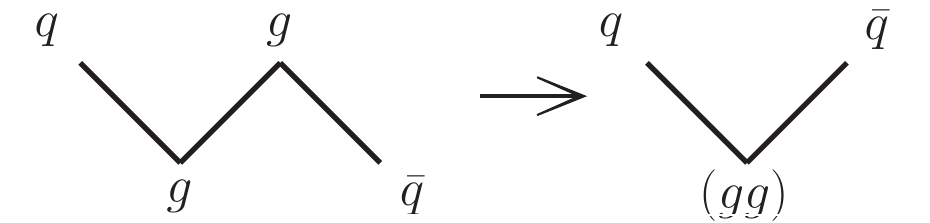}}
  \hspace{30pt}
  \subcaptionbox{Junction pseudoparticle\label{fig:pseudob}}{\includegraphics[width=0.4\textwidth]{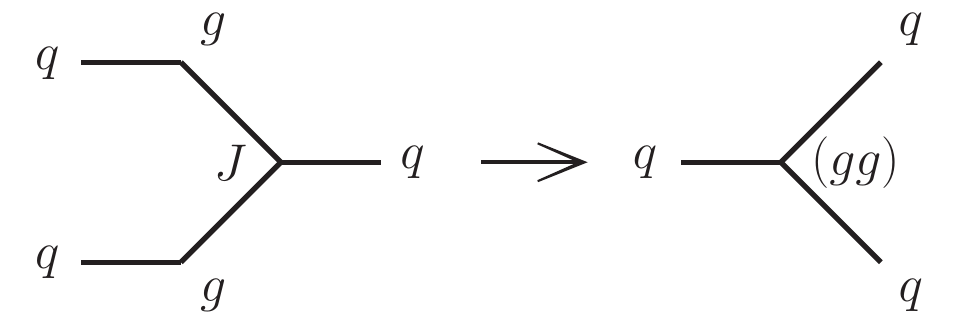}}
  \caption{
    \label{fig:pseudo} The figure shows how two gluons are turned into a
    pseudoparticle depending on whether they are connected via an ordinary
    string (\subref{fig:pseudoa}) or a junction (\subref{fig:pseudob}). The
    (gg) represents the formed pseudoparticle.}
\end{figure}

In the context of CR, it is generally the dipoles with the largest
invariant masses which are the most interesting; they are the ones for
which reconnections can produce the largest reductions of the
$\lambda$ measure. However, as evident from the above discussion,
dipoles with \emph{small} invariant masses can actually be the most
technically problematic to deal with. It was therefore decided to remove
dipoles with an invariant mass below $m_0$ from the colour reconnection. 
Technically this is achieved by combining the small-mass pair into a new 
pseudo-particle. For an ordinary dipole this is a trivial task (\figRef{fig:pseudoa}), but if the dipole 
is connected to a junction the technical
aspects becomes more complicated (\figRef{fig:pseudob}). The easiest way to think of this is as an 
ordinary diquark, but in addition to these we can have digluons, which will have 
three ordinary (anti-)colour tags. Note that we do not intend these to
represent any sort of weakly bound state; we merely use them to
represent a low-invariant-mass collection of partons whose internal
structure we consider uninteresting for the purpose of CR. The pseudo-particles are 
formed after the LC dipoles are formed, and also after any colour reconnections if the
new dipoles have a mass below $m_0$. Increasing $m_0$ will therefore 
lower the amount of CR. Only small effects occur for variations around 
the $\Lambda_{\text{QCD}}$ scale, however increasing $m_0$ beyond 1 GeV 
introduces a significant reduction of CR.

The complete algorithm for the colour reconnection can be summarised as below.
\begin{enumerate}
\item Form dipoles from the LC configuration.
\item Make pseudoparticles of all dipoles with mass below $m_0$.
\item Minimise $\lambda$-measure by normal string reconnections.
\item Minimise $\lambda$-measure by junction reconnections.
\item If any junction reconnections happened return to point 3.
\end{enumerate}
The choice to first do the normal string reconnections before trying to form any
junctions is due to the algorithm not allowing to remove any junction pairs. 

Since each reconnection is required to result in a lower $\lambda$-measure than
the previous one, the minimisation procedure is only expected to reach a local 
minimum. A possible extension to reach the global minimum would be to use
simulated annealing~\cite{1983Sci...220..671K}. This is, e.g., the
approach adopted in the HERWIG++ CR model~\cite{Gieseke:2012ft}. 
However this would also require the 
implementation of inverse reconnections (i.e. a junction and an 
antijunction collapsing to
form strings, and the unfolding of pseudo-particles.). Secondly the 
computational time needed to find the global minimum would slow down 
the event generation speed very significantly. For purposes of this
implementation, we therefore restrict ourselves to a
 local deterministic minimisation here, noting that an algorithm
 capable of reproducing the full expected area-law exponential would
 be a desirable future refinement. 

\subsubsection{Hadronisation of Multi-Junction Systems}
The existing junction hadronisation model~\cite{Sjostrand:2004pf} was
developed mainly for the case of string systems containing a single
junction (in the context of baryon-number violating SUSY decays like
$\tilde{\chi}^0 \to qqq$). 
For such systems, the strategy of  is to
take the two legs with lowest energy in the junction rest-frame and hadronise them 
from their respective quark ends inwards towards the junction, until
a (low) energy threshold is reached, at which point the two endpoints are
combined into a diquark (which contains the junction inside). This
diquark then becomes the new endpoint of the last string piece, which
can then be fragmented as usual. 

The case of a junction-antijunction system was also addressed
in~\cite{Sjostrand:2004pf} (arising e.g., in the case of $e^+e^-\to
\tilde{t}\,\tilde{t}^* \to \bar{q}\bar{q}\,qq$), but the new treatment of the beam
remnants presented here, as well as the  
new CR model, can produce configurations with any number of
colour-connected junctions and antijunctions. 
This goes beyond what the existing model can handle. 

The systems of equations describing such arbitrarily complicated string
topologies are likely to be quite involved, with associated risks of
instabilities and pathological cases. Rather than attempting to address
these issues in full gory detail, we here adopt a simple
``divide-and-conquer'' strategy, slicing the full system into
individual pieces that contain only one junction each, via the following 3 steps:
\begin{figure}[t]
  \centering
  \subcaptionbox{Single gluon split\label{fig:junSplit1a}}{\includegraphics[width=0.7\textwidth]{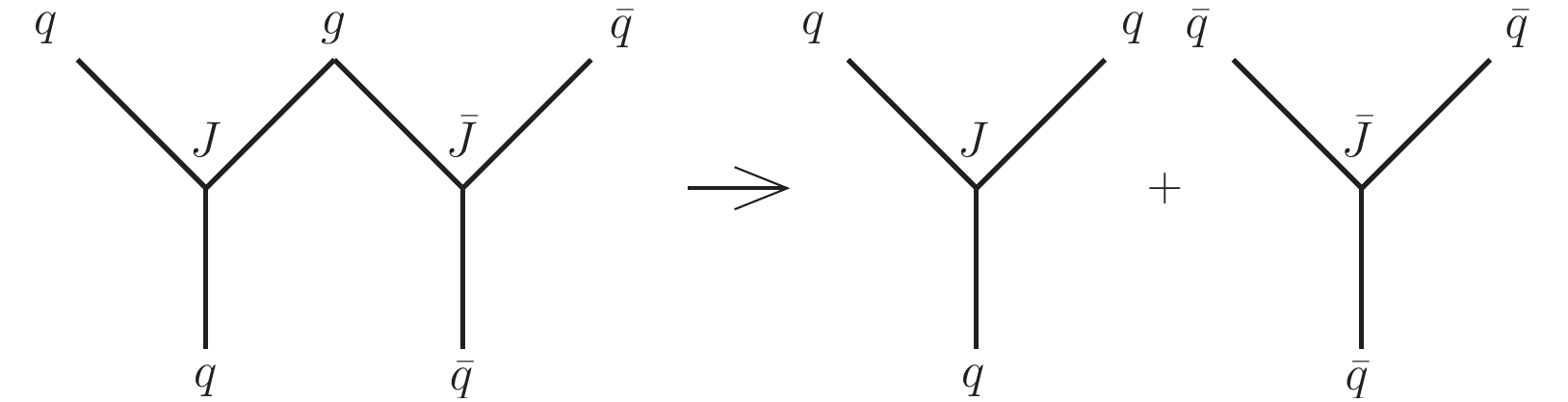}}

  \vspace{8pt}

  \subcaptionbox{Multi gluon split\label{fig:junSplit1b}}{\includegraphics[width=0.7\textwidth]{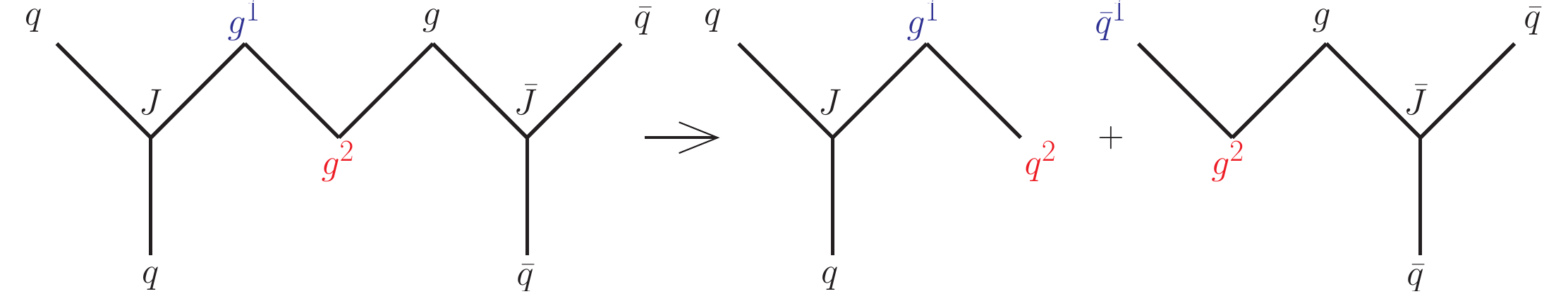}}
  \caption{
    \label{fig:junSplit1} The figure shows how connected junctions are separated
    if they are connected by respectively a single gluon (\subref{fig:junSplit1a}) or multiple 
    gluons (\subref{fig:junSplit1b}). The indices indicate where the split happens and which
    particles each gluon splits into.}
\end{figure}

\begin{itemize}
\item  If a junction and an antijunction are connected with a single gluon 
  between them, that gluon is forced to split into two light quarks (u,d and s) that each equally share the
  4-momentum of the gluon (corresponding to $z=\frac12$). Since the
  gluon is massless, the two quarks will have to be parallel (\figRef{fig:junSplit1a}).
\item If a junction and an antijunction are connected with at least 2
  gluons in between, the gluon pair with the highest invariant mass is found, and
  is split according to the string-fragmentation function. The highest invariant
  mass is chosen due to it having the largest phase space and being the most likely to have a string
  breakup occur. The split is done in such a way that the two gluons
  are preserved but each of them give up part of their 4-momentum to the new
  quark pair. The new quark that is colour-connected to one of the gluons will
  be parallel to the other gluon (\figRef{fig:junSplit1b}, where the
  indices indicate who is parallel with whom.).
\item After the two rules above have been applied, only directly-connected
  junction-antijunctions are left. 
  If all three legs of both junctions
  are connected to each other, the system contains no partons and can
  be thrown away. 
  If two of the legs are directly connected, the
  junction-antijunction system
  is equivalent to a single string piece and is replaced by such, see
  \figRef{fig:junSplit2a}. 
  Finally, the case of a single direct
  junction-antijunction connection is dealt with differently,
  depending on whether the system contains further
  junction-antijunction connections 
  or not. In the former case, illustrated in \figRef{fig:junSplit2b}, 
  the maximum number of junctions 
      are formed from the partons directly connected to the junction system. 
      The remaining particles are formed into
      normal strings. In the example of \figRef{fig:junSplit2b}, three
      quarks are first removed to form a junction system; the
      remaining $q$ and $\bar{q}$ then have no option but to form a normal string. The current method randomly
      selects which outgoing particles to connect with junctions. One
      extension would be to use the string measure to decide who combines with
      whom. (However the effect of this might be smaller than expected, since
      the majority of the multi-junction configuration comes from the beam remnant
      treatment, which later undergoes CR.) 

      For cases with a single
      direct junction-antijunction string piece and no further
      junctions in the system,
     illustrated in \figRef{fig:junSplit2c}, the $\lambda$-measure is
     used to determine whether the two junctions should annihilate or
     be kept~\cite{Sjostrand:2004pf} (essentially by determining
     whether the strings pulling on the two junctions cause them to
     move towards each other, towards annihilation, or away from
     each other). If the junctions survive, a new $q\bar{q}$ pair is formed by
      taking momentum from the other legs of the junction. Otherwise the
      junction topology is replaced by two ordinary strings. An option to always keep 
      the junctions also exists.
\end{itemize}
\begin{figure}[t]
  \centering
\subcaptionbox{Doubly-connected $J\bar{J}$ system\label{fig:junSplit2a}}{
  \includegraphics[width=0.7\textwidth]{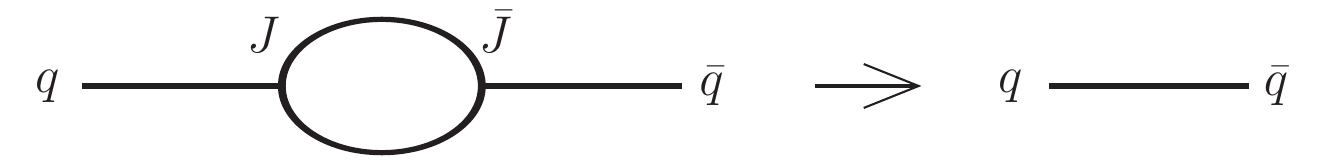}}\\[5mm]
\subcaptionbox{Multiple $J\bar{J}J\cdots$ Connections\label{fig:junSplit2b}}{
  \includegraphics[width=0.7\textwidth]{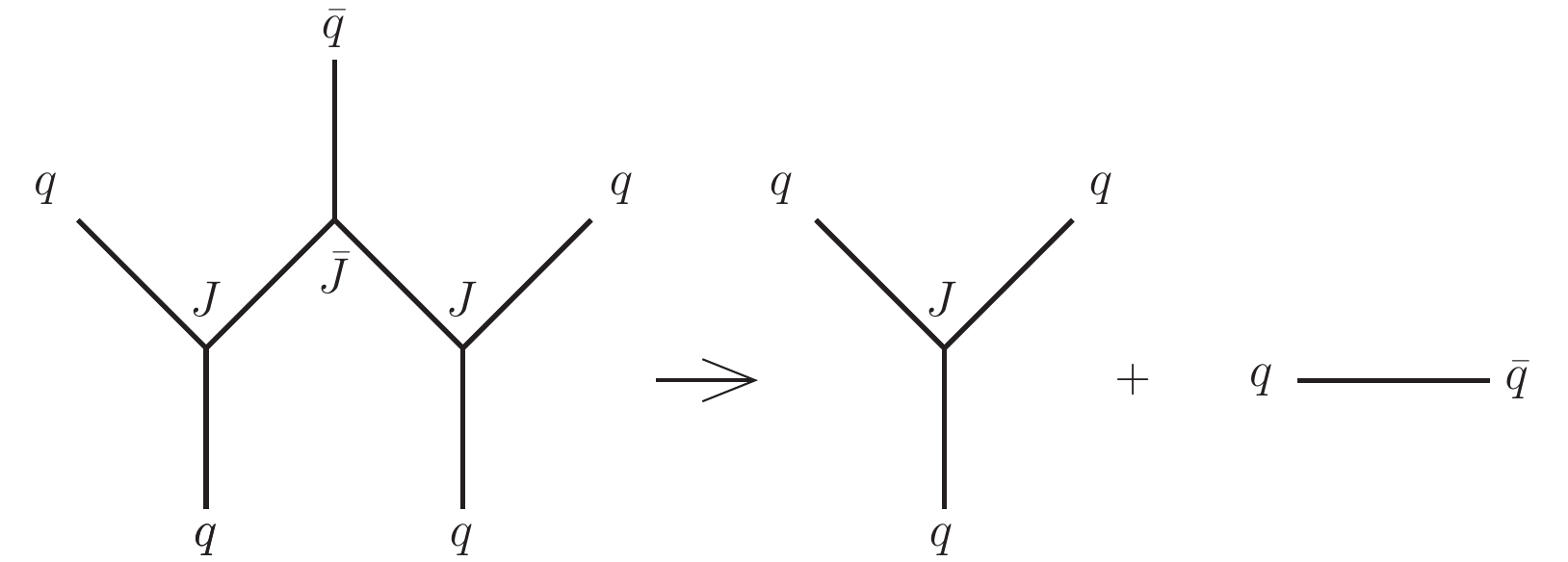}}\\[5mm]
\subcaptionbox{A single $J\bar{J}$ Connection\label{fig:junSplit2c}}{
  \includegraphics[width=0.7\textwidth]{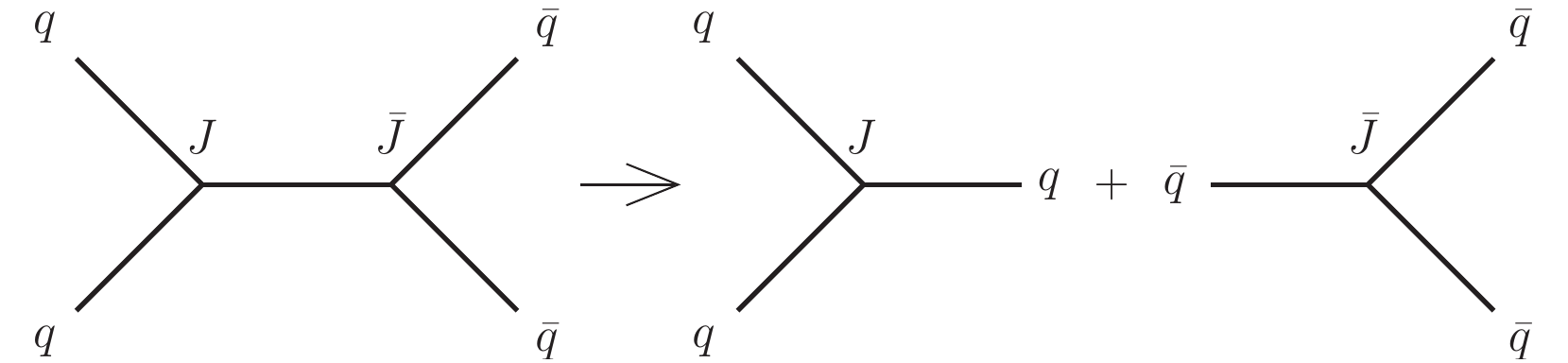}}
  \caption{
    \label{fig:junSplit2} The figure shows how directly connected junctions are
    separated. Fig.~{\subref{fig:junSplit2a}} shows the replacement of a
    doubly-connected junction-antijunction system by an ordinary
    string piece. Fig.~{\subref{fig:junSplit2b}} shows the method used to reduce
    systems with more than two interconnected junctions. Fig.~{\subref{fig:junSplit2c}}
    shows the split-up of a system containing exactly one
    junction-antijunction connection, into two separate junction
    systems.}
\end{figure}

\subsubsection{Space-Time Structure \label{sec:spacetime}}

By default, we do not account for any space-time separation between different
MPI systems. This is motivated by the observation
that, physically, the individual MPI vertices can at most be separated
by transverse distances of order the proton radius, which by
definition is small compared with the length of any string 
long enough to fragment into multiple particles.

We do note, however, that in order for reconnections to occur between
two string pieces, they should be in \emph{causal contact}; if either
string has already hadronised before the other forms, there is no
space-time region in which reconnections between them 
could physically occur. In the rest frame of a hadronising string
piece, we take the formation time of the corresponding QCD dipole 
to be given roughly by the inverse of its invariant mass, $\tau_\mrm{form} \sim
1/m_\mrm{string}$. Alternative measures (e.g., the $k_\perp$ evolution
variable of the PS) could also have 
been used, and to allow at least a range of variations of the exact
definition, a free parameter is introduced. The time at which the
string piece begins to hadronise is related to  
the inverse of $\Lambda_{\text{QCD}}$, $\tau_\mrm{had} \sim 1/
\Lambda_{\text{QCD}}$. In order for reconnections 
to be possible between two string pieces, we require that they must be
able to resolve each other during the time between formation and
hadronisation, taking time-dilation effects caused by relative boosts
into account. There are several ways in which this requirement can be
formulated at the technical level, and accordingly we have implemented
a few different options in the code. In principle, the two strings can
be defined to be in causal contact if the relative boost parameter fulfils:
\begin{equation}
\gamma\, \tau_\mrm{form}  < C_\text{time} \, \tau_\mrm{had} \Rightarrow 
\frac{\gamma
\,  \, c}{m_\mrm{string}r_\mrm{had}}  < C_\text{time} 
\end{equation}
where $C_\text{time}$ is a tuneable parameter and  $r_\mrm{had}(=\,
\tau_\mrm{had}\, c \equiv 1~\mrm{fm})$ is a fixed constant given by the typical
hadronisation scale. There are however two major problems  
with this definition: first it is not Lorentz invariant; the two dipoles will not always agree 
on whether they are in causal contact or not. This can be circumvented, 
by either requiring both to be able to resolve each other (strict) or just either of them to be 
able to resolve the other (loose). Secondly, the emission of a soft gluon from an otherwise 
high-mass string changes $m_\mrm{string}$ significantly for each of
the produced string pieces, which gives an undesirable infrared
sensitivity to this measure, reminiscent of the problems associated
with defining the $\lambda$ string-length measure itself. 
One way to avoid this problem is to consider the \emph{first} 
formation time of each colour line, i.e.\ the dipole mass at the time the
corresponding colour line was first created in the shower, which we
have implemented as an alternative option. No matter the exact 
definition of formation time and hadronisation time, all models agree that 
reconnection between boosted strings should be suppressed. A final
extremely simple way to capture this in a Lorentz-invariant way is to
apply a cut-off directly on the boost factor $\gamma$, which thus provides
a simple alternative to the other models.  

These different methods have all been implemented and are
available in PYTHIA, via the mode
\texttt{ColourReconnection:timeDilationMode}. 
The $C_\text{time}$ parameter introduced above is specified by
\texttt{ColourReconnection:timeDilationPar} and controls the size of
the allowed relative boost factor for reconnections to occur. As such
it can be  used to tune the amount of CR. Its optimal value will vary depending
on the method used, but after  the methods are tuned they produce
similar results (see section \ref{sec:tuning}  for details). 

A final aspect related to space-time structure that deserves special
mention is resonance decays. By default, these are treated separately
from the rest of the event. Physically, this is well  
motivated for longer-lived particles (e.g., Higgs bosons), which are expected 
to decay and hadronise separately. For shorter-lived resonances the separation of the
MPI systems and resonance decays 
is physically not so well motivated. E.g., most $Z/W$ bosons and top quarks will decay 
before hadronisation takes place, $\Gamma \gg \Lambda_\mrm{QCD}$, and as such should 
be allowed to interact with the particles from the MPI systems, ideally with a slightly 
suppressed probability due to the decay time.

Currently, only two extreme cases are implemented, corresponding to
letting CR occur \emph{before} or \emph{after} (all) resonance decays. The
corresponding flag in PYTHIA is called
\texttt{PartonLevel:earlyResDec}. When switched on, CR is performed
\emph{after} all resonance decays have occurred, and all 
final-state partons therefore participate fully in the CR. Since no
suppression with resonance lifetime is applied, this
gauges the largest possible impact on resonance decays from CR. 
When switched off, CR is performed \emph{before} resonance decays,
hence involving only the beam remnant and MPI systems. It is equivalent to
assuming an infinite lifetime for the resonances, and hence estimates the
smallest possible impact on resonance decays from CR. An optional additional
CR can be performed between the decay products of the resonance decays, with the
physics motivation being $H\rightarrow WW \rightarrow
q\overline{q}q\overline{q}$ studies.

To summarise, we acknowledge that the treatment of space-time
separation effects and causality is
still rather primitive in this model. The derivation of a more
detailed formalism for these aspects would therefore be a welcome and interesting
future development.

\section{Constraints and Tuning \label{sec:tuning}}
The tuning scheme follows the same procedure as for the 
Monash 2013 tune~\cite{Skands:2014pea}. However at a more limited scope,
since only CR parameters, and ones strongly correlated with them, are tuned. As
a natural consequence of this, the Monash tune was chosen as the
baseline. As discussed in \secRef{sec:spacetime}, several options are
available for the choice of CR
time-dilation method, which naturally results in slightly different
preferred parameter sets. Here, we consider the following three modes:
\begin{itemize}
\item Mode 0: no  time-dilation constraints. $m_0$ controls the amount
of CR (mode 0);
\item Mode 2: time dilation using the boost factor obtained from the 
final-state mass of the dipoles, requiring all dipoles involved in a
reconnection to be causally connected (strict);
\item Mode 3: time dilation as in Mode 2, but requiring only a  
single connection to be causally connected (loose). 
\end{itemize}
This allows to investigate the consequences of 
some of the ambiguities in the implementation of the model. 
For the purpose of later studies that may want to focus on a single model, we suggest to
use mode 2 as the ``standard'' one for the new CR. The parameters described in
this section will therefore correspond to that particular model, with
parameters for the others given in appendix A. Note that this section only
contains the main physical parameters; for a complete list we again 
refer to appendix A.

\subsection{Lepton Colliders \label{sec:lep}}

We begin with $e^+e^-$ collisions. Only small effects are expected in
this environment, due to the $p_\perp$-ordering of the shower and the absence
of MPIs. Only CR and string-fragmentation variables were studied,  
since the shower was left untouched. The fragmentation model contains three main 
parameters governing the kinematics of the produced hadrons: the
non-perturbative $p_\perp$ produced in string breaks, controlled by the
$\sigma_\perp$ parameter (\ttt{StringPT:sigma}), and the two parameters, $a$ and $b$, which
control the shape of the longitudinal ($z$) fragmentation
function. For pedagogical descriptions,
see e.g.~\cite{Andersson:1998tv,Skands:2012ts,Skands:2014pea,Sjostrand:2014zea}.
Since the effects are expected to be small, we made the choice of
keeping $\sigma_\perp = 0.335~\mathrm{GeV}$ unchanged, adjusting only the longitudinal ($a$
and $b$) parameters. Changing the minimal number of parameters
 also helps to disentangle the effects of CR from the retuning. 
As a verification, a tune with a smaller $\sigma_\perp$ ($0.305~\mathrm{GeV}$) was
considered, however after retuning $a$ and $b$ the two tunes described the LEP
data with a similar fidelity. (The choice of testing a lower $\sigma_\perp$
was made since the CR model tends connect more collinear partons leading to shorter strings, but a harder 
$p_\perp$ spectrum of the produced hadrons~\cite{Ortiz:2013yxa}.) \\  


\begin{figure}[t]
  \captionsetup[subfigure]{skip=3pt}
    \subcaptionbox{\label{fig:lep_Nch1}}{\includegraphics*[scale=0.4]{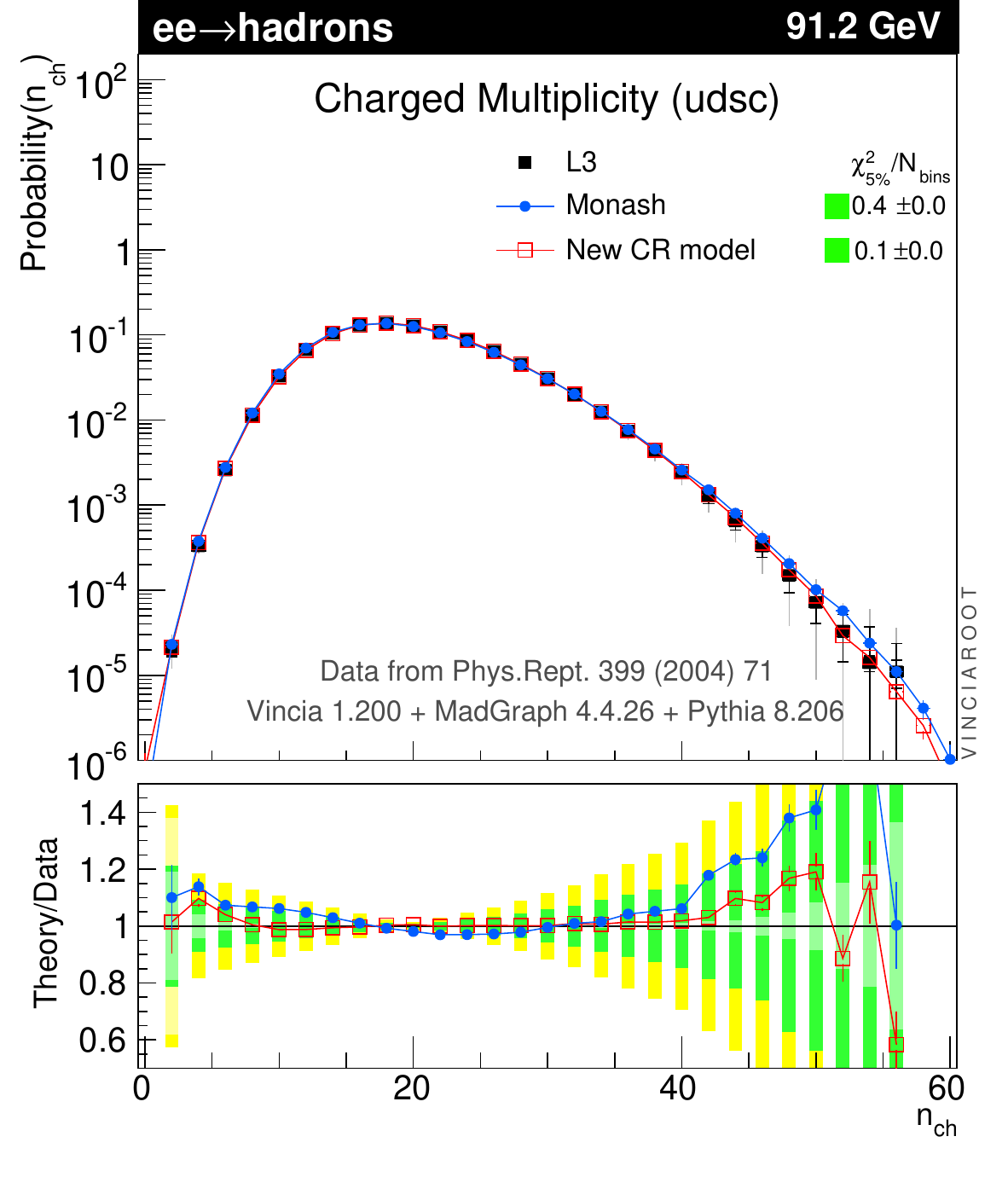}}
    \subcaptionbox{\label{fig:lep_Lnx}}{\includegraphics*[scale=0.4]{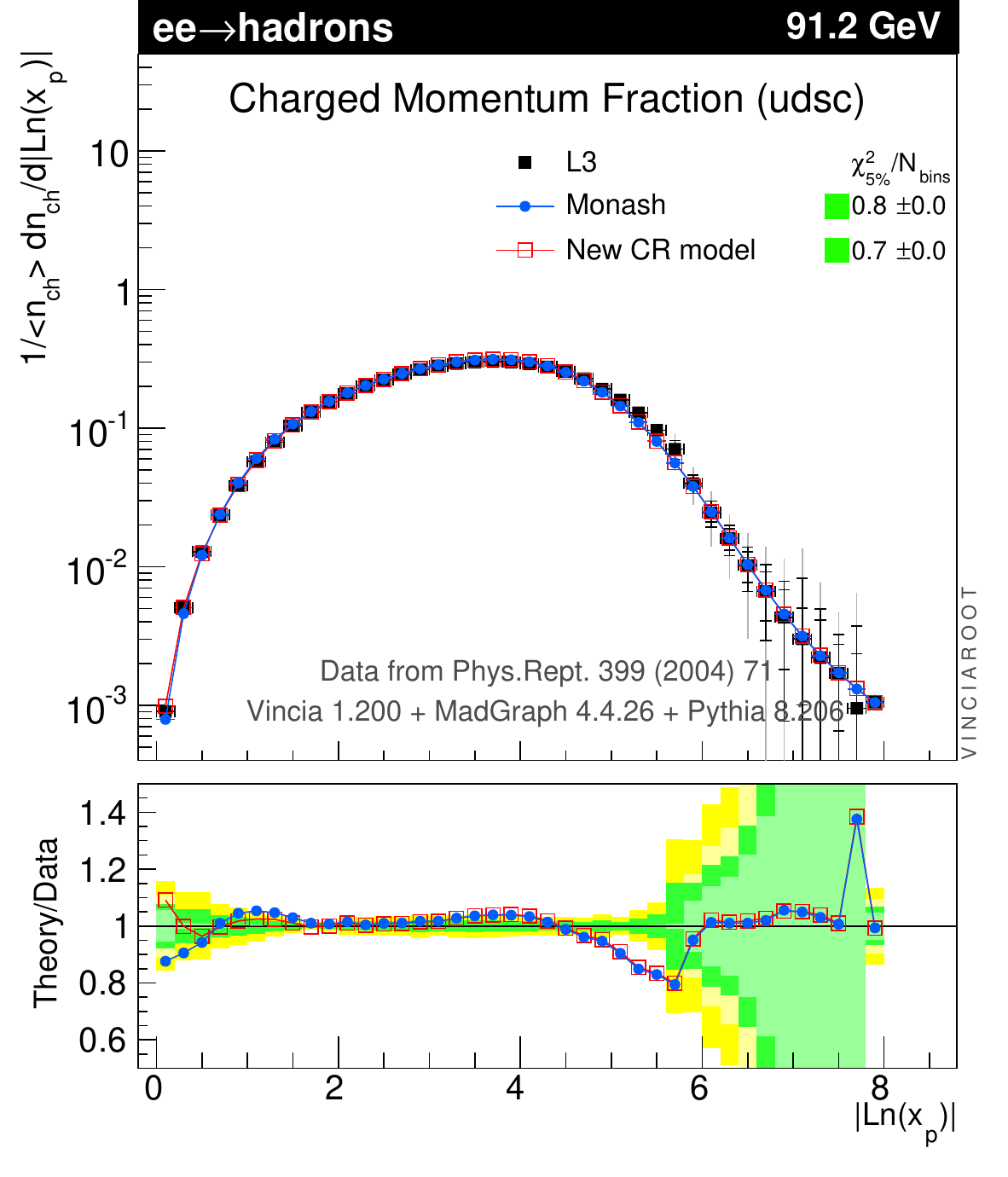}}
    \caption{
      \label{fig:lep_Nch} Charged-particle multiplicity (\subref{fig:lep_Nch1}) and momentum fraction 
      (\subref{fig:lep_Lnx}) spectra, in light-flavour tagged data from the L3
      collaboration~\cite{Achard:2004sv}. (Plots made with 
      \textsc{VinciaRoot}~\cite{Giele:2011cb}. The ratio panes follow
      the now-standard ``Brazilian'' colour conventions, with outer
      (yellow) bands corresponding 
      to $2\sigma$ deviations and inner (green) bands corresponding to
      $1\sigma$ deviations.)
    }
\end{figure}

The determination of the two parameters of the Lund fragmentation
function, $a$ and $b$, is complicated slightly by the fact that they
are highly correlated; choosing both of them to be quite small often
produces equally good descriptions of fragmentation spectra as
choosing both of them large, corresponding to a relatively elongated
and correlated $\chi^2$ ``valley''.
By simultaneously considering both variables and comparing them to both multiplicity 
and momentum spectra, cf.~\figRef{fig:lep_Nch} (with the
{\sl``New CR model''} curve showing
our new model, and {\sl ``Monash''} the baseline Monash 2013 tune), we
here settled on a   
low-valued pair, as compared with the default Monash values:\\

\code{StringZ:aLund = 0.38 \# was 0.68}

\code{StringZ:bLund = 0.64 \# was 0.98} \\

The new CR also alters the ratio between the identified-particle yields, 
especially so for baryon production due to the introduction of additional junctions.
Thus the flavour-selection parameters of the string model also need to
be retuned, by comparing with the total identified-particle yields,
see e.g.~\cite{Skands:2014pea}. As
expected the effects are minimal in $e^+e^-$ collisions, and only 
small changes are required. The modifications were therefore done with
a view to providing a better description also for $pp$ colliders, but
staying within the uncertainties allowed by the LEP data. This resulted in an
adjustment of the parameters for the diquark over quark fragmentation
probability and the strangeness suppression: \\

\code{StringFlav:probQQtoQ = 0.078 \# was 0.081}

\code{StringFlav:probStoUD = 0.2 \# was 0.217} \\

\noindent As expected the diquark over quark probability is reduced due to
the introduction of junctions. More surprisingly is the increased suppression
of strange quarks, since the model a priori should not influence flavour
selection. The technical implementation of the junction hadronisation does,
however,  
introduce a slight enhancement of the strangeness production, due to an even
probability for a gluon to split into an u, d or s quarks when separating
junction systems. This is not 
visible at LEP, but at $pp$ colliders the slightly lower strangeness
fragmentation is favoured.

The final set of fragmentation parameters we define is more technical. 
For junction systems and beam remnants, a separate set of
parameters controls the choice of total spin when two, already
produced, quarks are combined into a diquark. Unlike 
 diquarks produced by ordinary string breaks (whose spin is controlled by the
 parameter \ttt{StringFlav:probQQ1toQQ0}), which
can only contain the light quark flavours (u, d, s), and for which the significant mass
splittings between the light-flavour spin-3/2 and spin-1/2 baryon
multiplets necessitates a rather strong suppression of spin-1 diquark
production (relative to the naive factor 3 enhancement from spin counting),
junction systems in particular can allow the formation of 
baryons involving heavy flavours, which have smaller mass splittings
and which therefore might require less suppression of spin-1 diquarks. 
We note also that diquarks produced in string breakups are produced within the linear
confinement of the string, whereas junction diquarks come from the
combination of two already uncorrelated quarks, so there is a priori
little physics reason to assume the parameters must be identical.  

With the limited amount of junctions in the old model, none for $ee$ and at
most two for $pp$, these parameters previously had almost no influence on measurable
observables and were therefore largely irrelevant  
for tuning. With the additional junctions produced by our model, these
parameters can now give larger effects. Measurements 
of higher-spin and heavy-flavour baryon states at $pp$ colliders are 
still rather limited though, and so far we are not aware of published directly
usable constraints from experiments.  
For the time being therefore, we choose to fix
the parameters to be identical to those for the production of ordinary
diquarks in string breakups: \\

\code{StringFlav:probQQ1toQQ0join = 0.027,0.027,0.027,0.027} ~\\

\noindent The four components give the suppression when the heaviest quark
is u/d, s, c or b, respectively. We stress that this is merely a starting point, 
hopefully to be revised soon by comparisons with new data from the LHC experiments.\\

\subsection{Hadron Colliders \label{sec:pp}}

The retuning to hadron colliders consisted of tuning three main 
parameters: 
\begin{itemize}
\item $C_{\text{time}}$ (\ttt{ColourReconnection:timeDilationPar}):
  controls the overall strength of the colour-reconnection effect via
  suppression of high-boost reconnections, see
  \secRef{sec:spacetime}. Can be tuned to the $\left 
    <p_\perp \right>$ vs $n_{\text{ch}}$ distribution. 
\item $C_\text{j}$ (\ttt{ColourReconnection:junctionCorrection}):
 multiplicative factor, $m_{0\mrm{j}}/m_{0}$, applied to the string-length
  measure for junction systems, thereby enhancing or suppressing the
  likelihood of junction reconnections. Controls the junction
  component of the baryon to meson fraction and is tuned to the
  $\Lambda/K_{s}^0$ ratio. 
\item $p_\perp^{\text{ref}}$ (\ttt{MultiPartonInteractions:pT0Ref}):
  lower (infrared) regularisation scale of the MPI framework. Controls
  the amount of low $p_\perp$ MPIs and is therefore closely 
related to the total multiplicity and can be tuned to the
$d\left<n_{\text{ch}}\right>/d\eta$ distribution. 
\end{itemize}
\begin{figure}[tp]
  \captionsetup[subfigure]{skip=3pt}
  \centering
  \subcaptionbox{\label{fig:meanpTvsNch}}{\!\includegraphics*[scale=0.64]{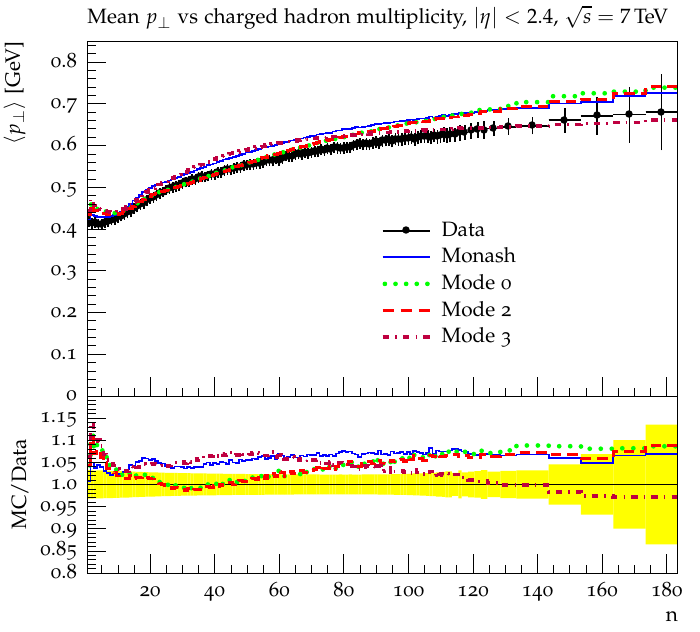}\!}
  \subcaptionbox{\label{fig:dndeta}}{\!\includegraphics*[scale=0.64]{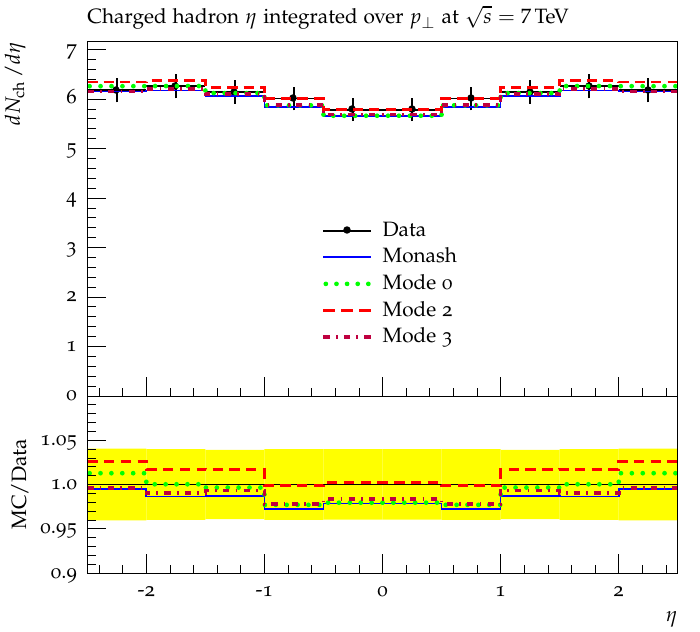}\!} \\
  \subcaptionbox{\label{fig:lambdaksratio}}{\!\includegraphics*[scale=0.64]{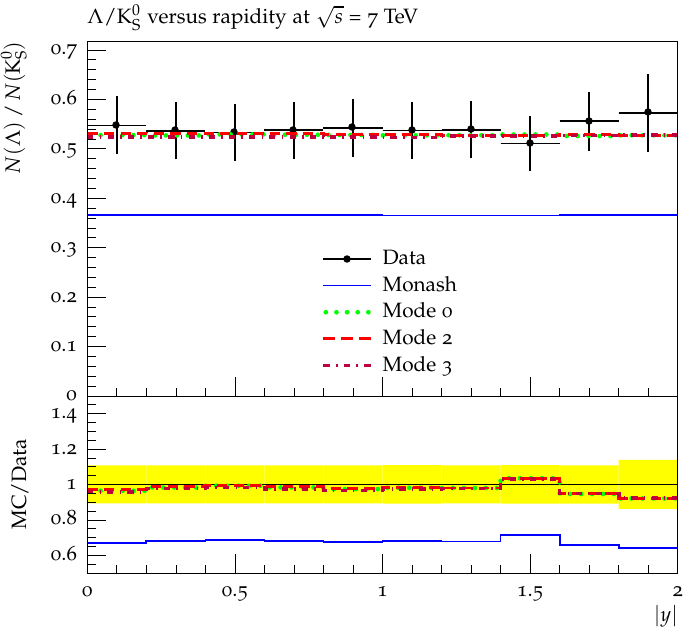}\!}
  \caption{
      \label{fig:lhc_constrains} The average $p_\perp$ as a function of 
      multiplicity~\cite{Khachatryan:2010nk} (\subref{fig:meanpTvsNch}), 
      the average charged multiplicity as a function of pseudorapidity~\cite{Khachatryan:2010us} 
      (\subref{fig:dndeta}), and the $\Lambda/K_s$
      ratio~\cite{Khachatryan:2011tm} (\subref{fig:lambdaksratio}). 
      All observables from the CMS collaboration and plotted with the Rivet 
      framework~\cite{Buckley:2010ar}. All PYTHIA simulations were non single
      diffractive (NSD)
      with a lifetime cut-off $\tau_\text{max} = 10$ mm/c and no
 $p_\perp$ cuts applied to
      the final state particles. The yellow error band represents the
      experimental $1\sigma$ deviation.
    }
\end{figure}
By iteratively fitting each parameter to  
its respective most sensitive curve an overall good agreement with data
was achieved  
(see \figRef{fig:lhc_constrains}) with the following parameters: 
\\

\code{ColourReconnection:junctionCorrection = 1.2~~\# new parameter} 

\code{ColourReconnection:timeDilationPar~~~~= 0.18~\# new parameter} 

\code{MultiPartonInteractions:pT0Ref~~~~~~~~= 2.15~\# was 2.28} \\

\noindent Note in particular this is
      the first time that PYTHIA has been able to describe the
      $\Lambda/K_s$ ratio in $pp$ collisions     
      while remaining consistent with LEP bounds. We explore this
      in more detail in \secRef{sec:newObs}. 

\begin{figure}[tp]
  \captionsetup[subfigure]{skip=3pt}
\subcaptionbox{\label{fig:mult1}$|\eta|<0.5$}{%
\!\includegraphics*[scale=0.64]{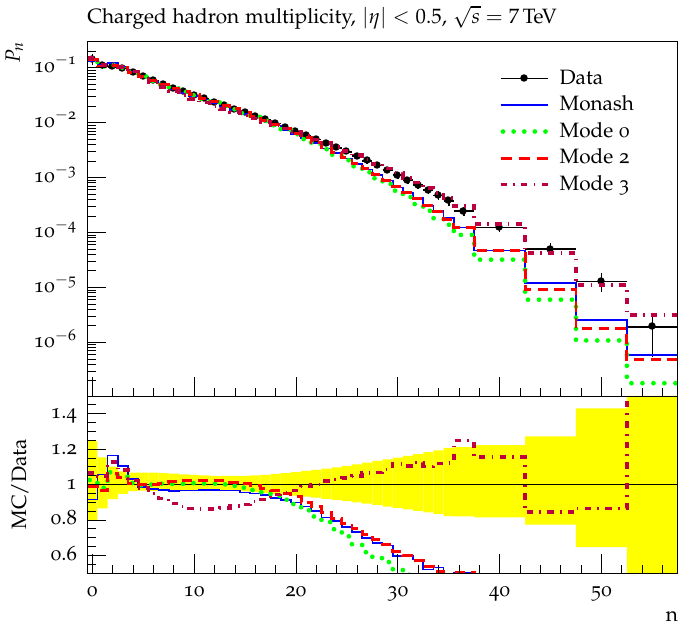}\!}
\subcaptionbox{\label{fig:mult2}$|\eta|<2.4$}{%
\!\includegraphics*[scale=0.64]{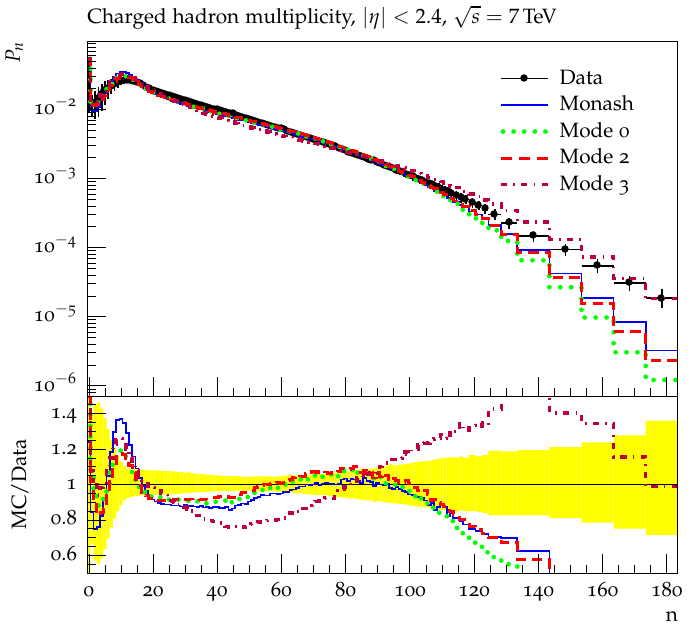}\!}
    \caption{
      \label{fig:multiplicity}
    The two plots show the multiplicity distributions for respectively very
    central tracks (\subref{fig:mult1}) and the full CMS tracker coverage
 (\subref{fig:mult2}), compared with CMS
 data~\cite{Khachatryan:2010nk}. All PYTHIA 
    simulations were NSD with a lifetime cut-off ($\tau_\text{max} = 10$
 mm/c) and no $p_\perp$ cuts were applied to the final state particles. The
 yellow error band represents the experimental $1\sigma$ deviation.
    }
\end{figure}
The $C_\text{j}=1.2$ parameter shows that a slight enhancement of 
junction reconnections (i.e., baryon production) is needed, relative
to mesonic ones. However, given
the approximations used in the implementation of especially the junction
structures, such a difference is not unreasonable. Small differences
between the modes can be seen in the $\left <p_\perp \right>$ vs
$n_{\text{ch}}$ and more significant differences for multiplicity
distributions , cf.~\figRef{fig:multiplicity}. 
With respect to the
latter, however, we note that the differences in the tails of the
multiplicity distributions can be tuned away by modifying the assumed
transverse matter density profile of the proton, which was kept fixed
here to highlight the differences with the minimal number of retuned
parameters. 

\begin{figure}[tp]
  \captionsetup[subfigure]{skip=3pt}
   \subcaptionbox{\label{fig:totema}TOTEM Comparison}{\!\includegraphics*[scale=0.64]{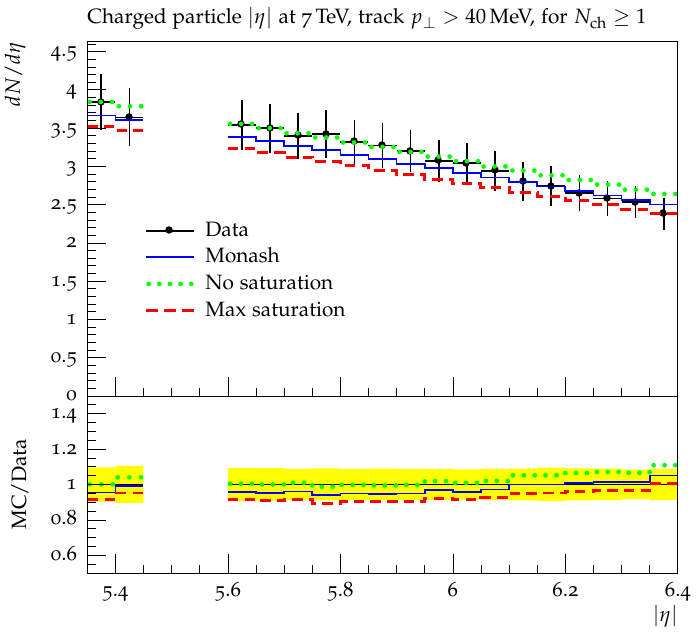}\!} 
   \subcaptionbox{\label{fig:totemb}Generator-Level}{\!\includegraphics*[scale=0.64]{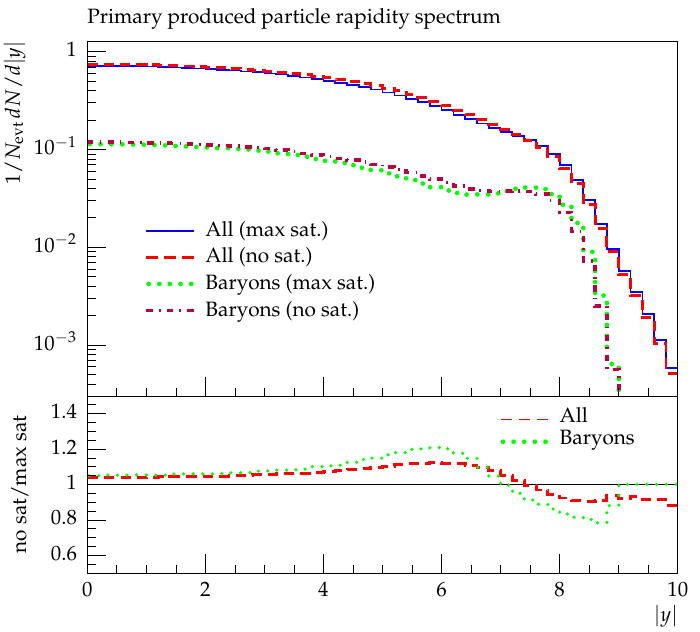}\!}
    \caption{
      \label{fig:Totem}
      (\subref{fig:totema}): different extreme saturation
      choices compared with the TOTEM forward multiplicity 
      data~\cite{Aspell:2012ux}. (The PYTHIA simulation includes all
 soft-QCD processes and a particle lifetime cut-off $\tau_\text{max} =
 10$ mm/c.) The yellow error band represents the
      experimental $1\sigma$ deviation.   
      (\subref{fig:totemb}): MC rapidity distributions for
 respectively all particles 
      and baryons only. (For simplicity only non diffractive (ND) events were used, hadron
      decays were turned off to reflect primary hadron production, and no
      $p_\perp$ cuts were imposed.)
    }
\end{figure}
The new colour treatment of the beam remnant (BR) introduces a single new parameter 
controlling the amount of saturation, cf.~the discussion in
\secRef{sec:remnant}. Due to the low $p_{\perp}$ of the BR particles, 
the effects are largest in the forward direction. We therefore use the 
forward charged multiplicity as measured by the TOTEM
experiment~\cite{Aspell:2012ux} to
compare different modelling choices of this aspect, see \figRef{fig:totema}.  
The difference between no saturation\footnote{Technically:
  \texttt{BeamRemnants:saturation = 1E9.}} $(k_\text{saturation} \to \infty)$ and
maximal saturation $(k_\text{saturation} = 0.1)$ is about $10\%$ and
exhibits no shape difference over the TOTEM  pseudo rapidity range. 
For illustration and completeness, we may also consider what happens over the full 
rapidity range, at least at the theory level. This is illustrated in 
\figRef{fig:totemb}. In the central region, the effect of applying 
saturation is a slight decrease of the particle yield, and thus would already
have been tuned away by $p_\perp^{\text{ref}}$. It was therefore chosen to use
a relative high saturation to mimic the effect of the earlier PYTHIA beam
remnant model:\\

\code{BeamRemnants:saturation = 5} \\

\noindent We emphasise however that this is merely a starting
point, and that a different balance between $p_\perp^{\text{ref}}$ and
$k_\text{saturation}$ may be preferred in future tuning efforts,
especially ones taking a more dedicated look at the forward region. In such a
study the sharing of momentum between the partons in the remnant should also
be considered, since it is known to alter both particle production in the
forward region and the multiplicity distributions.

An interesting signal that may help to break the relative degeneracy between
$p_\perp^{\text{ref}}$ and $k_\text{saturation}$, 
is to look for baryons at high rapidities, which, due to the introduction of junction
structures in the BR can act as further tracers of the degree to which
the BR has been disturbed. This is illustrated by the lower set of
curves in \figRef{fig:totemb}. The effects are indeed seen to be slightly larger for
baryons, however the total cross section is also significantly lower. From
this simple MC study we are not able to say clearly whether such a
measurement, which requires the additional non-trivial ingredient of
baryon identification, would be experimentally feasible.  

\subsection{Direct CR Constraints?}

In the preceding sections, we constrained (``retuned'') the
fragmentation parameters using observables like the charged-particle
multiplicity and fragmentation spectrum, which are indirectly
sensitive to CR effects via the modifications caused to these spectra 
by the minimisation of string lengths. 
But what about observables with more direct sensitivity to CR effects? 
There are two main categories of dedicated CR studies at LEP:
Fully hadronically decaying $WW$ events (looking for reconnections
between the two $W$ systems), and colour-flow sensitive observables in
three-jet events. In this study we restrict our attention to the
latter of these. A follow-up dedicated study of CR effects at $e^+e^-$
colliders, both earlier as well as possible future colliders, is planned. 

\begin{figure}[tp]
\centering
\setlength{\unitlength}{0.83mm}
\captionsetup[subfigure]{skip=20pt}
\vspace*{5mm}
\subcaptionbox{\label{fig:3jeta}}{
\begin{fmffile}{fmf3jeta}
\begin{fmfgraph*}(85,35)
\fmfleft{l1,l2}
\fmfright{r1,r2}
\fmftop{t1,t2,t3}
\fmf{plain}{v,l1}
\fmf{plain}{r1,v}
\fmf{gluon}{v,t2}
\fmffreeze
\fmfv{d.sh=circ,d.siz=5}{v}
\fmfv{label=$q^{1}$}{l1}
\fmfv{label=$\bar{q}^{2}$}{r1}
\fmfv{label=$g^{12}$}{t2}
\fmf{dashes,right=0.05,fore=red,label=$\small\text{string}$,l.side=left}{l1,t2}
\fmf{dashes,right=0.05,fore=red,label=$\small\text{string}$,l.side=left}{t2,r1}
\end{fmfgraph*}
\end{fmffile}
}
\hspace*{-0.5cm}\raisebox{2cm}{\large$\stackrel{1/N_C^2}{\bf\rightarrow}$}\hspace*{-0.5cm}
\captionsetup[subfigure]{skip=20pt}
\subcaptionbox{\label{fig:3jetb}}{
\begin{fmffile}{fmf3jetb}
\begin{fmfgraph*}(85,35)
\fmfleft{l1,l2}
\fmfright{r1,r2}
\fmftop{t1,t2,t3}
\fmfforce{0.44w,0.98h}{gg1}
\fmfforce{0.56w,0.75h}{gg2}
\fmf{plain}{v,l1}
\fmf{plain}{r1,v}
\fmf{phantom}{v,t2}
\fmffreeze
\fmf{gluon}{v,gg1}
\fmf{gluon}{v,gg2}
\fmfv{d.sh=circ,d.siz=5}{v}
\fmfv{label=$q^{1}$}{l1}
\fmfv{label=$\bar{q}^{1}$}{r1}
\fmfv{label=$g^{12}$}{gg1}
\fmfv{label=$g^{21}$}{gg2}
\fmf{dashes,right=0.1,fore=red}{gg1,gg2}
\fmf{dashes,right=0.1,fore=red}{gg2,gg1}
\fmf{dashes,left=0.05,fore=red,label=$\small\text{string}$,l.side=right}{l1,r1}
\end{fmfgraph*}

\end{fmffile}
}
\caption{(\subref{fig:3jeta}): illustration of the correspondence between colour
  connections and string pieces in an ordinary (LC) 3-jet 
  topology. \subref{fig:3jetb}): if the gluon jet is composed of (at least)
  two gluons, there is probability for the
  $q\bar{q}$ system to be in an overall singlet. (The LC string topology remains a possibility as well,
  with the string-length measure $\lambda$ used to decide between them.)
  Notation: $g^{ac}$ denotes a gluon with colour (anticolour) index
  $c$ ($a$). For (anti)quarks, we use $q^{c} \equiv q^{0c}$
  and $\bar{q}^{a} \equiv \bar{q}^{a0}$. \label{fig:3jet}}
\end{figure}
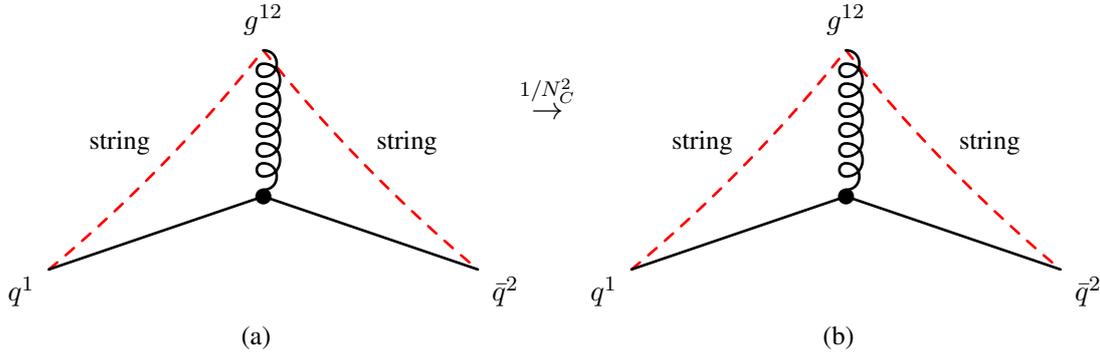

Without CR, the three produced jets will in general be represented 
by a colour string stretched from the quark via the
gluon to the antiquark, illustrated in~\figRef{fig:3jeta}. The string
pieces spanned between the quark and gluon jets lead to a relatively
large particle production between these jets, while there is a
suppression of particle production in the phase-space region directly
between the quark and antiquark jets. However,  
if CR is allowed, there is a $1/N_C^2$ chance that two (or more)
sequentially emitted gluons end up cancelling each others' colour
charges. Thus, if at least one additional gluon was produced in the
FSR, the ``gluon jet'' may effectively become overall colour neutral,
allowing it to decouple from the quark-antiquark system  in colour space. This is
illustrated in~\figRef{fig:3jetb}. There is a caveat to the above, namely,
if the two gluons originate from a single gluon, ie., $g\rightarrow gg$, the
two gluons must form a colour octet. In the gluon-collinear limit, this
colour structure dominates and the probability for the jet to end up
colour neutral should therefore be strongly suppressed, below
$1/N_C^2$. This aspect is not included in our 
model, since the history of the final state particles is not
considered. The model may therefore overestimate the CR effect on 
three-jet events.  

An additional consequence is that the jet will also have a total
electric charge of zero, if all particles from the 
fragmentation fall within the jet. The best observable uses both of these ideas.
Firstly a rapidity gap is required in order to select events with minimum radiation 
between the jets. This also enhances the probability that all the particles from the 
fragmentation falls within the jet. Secondly the jet is required to have a total 
charge of zero. This observable was first proposed during the LEP runs~\cite{Friberg:1996xc} 
and successively followed up by several of the experiments~
\cite{Abbiendi:2003ri,Schael:2006ns,Achard:2003ik}. The studies were limited
to excluding only the most extreme  
CR models, with no conclusions drawn between more moderate CR models and no CR;
the data was located in between the two predictions.

\begin{figure}[tp]
\centering
\captionsetup[subfigure]{skip=3pt}
\subcaptionbox{Rapidity along jet axis\label{fig:lepa}}{\!\includegraphics*[scale=0.64]{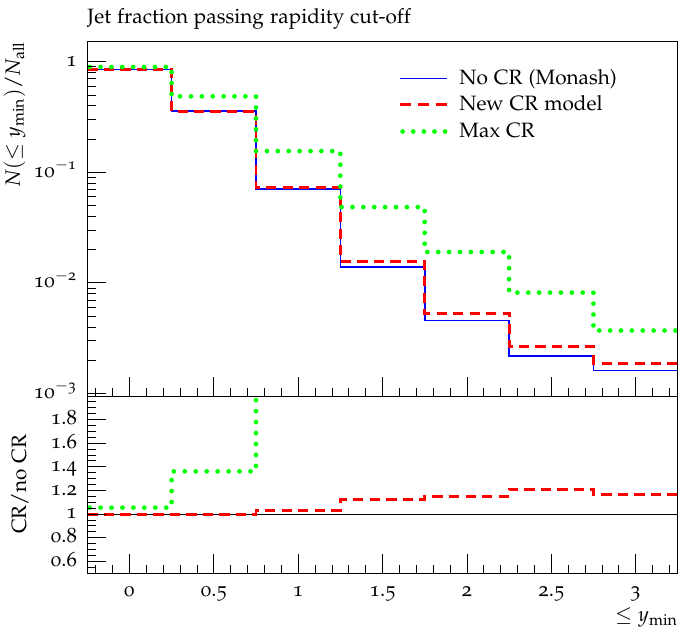}\!} 
\subcaptionbox{Neutral fraction\label{fig:lepb}}{\!\includegraphics*[scale=0.64]{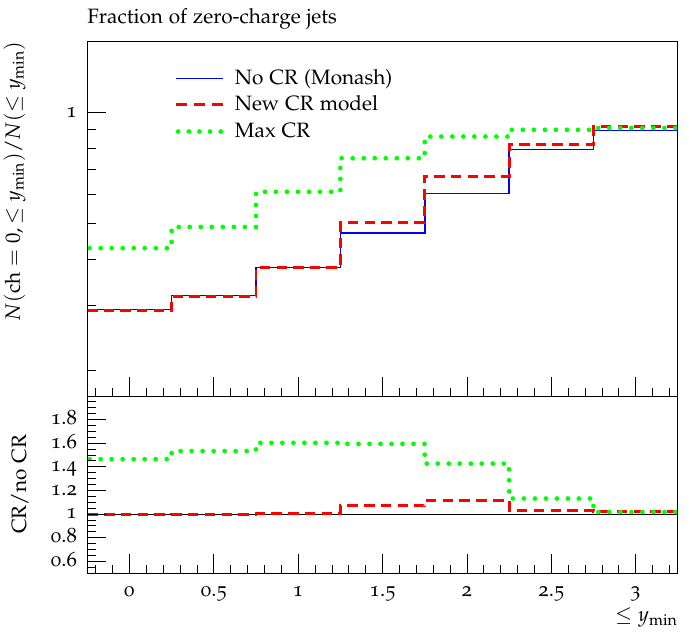}\!}
\caption{\label{fig:lep}Observables constraining CR in 3-jet events at
  LEP. (\subref{fig:lepa}): 
  the minimum rapidity of the constituents of the third jet with
  respect 
  to its jet axis. (\subref{fig:lepb}): the 
  fraction of the third jet with total charge equal to zero as a function of
  minimum rapidity of particles in the third jet.}
\end{figure}

Rather than comparing our model directly on the LEP data, we took a slightly simpler approach by 
only considering the difference between no CR and the new CR model in the relevant observables.
The difference is found to be negligible on those observables, cf.~the
``No CR'' and ``New CR'' histograms in \figRef{fig:lep}. Thus
we do not expect that the new CR model could be ruled out by these LEP
constrains. We note that the small difference between with and without   
CR can be understood, by remembering the large focus on junction structures in the new CR model.
Junction structures do not produce colour-singlet jets in the same manner as ordinary strings,
and thus are not sensitive to this observable in the same way as ordinary strings reconnections.
It is possible to consider a more extreme version of the new CR model where all dipoles are allowed 
to connect with each other, i.e.\
effectively replacing $1/N_C^2$ by unity! This is illustrated by the
``max CR'' histogram in \figRef{fig:lep}. For this unphysically extreme case the difference between the two models 
becomes so large it most likely would have been ruled out by the experiments. 
However such an extreme case would also be eliminated by just considering LHC measurements 
(e.g. $\langle p_\perp \rangle$ vs $n_{\text{ch}}$). 

\subsection{Suggestions for New Observables\label{sec:newObs}}

As discussed in \secRef{sec:pp}, the new CR model is able to reach
agreement with some key observables that have otherwise proved
difficult for the string model (as implemented in PYTHIA) in the $pp$
environment. In particular, subleading dipole connections that
minimise the string-length measure can account for the rise in the
$\left<p_\perp\right>(n_\text{ch})$ distribution (a feature also
present in earlier CR models in PYTHIA, though without the connection with
subleading colour), and subleading string-junction connections can 
account for the observed increase in e.g.\ the $\Lambda/K$ ratio between
$ee$ and $pp$ collisions. 

\begin{figure}[tp]
\centering
\!\includegraphics*[scale=0.78]{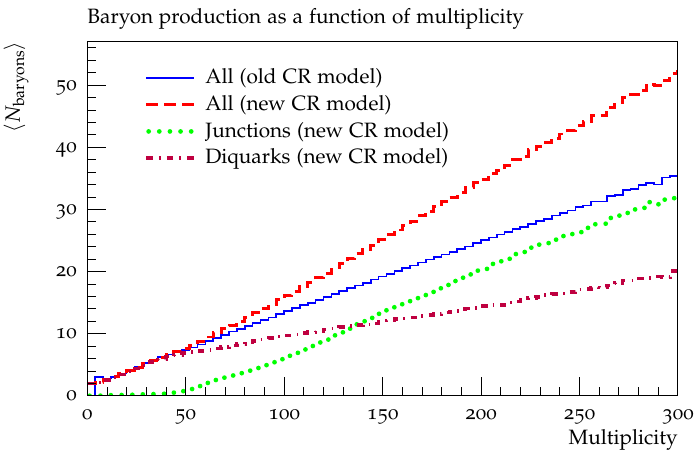}\!
\caption{\label{fig:baryons} average baryon multiplicity as a
  function of hadron
  multiplicity (generator-level, including all particles, hadronic
  decays switched off).}
\end{figure}
The price is (a few) new free parameters governing the CR modelling, so the question naturally arises to what extent
this type of model can be distinguished clearly from other other
phenomenological modelling attempts to describe the same
data. All models we are aware of that simultaneously aim to describe both the
LEP data and the LHC data (or $ee$ and $pp$ data more generally)
rely on the higher colour/energy densities present in $pp$ collisions
to provide the extra baryons. 
The multiplicity scaling of the baryon production is therefore expected
to be higher  
than the linear scaling of the diquark model. This is also what we
observe, cf.~\figRef{fig:baryons}.
For low multiplicity, both of the CR models agree with each other, however the increase 
happens faster for the new model. This shape difference in the scaling
with particle multiplicity could provide an additional probe to test the new model. 

One also notices that there is a significant difference between the baryon production of the old model and 
baryon production from diquarks in the new model. This is somewhat surprising since 
the hadronisation model is essentially left untouched. The explanation for this is two-fold: 
1) the new CR model produces a different mass spectrum of strings
(with generally lower invariant masses), and 2) low-mass strings and junction structures 
 produce fewer additional diquarks. 

The first point is illustrated in 
\figRef{fig:stringMass}, which shows the invariant-mass distribution
of strings in the new and old model. In the old model, the
distribution is essentially flat, and includes a significant plateau 
towards very large invariant masses, whereas the distribution is
strongly peaked at small invariant masses in the new model.
The differences arise  
both from the junction cleanup procedure (by which longer strings can be
split by insertion of an additional quark-antiquark pair), and from the
minimisation of the $\lambda$-measure.  
The old model also minimised the $\lambda$-measure, however this is
achieved by combining strings, giving fewer but higher-mass string systems
than before the CR.

\begin{figure}[tp]
\centering
\captionsetup[subfigure]{skip=3pt}
\subcaptionbox{\label{fig:stringMass}}{\!\includegraphics*[scale=0.64]{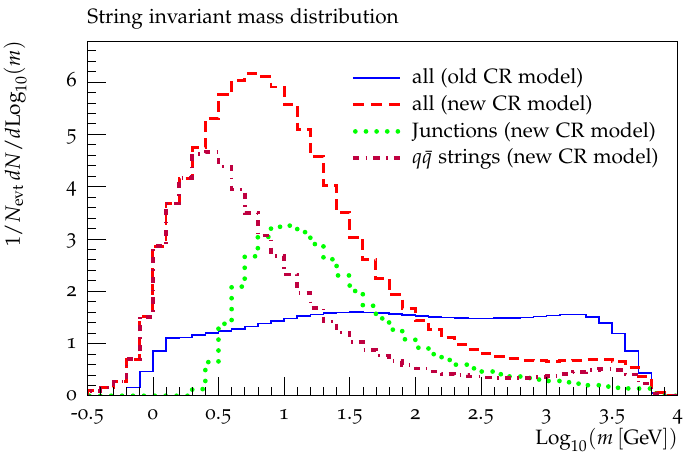}\!}
\subcaptionbox{\label{fig:diquarkProb}}{\!\includegraphics*[scale=0.64]{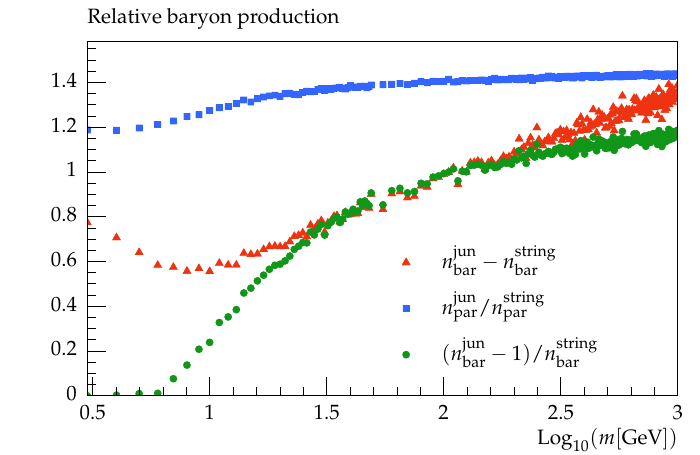}\!}
\caption{(\subref{fig:stringMass}): invariant mass distribution of string systems
 (note logarithmic $x$ axis). (\subref{fig:diquarkProb}): the production of baryons 
 from respectively a $q\bar{q}$ system (``string'') 
and a $qqq$ system (``junction'') 
with the same total invariant mass. The red triangles show 
 the difference of total primary-baryon multiplicities, the blue squares show
 the ratio of total primary-hadron multiplicities (mesons+baryons), and the 
 green circles show the ratio for the primary-baryons multiplicities,
 subtracting off the extra baryon that the junction topology always produces.}
\end{figure}
 Due to energy-momentum  
conservation (and a greater relative importance of the quark endpoints), 
low-invariant-mass strings produce fewer baryons. 
Despite the fact that each junction system produces at least one baryon,
we therefore note that this does  
not automatically lead to an increase in the total number of produced
baryons. Since the invariant mass  
of a $qqq$ junction system is distributed on three string pieces, whereas
that of a $q\bar{q}$ system is carried  
by a single string, diquarks, which are relatively heavy, can actually
be quite strongly phase-space suppressed  
in junction topologies, especially at low invariant masses of the string system (where the majority of the 
junction topologies lie cf.~\figRef{fig:stringMass}). In addition the diquarks
need to be pair produced to conserve baryon number, and the current
implementation requires the pair to be on the same junction leg, leading to an even
larger phase-space suppression.
This effect is illustrated in \figRef{fig:diquarkProb}, 
where for instance a junction system with $E_\text{CM}= 10~\text{GeV}$
(in the peak of the mass distribution) has a five
times lower  
probability to produce an (extra) diquark pair compared to a 10-GeV dipole
string. At fixed multiplicity this effect is hidden in the tuned
parameters, but can be observed by the different scaling. 

Considering baryon production in more detail, the relative yield of
different types of baryons is highly revealing. A
collection of such yields for the different models are listed in
\tabRef{tab:pID}. For most of the 
baryons, the new CR model predicts about $20-50\%$ above the old model. 
This is in agreement with result for $\Lambda$ production shown earlier
(\figRef{fig:lhc_constrains}). 
There are however also some clear order-of-magnitude differences,
for charm and bottom baryons. 
One example is $\Sigma^{0}_c$ production\footnote{The  $\Sigma^{0}_c$ is
a $cdd$ state with spin $S=1/2$, mass $\sim~2.5~\text{GeV}$ and PDG code
4112~\cite{pdg2012}.}, for which the new CR model 
predicts a rate more than a factor of 20 above that of the old model!   

\begin{table}[tp]
  \centering
  \begin{tabular}{c|ccc|c}
    Particle & \multicolumn{3}{|c|}{New CR model ($N_{\text{par}} / N_{\text{events}}$)} 
    & Old CR model \\
    & string & junction & all & $N_{\text{par}} / N_{\text{events}}$ (all) \\
    \hline
    $\pi^+$ & $2.5 \cdot 10^{1} $ & $ 0 $ & $ 2.5 \cdot 10^{1} $ & $ 2.4 \cdot 10^{1} $ \\ 
    $p$ & $2.5 $ & $ 1.4 $ & $ 3.8 $ & $ 3.2 $ \\ 
    $n$ & $2.4 $ & $ 1.3 $ & $ 3.7 $ & $ 3.2 $ \\ 
    $\Delta^{++}$ & $6.1 \cdot 10^{-1} $ & $ 4.5 \cdot 10^{-1} $ & $ 1.1 $ & $ 8.9 \cdot 10^{-1} $ \\ 
    $\Delta^+$ & $6.0 \cdot 10^{-1} $ & $ 4.0 \cdot 10^{-1} $ & $ 1.0 $ & $ 8.6 \cdot 10^{-1} $ \\ 
    $\Delta^{0}$ & $5.5 \cdot 10^{-1} $ & $ 4.0 \cdot 10^{-1} $ & $ 9.4 \cdot 10^{-1} $ & $ 7.9 \cdot 10^{-1} $ \\ 
    $\Delta^{-}$ & $4.7 \cdot 10^{-1} $ & $ 4.4 \cdot 10^{-1} $ & $ 9.1 \cdot 10^{-1} $ & $ 7.1 \cdot 10^{-1} $ \\ 
    \hline
    $K^+$ & $5.2  $ & $ 0 $ & $ 5.2  $ & $ 5.1 $ \\ 
    $\Lambda$ & $4.7 \cdot 10^{-1} $ & $ 3.9 \cdot 10^{-1} $ & $ 8.6 \cdot 10^{-1} $ & $ 6.5 \cdot 10^{-1} $ \\ 
    $\Sigma^+$ & $3.4 \cdot 10^{-1} $ & $ 4.2 \cdot 10^{-1} $ & $ 7.6 \cdot 10^{-1} $ & $ 5.1 \cdot 10^{-1} $ \\ 
    $\Sigma^0$ & $3.5 \cdot 10^{-1} $ & $ 4.5 \cdot 10^{-1} $ & $ 7.9 \cdot 10^{-1} $ & $ 5.1 \cdot 10^{-1} $ \\ 
    $\Sigma^-$ & $3.2 \cdot 10^{-1} $ & $ 4.2 \cdot 10^{-1} $ & $ 7.4 \cdot 10^{-1} $ & $ 4.9 \cdot 10^{-1} $ \\ 
    $\Sigma^{\ast +}$ & $9.6 \cdot 10^{-2} $ & $ 8.9 \cdot 10^{-2} $ & $ 1.9 \cdot 10^{-1} $ & $ 1.5 \cdot 10^{-1} $ \\ 
    $\Sigma^{\ast 0}$ & $9.2 \cdot 10^{-2} $ & $ 7.7 \cdot 10^{-2} $ & $ 1.7 \cdot 10^{-1} $ & $ 1.4 \cdot 10^{-1} $ \\ 
    $\Sigma^{\ast -}$ & $8.3 \cdot 10^{-2} $ & $ 8.7 \cdot 10^{-2} $ & $ 1.7 \cdot 10^{-1} $ & $ 1.3 \cdot 10^{-1} $ \\ 
    $\Xi^-$ & $6.9 \cdot 10^{-2} $ & $ 1.1 \cdot 10^{-1} $ & $ 1.8 \cdot 10^{-1} $ & $ 1.1 \cdot 10^{-1} $ \\ 
    $\Omega^-$ & $2.0 \cdot 10^{-3} $ & $ 1.3 \cdot 10^{-2} $ & $ 1.5 \cdot 10^{-2} $ & $ 3.9 \cdot 10^{-3} $ \\ 
    \hline
    $D^+$ & $5.3 \cdot 10^{-2} $ & $ 0 $ & $ 5.3 \cdot 10^{-2} $ & $ 6.5 \cdot 10^{-2} $ \\ 
    $\Lambda^+_c$ & $4.0 \cdot 10^{-3} $ & $ 7.9 \cdot 10^{-3} $ & $ 1.2 \cdot 10^{-2} $ & $ 6.6 \cdot 10^{-3} $ \\ 
    $\Sigma^{++}_c$ & $2.7 \cdot 10^{-4} $ & $ 1.3 \cdot 10^{-2} $ & $ 1.3 \cdot 10^{-2} $ & $ 5.4 \cdot 10^{-4} $ \\ 
    $\Sigma^{+}_c$ & $2.5 \cdot 10^{-4} $ & $ 1.5 \cdot 10^{-2} $ & $ 1.5 \cdot 10^{-2} $ & $ 5.2 \cdot 10^{-4} $ \\ 
    $\Sigma^{0}_c$ & $2.5 \cdot 10^{-4} $ & $ 1.3 \cdot 10^{-2} $ & $ 1.3 \cdot 10^{-2} $ & $ 5.1 \cdot 10^{-4} $ \\ 
    $\Sigma^{\ast ++}_c$ & $5.1 \cdot 10^{-4} $ & $ 1.7 \cdot 10^{-3} $ & $ 2.2 \cdot 10^{-3} $ & $ 9.5 \cdot 10^{-4} $ \\ 
    $\Sigma^{\ast +}_c$ & $4.9 \cdot 10^{-4} $ & $ 1.9 \cdot 10^{-3} $ & $ 2.4 \cdot 10^{-3} $ & $ 9.4 \cdot 10^{-4} $ \\ 
    $\Sigma^{\ast 0}_c$ & $4.8 \cdot 10^{-4} $ & $ 1.7 \cdot 10^{-3} $ & $ 2.2 \cdot 10^{-3} $ & $ 9.1 \cdot 10^{-4} $\\ 
    $ccq^7$ & $ 0 $ & $ 2.1 \cdot 10^{-4} $ & $ 2.1 \cdot 10^{-4} $ & $ 1.0 \cdot 10^{-7} $ \\
    \hline
    $B^+$ & $1.6 \cdot 10^{-3} $ & $ 0 $ & $ 1.6 \cdot 10^{-3} $ & $ 2.3 \cdot 10^{-3} $ \\ 
    $\Lambda^0_b$ & $1.9 \cdot 10^{-4} $ & $ 6.3 \cdot 10^{-4} $ & $ 8.2 \cdot 10^{-4} $ & $ 3.9 \cdot 10^{-4} $ \\ 
    $\Sigma^+_b$ & $1.1 \cdot 10^{-5} $ & $ 9.3 \cdot 10^{-4} $ & $ 9.5 \cdot 10^{-4} $ & $ 3.1 \cdot 10^{-5} $ \\ 
    $\Sigma^0_b$ & $1.2 \cdot 10^{-5} $ & $ 1.0 \cdot 10^{-3} $ & $ 1.0 \cdot 10^{-3} $ & $ 3.7 \cdot 10^{-5} $ \\ 
    $\Sigma^-_b$ & $1.1 \cdot 10^{-5} $ & $ 9.3 \cdot 10^{-4} $ & $ 9.4 \cdot 10^{-4} $ & $ 3.2 \cdot 10^{-5} $ \\ 
    $\Sigma^{\ast +}_b$ & $1.1 \cdot 10^{-5} $ & $ 9.3 \cdot 10^{-4} $ & $ 9.5 \cdot 10^{-4} $ & $ 3.1 \cdot 10^{-5} $ \\ 
    $\Sigma^{\ast 0}_b$ & $1.2 \cdot 10^{-5} $ & $ 1.0 \cdot 10^{-3} $ & $ 1.0 \cdot 10^{-3} $ & $ 3.7 \cdot 10^{-5} $ \\ 
    $\Sigma^{\ast -}_b$ & $1.1 \cdot 10^{-5} $ & $ 9.3 \cdot 10^{-4} $ & $ 9.4 \cdot 10^{-4} $ & $ 3.2 \cdot 10^{-5} $ \\ 
    $bcq^7$ & $ 0 $ & $ 1.8 \cdot 10^{-5} $ & $ 1.8 \cdot 10^{-5} $ & $ 0 $ \\
    $bbq^7$ & $ 0 $ & $ 1.1 \cdot 10^{-6} $ & $ 1.1 \cdot 10^{-6} $ & $ 0 $ 
    
  \end{tabular}
  \caption{\label{tab:pID}Primary particle production
    of identified hadrons at 7 TeV over the full $y$ and $p_\perp$
    range, summed over particles and antiparticles. 
    Ten million ND events were simulated and hadron decays
    were switched off to focus on the primary 
    production.  \\$^7$Double heavy baryons where the last q can be any quark. } 
\end{table}
\afterpage{\clearpage}
To understand how such large differences can occur, we need to recall 
how baryons are produced in ordinary string fragmentation. 
Since no charm is produced in string break-ups, the only way to produce 
a $\Sigma^{0}_c$ is to produce a $dd$-diquark and combine it with $c$ quark
from the shower.  
 However, since $dd$ diquarks must have spin 1 (due to symmetry), their
 production is heavily suppressed relative to $ud$ ones which can also
 exist in the far lighter spin-0 state. Enter junctions, for which there
 is a priori no specific penalty associated with having two legs end on
 same-flavour quarks as compared to different flavours. Up to some
 combinatorics   
and symmetry factors the production of $ud$ and $dd$ is therefore expected to be
the same for junction systems, in good agreement with the
observed predictions of the new CR model. As such the production of
baryons like $\Sigma^{0}_c$ theoretically provides an excellent probe to
study the relative importance between diquark- and junction-driven
baryon production. 

Similar results are also observed for $\Sigma^{++}_c$ as well as for the
production of the analogous $b$-baryons. And with the large $b$-physics
programme at the LHC, we believe this could be an interesting study, both
for its physics value as well as a possible source of background for  
other measurements. The effect can also be seen for $\Omega^-$, however
not as clearly as for charm and bottom baryons. Moreover the presence of
additional suppression of strange diquarks in the string fragmentation
model makes $\Omega^-$ connection more complicated. We note that it can
be an important validation channel however.  

Due to the majority of baryons being produced by the junction
mechanism in the new CR model, the baryon yields also provides a clear
probe to test the spin structure of the diquarks formed from the
junctions. The large difference in yield between $\Sigma^{\ast +}_c $
and $\Sigma^{0}_c$ is due to the choice of spin suppression
mentioned earlier in the tuning section. An actual measurement could be 
directly applied as a constraint for this variable. At least with the
development of the present model, we now have a vehicle that allows to 
explore this type of phenomenology and interpret the findings. 

We should note that the  baryons considered above are excited states
that rapidly decay through the emission of a pion, e.g.\
$\Sigma_c^0\to\Lambda_c^+ \pi^-$. As such they may be quite challenging
to observe experimentally. It is therefore not a given that it will be
easy to utilise measurements of these yields. But it does provide a
theoretical motivation for studying the production and measurement of
heavy baryons. 

Another special class of baryons is the double (or triple) heavy baryons
containing at least two c or b quarks. These baryons can not be formed in
ordinary string fragmentation and is therefore almost non-existent in the old
CR scheme. The only production mechanism is via the junctions from the beams (which
also means that for $p^+p^+$ collisions no double-heavy antibaryons are predicted). This
is also observed in \tabRef{tab:pID}, where only a single double-heavy baryon
is produced in the 10 million events. With the large amount of junctions, the
new CR model provides a natural production mechanism for 
double-heavy baryons, and as such the expected amount is also significantly
higher than for the old CR model. The effect of massive quarks in the
$\lambda$-measure is not well understood, however, and possible other production
mechanism might also contribute, thus the estimate is most likely rather
crude. Irrespectively, a measurement of double-heavy baryons
probes a region of hadronisation that the current models do not
describe. And it could possible also shed some light on whether the junction
mechanism might be a reasonable production mechanism.

\begin{figure}[tp]
  \centering
  \includegraphics[scale=1.0]{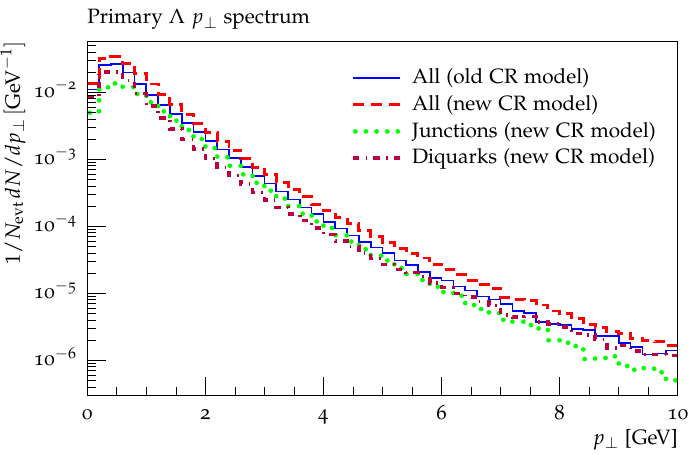}
  \caption{\label{fig:lambdapT}The $\Lambda$ $p_\perp$-distribution separated 
    by production mechanism. Only ND events were included and hadron decays
    were switched off.}
\end{figure}

So far, we only considered total particle yields; more  
knowledge is available by studying more differential distributions. A
natural next extension is  
the transverse momentum distributions. As the junctions are formed by
minimising the  
$\lambda$-measure, the particles defining the junction may be expected to
preferentially be moving in the same direction and thereby create a
boosted baryon. This in turn leads to an expected  
increase in transverse momentum for such junction baryons. This is also
observed in the low-$p_\perp$ region (below roughly $p_\perp\sim
4~\text{GeV}$), where the particle production 
peak is higher for junction baryons (\figRef{fig:lambdapT}).
In the region of very high $p_\perp$  (above roughly $p_\perp\sim
4~\text{GeV}$) the particle production is dominated by jets, for which
the hard high-$p_\perp$ partons are more important than the overall
boost. In addition, the perturbative gluons  
associated with the jet already provides a low $\lambda$-measure and as
such limited CR is expected inside the jet regions. This leads to the
high-$p_\perp$ region being occupied predominantly by baryons
produced in ordinary (diquark) string-breaks.

\begin{figure}[tp]
  \centering
  \captionsetup[subfigure]{skip=3pt}
  \subcaptionbox{\label{fig:lambdapTa}}{\!\includegraphics[scale=0.64]{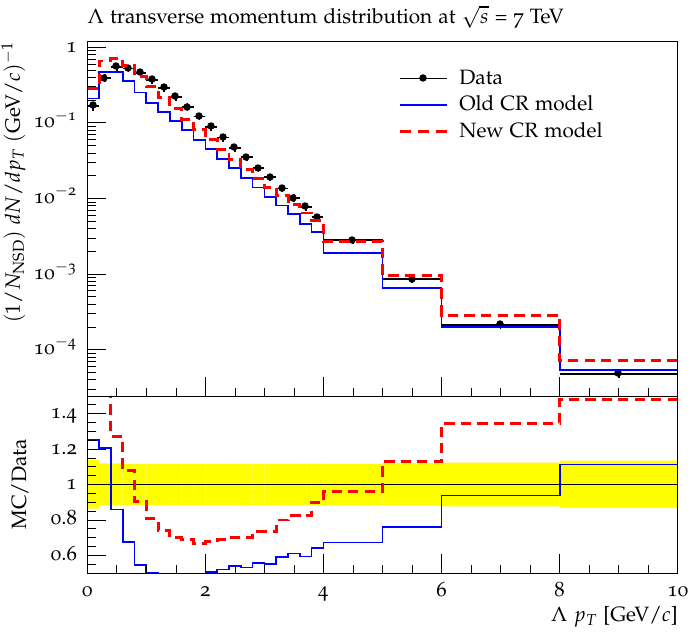}\!}
  \subcaptionbox{\label{fig:lambdapTb}}{\!\includegraphics[scale=0.64]{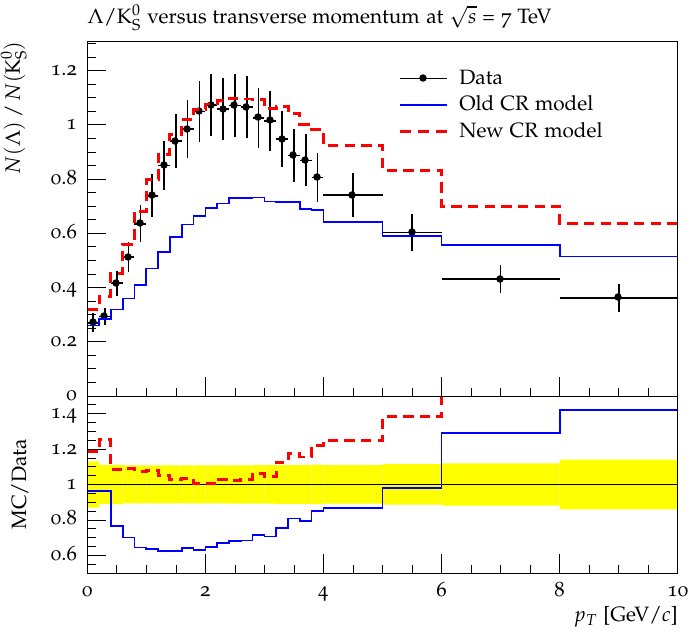}\!}
  \caption{\label{fig:lambdapTdata} The (\subref{fig:lambdapTa})
    $\Lambda$ $p_\perp$-distribution  and
    (\subref{fig:lambdapTb}) the $\Lambda / K_s^0$ $p_\perp$-distribution as
    measured by the CMS experiment~\cite{Khachatryan:2011tm}. All PYTHIA 
    simulations were NSD with a lifetime cut-off ($\tau_\text{max} = 10$
 mm/c) and a rapidity cut on 2 ($|y| < 2$). The yellow error band represents the
      experimental $1\sigma$ deviation.}
\end{figure}

Transverse momentum spectra have already been measured for some of the
more common  baryon species and a comparison with the $\Lambda$
$p_\perp$ spectrum measured by CMS is given in in
\figRef{fig:lambdapTdata}. Sadly, the improvement is far less
satisfactory here. The new CR model (as well as the old model)
overshoots the production in the very low $p_\perp$ and the high
$p_\perp$ region, whereas too few $\Lambda$ baryons are predicted in
between. Thus the $\Lambda$ baryons from junctions 
tend to fall in the right region, however the effect is not large enough.
An interesting observation is that the \emph{ratio} $\Lambda / K_s$ is now well
described in the low $p_\perp$ region. This shows that the problem with
the $p_\perp$ distribution is not specific to baryons but is more
generic. The discrepancy between data and the model for large $p_\perp$
still exists, however 
the baryon production in this region is primarily from diquark string
breaks in jets and as such is not really unique for the new CR model. It
may point to a 
revision needed of the spectrum of hard (leading?) baryon production in jets,
which may not be unique to the $pp$ environment, see~\cite{Skands:2014pea}.

\begin{figure}[tp]
  \centering
  \includegraphics[scale=0.655]{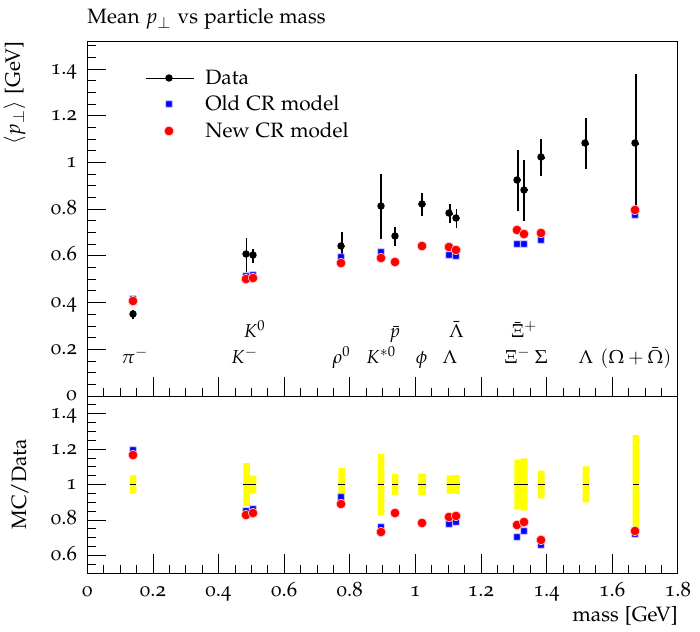}
  \caption{\label{fig:avgpTvsMass} Average $p_\perp$ 
    versus hadron mass at $E_{\mrm{CM}}\,=\,200~\mathrm{GeV}$, compared
    with STAR data~\cite{Abelev:2006cs}. The yellow error band represents the
      experimental $1\sigma$ deviation.}
\end{figure}

The problem in the low $p_\perp$ domain is a common theme for all heavier 
hadrons (i.e.\ anything but pions) and would be interesting to explore
further. (E.g., a measurement of $\rho$ spectra could reveal whether it
depends on the presence of strange quarks.) The PYTHIA models predict a 
$p_\perp$-distribution that peaks at lower values than what is actually
observed. To study this in more detail, one can  
calculate the average $p_\perp$ for the identified hadrons and plot it a
function of their mass, as done e.g.\ by the STAR collaboration for $pp$
collisions at 
$E_\text{CM}=200~\text{GeV}$~\cite{Abelev:2006cs}. In purely
longitudinal string fragmentation the 
expected result is a roughly flat curve, since no correlation between  
the mass of the particle and $p_\perp$ is present. The flat prediction is
altered when hadron decays and jet physics are included, leading to the curve
seen in \figRef{fig:avgpTvsMass}. The prediction is also altered if the string
is boosted (e.g., by partonic string endpoints), the boost is transferred to the
final particles and for the same boost velocity a heavy particle will
gain more $p_\perp$ than a light one. This effect can be enhanced by CR,
since minimisation of the 
$\lambda$-measure prefers reconnections among partons moving in the same
direction, thus creating boosted strings~\cite{Ortiz:2013yxa}. CR is
therefore expected to give a sharper rise of the $\langle p_\perp
\rangle$ vs mass distribution. 
Unfortunately, we do not observe this expected effect at any significant
level (\figRef{fig:avgpTvsMass}). To be candid, it is disappointing 
that the new model does not appear to address this
problem at all. At the very least, it leaves room open for criticism and
possibly additional new physics. Of special interest in this context are
possible collective phenomena, such as (gas-like) hadron reinteractions
or (hydro-like) flow, either of which could provide a (weak or strong,
respectively) velocity-equalising component, and at least the latter has
been applied successfully in the context of the EPOS
model~\cite{Pierog:2013ria}. So-called ``colour ropes'' (strings
carrying more than one unit of charge and hence having a higher tension)
can also generate harder momentum spectra, while remaining within a
string context, as was recently explored in~\cite{Bierlich:2014xba}.
These effects are generally not  
expected to be present at the relatively low energy density in $pp$
collisions at a centre-of-mass energy of 200 GeV, however. It would therefore be
of great interest to redo the measurement in detail at LHC energies, for high-
and low-multiplicity samples and/or in the underlying event, to study
whether the slope is steeper at the higher
energy densities. 

\begin{figure}[tp]
  \centering
 \captionsetup[subfigure]{skip=3pt}
 \subcaptionbox{\label{fig:rapGapa}}{\!\includegraphics[scale=0.64]{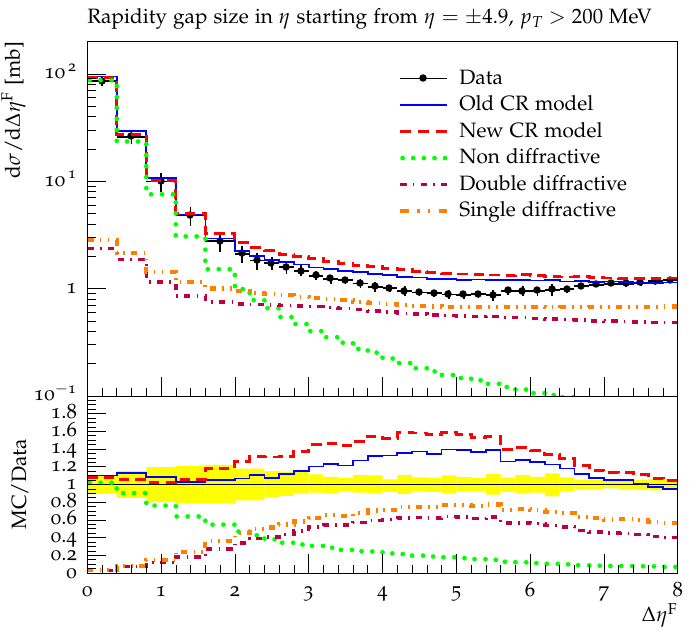}\!}
 \subcaptionbox{\label{fig:rapGapb}}{\!\includegraphics[scale=0.64]{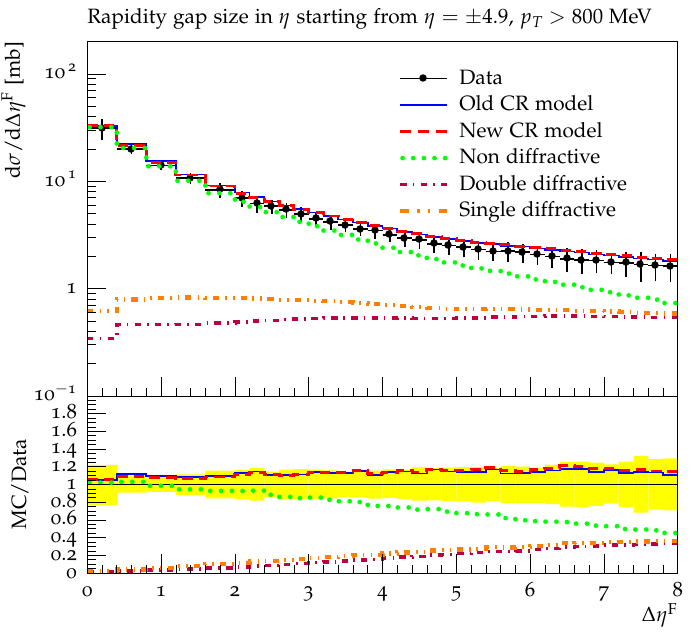}\!}
  \caption{\label{fig:rapGap} The rapidity gap survival for a low $p_\perp$ cut (\subref{fig:rapGapa}) and a 
    high $p_\perp$ cut (\subref{fig:rapGapb}). The different components are also shown for the new CR model 
    (DD/SD/ND). The gathered data is from the ATLAS
    experiment~\cite{Aad:2012pw}. The yellow error band represents the
      experimental $1\sigma$ deviation.}
\end{figure}

Finally, we emphasise that the rapidity gap survival is heavily affected
by the choice of CR model~\cite{Edin:1995gi,Rathsman:1998tp}. 
The explanation is similar to that of the three-jet LEP measurement, 
where the reconnected colour-singlet jet produces a rapidity gap to the
other jets. 
The old CR model combines different strings into a single larger string,
and thereby covers the same rapidity span, essentially adding kinks to
an already existing string topology. Instead the new model can
produce the colour-singlet union of particles in the same rapidity
region. The new model is therefore expected to produce  
more rapidity gaps compared with the old model, which is also what we
observe in \figRef{fig:rapGap}.  
The new model predictions are significantly above the data in the mid-range rapidity 
region for low $p_\perp$ cut-offs. (For higher $p_\perp$ cuts the effect 
vanishes, due to the partonic description being more influential on the
rapidity gap survival.) 
We therefore emphasise that the new CR scenario should in principle be
accompanied by a retuning of diffractive parameters. This
is not straightforward however, and involves not only the shown
rapidity-gap survival distributions, but several other measurements at
different energies. It was therefore deemed beyond the scope of this study 
to perform a retuning of the diffractive components and we limit
ourselves to pointing out the interplay.  
A future dedicated study of this aspect could also well incorporate a
study of the interference between the new BR model and diffractive
events.  

\section{Application to Top Mass Measurements at Hadron Colliders \label{sec:top}}
Colour reconnections contribute one of the dominant uncertainties on current
experimental top-mass extractions in hadronic
channels (see e.g.\ the mini-review in \cite{Juste:2013dsa}), and their size
was recently reexamined in the context of several simplified CR
schemes~\cite{Argyropoulos:2014zoa}. 
It is thus interesting to consider our new CR
model in the same framework. We follow ref.~\cite{Argyropoulos:2014zoa},
to which we refer for details on selection of the events and top mass
reconstruction procedure.  

Briefly stated, the idea is to select semi-leptonic top events, using
the charged lepton and escaping neutrino mainly for tagging and then
reconstruct the top mass from the hadronic decay.    
A mass window around the W mass is required and the raw top mass is extracted
by fitting the invariant mass distribution of the three jets with a
skewed Gaussian distribution, which fits the distributions better than a
standard Gaussian~\cite{Argyropoulos:2014zoa}. For the
models/tunes considered in this work (details below), the resulting shift on
the calibrated top mass is below 200 MeV,
which is comparable with the current level of CR uncertainty on the
measurements~\cite{Aaltonen:2012va,Chatrchyan:2012cz,Abazov:2014dpa,Aad:2015nba} 
and far below some of the ``devil's-advocate'' toy models
considered in~\cite{Argyropoulos:2014zoa}.  Characteristic for those models is
that they allow some fraction of reconnections where the $\lambda$ measure is
increased, whereas all the models considered here involve a minimisation of
$\lambda$, one way or another. 

We compare six models: no CR, the existing (default) PYTHIA 8 CR model with and without
early resonance decays (ERD), the new baseline CR model with and without ERD,
and the new CR with maximal CR and ERD.   
Since the largest effect is expected for ERD, the maximal CR scenario is
only considered with ERD switched on. Two versions of the ``no CR''
scenario were considered, one in which CR was simply switched off
without any further changes (resulting in significant increase in central hadron
multiplicity) and the other a semi-tuned version, where the activity ($\sum
E_\perp$) in the central region ($|\eta| < 3$) for ND events was retuned. A
similar approach is used for the maximal CR models, where again a non-tuned
version and retuned version are considered. It should be noted that neither of
the retuned models provide a good description of all MB data (for instance
$\left<p_\perp\right>(n_\mathrm{charged})$ is described by neither tune).
Since none of the models considered exhibit any top-specific behaviour, no
additional retuning to top events was needed. The results are collected in
\tabRef{tab:topMass}. 

\begin{table}[!tp]
  \centering
  \begin{tabular}{|cccc|}
    \hline
    Model & $\hat{m}_{\text{top}}$ [GeV] & $\Delta\hat{m}_{\text{top}}$ [GeV] &
    $\Delta\hat{m}_{\text{top}}^{\text{rescaled}}$ [GeV] \\
    \hline
    no CR         & $169.57 \pm 0.06$ & 0 & 0 \\
    default ERD   & $169.26 \pm 0.06$ & $-0.36 \pm 0.09 $ & $-0.04 \pm 0.10$ \\
    default       & $168.95 \pm 0.06$ & $-0.67 \pm 0.09 $ & $+0.17 \pm 0.10$ \\
    new model ERD & $169.18 \pm 0.06$ & $-0.45 \pm 0.09 $ & $+0.03 \pm 0.10$ \\
    new model     & $168.97 \pm 0.06$ & $-0.66 \pm 0.09 $ & $+0.14 \pm 0.11$ \\
    max CR ERD    & $169.41 \pm 0.06$ & $-0.22 \pm 0.09 $ & $-0.05 \pm 0.10$ \\
    no CR (tuned) & $168.99 \pm 0.06$ & $-0.63 \pm 0.09 $ & $-0.06 \pm 0.10$ \\
    max CR (tuned)& $170.28 \pm 0.07$ & $+0.66 \pm 0.09 $ & $+0.06 \pm 0.11$ \\

    \hline

  \end{tabular}
  \caption{\label{tab:topMass}Values of $m_{\text{top}}$ as predicted by the different CR models. 
    The rescaled top mass is obtained by 
    $\hat{m}_{\text{top}}^{\text{rescaled}} =\frac{80.385}{\hat{m}_{W}} \hat{m}_{\text{top}}$.}
\end{table}

The first observation is
the large difference between the no CR model and the non ERD models for
$\Delta\hat{m}_{\text{top}}$ ($\sim 650$ MeV). Since no CR is performed for the top
decay products, not much of difference was expected, and any difference observed has to reside
entirely in the underlying event (UE). The CR  
models considered lower the total string length, and thereby the
activity in the UE, which directly influences the non-scaled top mass. Since
the ``no CR'' model uses the same tune parameters as the CR models,
it has a too high activity in the UE, leading to the negative mass
shifts seen. After retuning to a similar activity in the central
region, the ``no CR (tuned)'' model agrees with the default CR
scenario within the statistical uncertainty.  This emphasises the importance of using
consistent UE tunes for this type of exercise, though we also note
that after recalibration by the hadronic $W$ mass, the rescaled top
mass $\hat{m}^\mrm{rescaled}_\mrm{top}$ is remarkably stable.  

For the ERD models the above shift in UE is still present, but the UE is now
also allowed to reconnect with the top decay products. Therefore, an UE parton
in close proximity to a jet from the top decay will have a large probability to be
reconnected with the jet. This will result in narrower jets, leading to less of
the energy falling outside the jet cone, and thereby a larger top mass. This is
in agreement with the simulations, where the ERD results are above the non-ERD
results. But the effect is smaller than that of the UE activity, thus the
overall shift is still negative compared to the non-tuned no CR model.

Both of the above effects are magnified in the maximal CR scenario. We
remind the reader that in this scenario, the $1/N_C^2$ suppression of
subleading connections is switched off, hence this should be
considered an unphysical extreme variation. 
Coincidentally, however, the two effects end up approximately canceling. A
similar retuning of the central activity as above, ``max CR (tuned)'',
shows a significant increase in the top mass shift of more than one
GeV with respect to the tuned no CR model, though again, the $W$ mass
calibration removes most of it.

The fact that the rescaled top mass ($\hat{m}_{\text{top}}^{\text{rescaled}}$) is less
sensitive to CR is due to a cancellation between 
CR in the W mass and the top mass. This is in perfect agreement with the
simulations, where the deviation for all rescaled masses are below their
respective non-rescaled deviations. On the other hand, this means that any
interpretation of where the variations between the models arise becomes
extremely difficult. We will therefore refrain from 
attempting this, and instead purely discuss the numerical values in term of the
uncertainty on the top mass. 

For the rescaled top mass, the differences between the models stay
below 200 MeV. This is slightly less than what was 
observed earlier even for identical CR models
(default)~\cite{Argyropoulos:2014zoa}. The variations in the 
results can be attributed to a new tune combined with a change in the PS for $t\bar{t}$
events. We regard the smaller differences as somewhat coincidental
however, and further work is needed to genuinely improve our
understanding of CR effects in the top mass measurement. What we can
say at least is that the results from the new model lie within those
from the default CR model, and therefore do not 
generate a need for larger uncertainties. Even the maximal CR, which is our
attempt to mimic the very large shifts seen for models that ignore
MB/UE constraints, does not change this picture. Instead a pattern emerges, namely
that whenever the minimisation of the $\lambda$ measure is used as a guideline
for the CR, the shifts stay below 300 MeV (taking the models studied
in~\cite{Argyropoulos:2014zoa} into account as well). The reason for this is
two-fold:   
firstly the coherence of the PS ensures that the jet structure is not too
significantly altered; secondly, the alterations are realised in a
systematic fashion leading to a similar shift in both the top and the W
mass, implying that the hadronic $W$ mass recalibration is highly
robust. An increased understanding of this interplay could potentially lower the
uncertainty even further. 

Since these models are mainly constrained by measurements, further
gains can also be achieved by improving and extending the programme of
measurements sensitive to CR effects. A few new observables targeting
top events specifically were already suggested
in~\cite{Argyropoulos:2014zoa}. In the context of top mass
uncertainties, such observables are of course especially relevant, as this 
is the closest to in-situ constraints as can be obtained, mimising the
``extrapolation'' that the model has to cover between the constraint and
measurement environments. 

In order to establish whether the small effects on $m_t$ predicted by
$\lambda$-minimising models are indeed 
conservative or not, it would be of crucial importance to test these models as directly as
possible in a variety of environments, top included. The fact that our new CR models
do not yet give good descriptions of identified-particle $p_\perp$
spectra should, in this context, be seen as a warning that there can be
additional non-perturbative uncertainties left unaccounted for,
possibly of a dynamical origin.

\section{Summary and Outlook \label{sec:summary}}
The question \emph{``between which partons do confining potentials
  arise?''} is a fundamental one in non-perturbative QCD, which any
attempt at modelling the process of hadronisation must address. In the
leading-colour approximation and neglecting beam-remnant correlations, this is
relatively simple:  
there is a one-to-one mapping between perturbative QCD dipoles and
string pieces / clusters. In this paper, we have attempted to take a
first step beyond leading colour, by including a
randomisation over the set of possible subleading-colour topologies,
with probabilities chosen according to a simplified version of the
$SU(3)$ colour algebra. 

We present the argument that while the LC approximation 
may be quite good in the environment of $e^+e^-$ collisions (more
specifically in the absence of multi-parton interactions), we expect
very significant deviations from it in $pp$ collisions, where the
survival of the strict LC topology should be heavily
suppressed. Although the probability for a subleading colour connection to be possible between any given pair
of (uncorrelated) partons is only roughly $1/N_C^2$, 
it becomes increasingly unlikely \emph{not} to have any such
connections as the number of uncorrelated partons increases, as e.g.\ 
in the case when considering MPI. 

This implies that a complex multi-parton system will in general have several
different string/cluster configurations open to it, at the time of
hadronisation; the LC one is only one among many possibilities. 
We invoke the string-length $\lambda$ measure to choose which one is
preferred, so far via a simple winner-takes-all algorithm that does not
purport to always find the global minimum. Nonetheless, we believe that this
model represents a 
significant step in the right direction, allowing us to probe for the
first time the effects of subleading colour on hadronisation in a way
that may be said to be systematic and consistent with (a simplified
version of) QCD. 

One noteworthy new aspect of our work is the use of string junctions to
represent antisymmetric colour combinations, such as two colours
combining to form an overall anticolour. This provides a new source of baryon
production, with properties qualitatively different from the standard
diquark scenario. We have shown that this aspect allows to reconcile 
measured baryon/meson ratios with the string model in both $pp$ and
$ee$ collisions simultaneously. However, we caution that the shapes of
the $p_\perp$ spectra are still not well described. We had anticipated
that the preference of reconnections to produce boosted string pieces
should lead to an enhancement of the $\left<p_\perp\right>$ especially
for heavier hadrons, but the magnitude of this effect observed in our
model is still far too low to explain the data. 

We emphasised that there is a conceptual difference between 
colour \emph{connections} and colour \emph{reconnections}. 
The former is related to colour-space \emph{ambiguities}, such as the
unknown colour 
correlations between different MPI initiators or the subleading-colour
connections explored in this work. Colour \emph{reconnections} are
related to \emph{dynamical} reconfigurations of the
colour/string space, via  perturbative gluon exchanges or
non-perturbative string interactions; i.e., they involve \emph{momentum}
exchange as well. We 
did not explore effects of the latter type directly in 
this work, though we note that the fairly realistic string-interaction
scenarios constructed by 
Khoze and Sj\"ostrand in the context of $e^+e^-\to W^+W^-$
studies~\cite{Sjostrand:1993hi,Khoze:1994fu,Khoze:1999up} also feature
string-length 
minimisation; hence it is possible that our tuned parameters effectively
attempt to cover both types. If so, the fact that the momentum spectra
remain discrepant may point to the need for dynamical CR. 
 
Finally, we presented a few suggestions for additional observables,
the measurement of which would give further insight and possibly help
to distinguish both physical and unphysical CR models, as well as
other ideas such as models based on colour
ropes~\cite{Andersson:1991er,Bierlich:2014xba}, 
hydrodynamics~\cite{Pierog:2013ria}, or (non-hydro) hadron rescattering. 
We ended by considering the simplified top-mass analysis
of~\cite{Argyropoulos:2014zoa} and conclude that the models presented
here lead to shifts in the top mass of order 200 MeV, which is within
the current level of non-perturbative uncertainties on the
measurements. 

\subsection*{Acknowledgements}
Many thanks to T.~Sj\"ostrand for his valuable comments on both our
physics and our code. JRC thanks the CERN theory unit for hospitality
during the main part of this study. Work supported in part by the
Swedish Research Council, contract 
number 621-2013-4287, in part by the MCnetITN FP7 Marie Curie
Initial Training Network, contract PITN- GA-2012-315877, and in part by
the Australian Research Council, contract FT130100744.

\appendix
\section{Model parameters}
A complete list of all the parameters that differ from the Monash tune for
the three different models are listed in the table below. 
\begin{table}[h!]
  \centering
  \begin{tabular}{lllll}
    Parameter & Monash & Mode 0 & Mode 2 & Mode 3 \\
    \hline
    StringPT:sigma                        & = 0.335 & = 0.335 & = 0.335 & =
    0.335 \\
    StringZ:aLund                         & = 0.68  & = 0.36  & = 0.36  & =
    0.36 \\
    StringZ:bLund                         & = 0.98  & = 0.56  & = 0.56  & = 0.56 \\
    StringFlav:probQQtoQ                  & = 0.081 & = 0.078 & = 0.078 & =
    0.078 \\
    StringFlav:ProbStoUD                  & = 0.217 & = 0.2   & = 0.2   & =
    0.2 \\
    \multirow{4}{*}{StringFlav:probQQ1toQQ0join} & = 0.5, & = 0.0275, & = 0.0275, & = 0.0275, \\
    &  0.7, &  0.0275, &  0.0275, &  0.0275, \\
    &  0.9, &  0.0275, &  0.0275, &  0.0275, \\
    &  1.0 &  0.0275 &  0.0275 &  0.0275 \\
    \hline
    MultiPartonInteractions:pT0Ref        & = 2.28  & = 2.12  & = 2.15  & =
    2.05 \\ 
    \hline
    BeamRemnants:remnantMode              & = 0     & = 1     & = 1     & = 1    \\
    BeamRemnants:saturation               & -       & = 5     & = 5     & = 5    \\
    \hline 
    ColourReconnection:mode               & = 0     & = 1     & = 1     & = 1 \\
    ColourReconnection:allowDoubleJunRem  & = on    & = off   & = off   & =
    off \\
    ColourReconnection:m0                 & -       & = 2.9   & = 0.3   & = 0.3 \\
    ColourReconnection:allowJunctions     & -       & = on    & = on    & = on \\
    ColourReconnection:junctionCorrection & -       & = 1.43  & = 1.20  & = 1.15 \\
    ColourReconnection:timeDilationMode   & -       & = 0     & = 2     & = 3    \\
    ColourReconnection:timeDilationPar    & -       & -       & = 0.18  & = 0.073 \\

  \end{tabular}
\end{table}

\bibliographystyle{utphys}
\bibliography{coherence}

\end{document}